\documentclass[11pt]{fizik}
\usepackage{hyperref}
\hypersetup{
colorlinks=true,
urlcolor=blue,
citecolor=blue}
\usepackage[all]{xy,xypic}
\usepackage{amsfonts,amssymb,amsmath,amsgen,amsopn,amsbsy,theorem,graphicx,epsfig}
\usepackage{eufrak,amscd,bezier,latexsym,mathrsfs,enumerate,multirow}
\usepackage[utf8]{inputenc}
\usepackage[english]{babel}
\usepackage[dvipsnames]{xcolor}
\usepackage{academicons}
\definecolor{orcidlogocol}{HTML}{A6CE39}
\usepackage[pagewise]{lineno}
\usepackage{epstopdf}
\usepackage{xspace}
\usepackage{color}
\usepackage{units}
\usepackage{slashed}
\usepackage{gensymb}
\usepackage{booktabs}
\usepackage{xcolor}
\usepackage{setspace}
\usepackage{physics}
\usepackage{nicefrac}
\usepackage{bbold}
\usepackage{accents}
\usepackage{mathtools}

\DeclareUnicodeCharacter{2009}{\,} 

\def\bt{\begin{equation}}
\def\bea{\begin{eqnarray}}
\def\ee{\end{equation}}
\def\eea{\end{eqnarray}}

\yil{}
\vol{}
\fpage{}
\lpage{}
\doi{}

\title{Exploring the high energy frontiers of the Milky Way \\ with ground-based gamma-ray astronomy: \\
PeVatrons and the quest for the origin of Galactic cosmic-rays} 

\author[E.O. Angüner - PeVatrons and the quest for the origin of Galactic cosmic rays]{                                                        												    
\textbf{Ekrem Oğuzhan ANGÜNER$^{1}$\thanks{oguzhan.anguner@tubitak.gov.tr}}\\ 
$^{1}$TÜBİTAK Research Institute for Fundamental Sciences, 41470 Gebze, Turkey \\               
\\ [1.8em]

\rec{...}
\acc{...}
\finv{...}
}

\newcommand{\bc}{\begin{center}}
\newcommand{\ec}{\end{center}}

\renewcommand{\phi}{\varphi}

\begin{document}

\maketitle
\begin{abstract} 
Cosmic rays (CRs) are charged particles that arrive at Earth isotropically from all directions and interact with the atmosphere. The presence of a spectral knee feature seen in the CR spectrum at $\sim$3~PeV energies is an evidence that astrophysical objects within our Galaxy, which are known as 'Galactic PeVatrons', are capable of accelerating particles to PeV energies. Scientists have been trying to identify the origin of Galactic CRs and have been looking for signatures of Galactic PeVatrons through neutral messengers. Recent advancements in ground-based $\gamma$-ray astronomy have led to the discovery of 12 Galactic sources emitting above 100~TeV energies, and even the first time detection of PeV photons from the direction of the Crab Nebula and the Cygnus region. These groundbreaking discoveries have opened up the field of ultra-high energy (UHE, E$>$100~TeV) $\gamma$-ray astronomy, which can help us explore the high energy frontiers of our Galaxy, hunt for PeVatron sources, and shed light on the century-old problem of the origin of CRs. This review article provides an overview of the current state of the art and potential future directions for the search for Galactic PeVatrons using ground-based $\gamma$-ray observations.

\end{abstract}

\newpage
{
  \hypersetup{linkcolor=black}
  \tableofcontents
}
\section{Introduction}
\label{intro} 

Cosmic rays (CR) are electrically charged particles, comprised mostly of atomic nuclei and electrons. These particles arrive uniformly from outer space and penetrate the Earth's atmosphere. The discovery of CRs dates back to 1912 when Viktor Hess made the first balloon measurements \citep{cr_hess}. The investigation of the CR flux on Earth has been ongoing for more than a century, and various experiments have provided detailed measurements. Today, it is a well known fact that protons are the dominant particles in the CR spectrum, followed by helium nuclei, as it was reviewed in detail in \citep{elena_rev, elena_rev2, blasi_rev}. A simple power-law model\footnote{A power-law distribution is defined as $\Phi$(E) = $\Phi_{0}$(E$_{0}$)~(E/E$_{0}$)$^{-\Gamma}$, where $\Phi_{0}$(E$_{0}$) is the normalization at the reference energy of E$_{0}$ and $\Gamma$ is the spectral index.} with a spectral index of --2.7 provides a good approximation of the all-particle CR spectrum from 30~GeV up to the 'knee' feature, which is a prominent spectral feature seen at 3~PeV (1~PeV = $10^{15}$~eV) energies. The CR spectrum steepens significantly above the knee energy and changes its index to a softer value of $\sim$--3.1. There exists experimental evidence suggesting that the knee feature for light nuclei, such as protons and helium, may be below 1~PeV, even around 700~TeV \citep{knee_below_1pev}. However, at least for the all-particle CR spectrum, which includes heavy elements, it is a well-established fact that the spectral steepening occurs well above 1~PeV energies \citep{z_dependence_knee}. Despite being observed for over a century, the exact sources, acceleration, and propagation mechanisms of CRs still remain unknown, and intense research is ongoing in this field. 

Two main physical interpretations of the knee feature observed in the CR spectrum are widely mentioned in the literature as detailed in \citep{blumer_rev}. The first interpretation describes the knee energy as the upper limit of achievable energy by particle accelerators located within our Galaxy. As a result, this limits the maximum energy that can be reached by particles in Galactic objects to 3~PeV. The second interpretation, which is more popular, suggests a connection between the knee energy and the maximum energy of charged particles that can be magnetically confined within our Galaxy due to the presence of the Galactic magnetic fields. This interpretation does not impose any restrictions on the maximum achievable energy of Galactic particle accelerators, but explains the spectral steepening by attributing it to a reduction in CR flux due to particles escaping from our Galaxy. Nevertheless, in both interpretations, it is evident that there must exist some Galactic objects capable of accelerating particles at least up to PeV energies.

The term 'PeVatron' is commonly used in astronomy to refer to astrophysical sources that can accelerate particles, such as electrons, protons, and nuclei, up to PeV energies and beyond. These astrophysical PeVatrons are among the most intriguing and enigmatic objects in the Universe. The growing interest in PeVatron sources is directly related to one of the most fundamental scientific questions, which has remained unresolved and mysterious since the first observation of CRs in 1912, namely 'the question of the origin of Galactic CRs'. Up to date, no Galactic source class has been identified with firm evidence of CR acceleration, more precisely hadronic acceleration, up to PeV energies and beyond. However, several sources, including supernova remnants (SNRs) \citep{bell1978}, massive stars and stellar clusters \citep{aharonian2019}, pulsars and pulsar winds \citep{pulsarsElena1,crab_rev2,pulsarsClaire3}, star formation regions (SFR) \citep{sfr_bykov}, super massive black holes (SMBH) \citep{hessGCNat}, microquasars \citep{microquasars}, and superbubbles \citep{higdon2003,binns2005} have been proposed as potential sources of Galactic PeV CRs, in other words, the Galactic PeVatrons.

One of the main challenges in the search for the Galactic sources of PeV CRs is related to the extremely high energies of these particles, which allows them to travel vast distances through space before reaching the Earth. Due to the deflection of these charged particles by Galactic magnetic fields, it is impossible to trace back their origins through CR measurements performed on Earth. Instead, astronomers must search for signatures of Galactic PeVatrons by means of neutral messengers, which are not affected by Galactic magnetic fields. Basically, these neutral messengers are $\gamma$ rays and neutrinos that are generated when high energy CRs collide with target material, e.g.~molecular clouds, present in the interstellar medium. Recent advancements in very high energy (VHE,~E$>$0.1~TeV) and ultrahigh energy (UHE,~E$>$0.1~PeV) $\gamma$-ray astronomy have provided unprecedented observational data to search for signatures of Galactic PeVatrons. However, further deeper investigation is required for their robust confirmation, to establish their characteristic properties, and comprehend their acceleration mechanisms. Indeed, resolving this century-old mystery would constitute a significant breakthrough in our understanding of the Universe, and the ongoing high energy processes in it.

This review article discusses the current state and potential future directions for investigating Galactic PeVatrons by means of ground-based $\gamma$-ray observations. The review article is organized as follows: Section~\ref{sec_cr} provides a brief introduction to the CR spectrum, and a discussion of its spectral characteristics. Section~\ref{vhe_uhe} covers the basics of ground-based $\gamma$-ray astronomy, and its relevance to the search of Galactic PeVatrons. The current understanding of Galactic PeVatrons is discussed in Sect.~\ref{pev_def_main}, providing a clear definition for the PeVatron concept, along with specific discussions on given examples. Section~\ref{sig_pev} provides a general review and wide discussion on the spectral signatures of PeVatron sources, together with analysis methods used to determine these unique signatures. In Section~\ref{source_conf}, the problem of source confusion is examined, with particular examples of Galactic sources such as HESS~J1641$-$463, HESS~J1825$-$137, HESS~J1702$-$420, and the $\gamma$-ray emission from the region including SNR~G106.3$+$2.7 and the Boomerang Nebula. Finally, Section~\ref{conclusions} presents the summary and conclusions of the review.
\section{Cosmic rays}
\label{sec_cr} 

A very brief and plain definition of CRs is given in \citep{rossi1964cosmic} as \textit{"A steady rain of charged particles, moving at nearly the speed of light, falls upon our planet at all times and from all directions in the sky. These particles are just the nuclei of ordinary atoms stripped of their electrons, for the most part nuclei of hydrogen"}. These 'rain of charged particles' was first discovered by Viktor Hess \citep{cr_hess} with several balloon experiments and was named as "cosmic rays" by Millikan in 1928. Later, Viktor Hess was awarded the Nobel Prize in Physics in 1936 for the discovery of CRs. The CRs are one of the fundamental aspect of the study of high energy astrophysics. Their spectrum provides extremely valuable information about the origin and acceleration of CRs, their propagation through the interstellar medium, and their role in shaping the evolution of the Universe.

\begin{figure}[ht!]
\centering
\includegraphics[width=17cm]{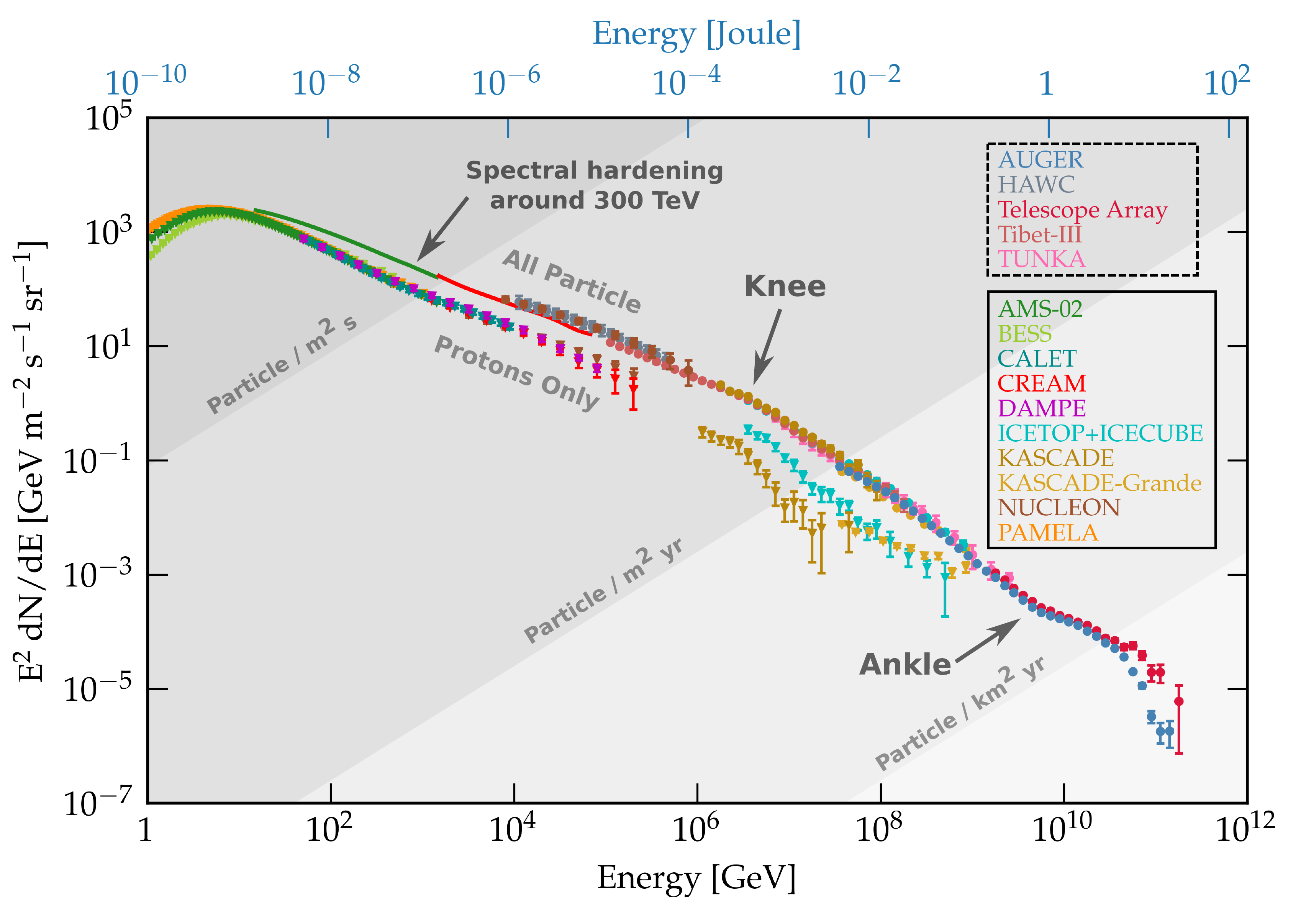}
\caption{The all particle cosmic-ray (top) and proton (bottom) energy spectra measured from earth. The figure is adopted from \cite{cr_spect_plot}. The all particle cosmic-ray spectrum data is taken from AUGER \citep{auger_cr}, HAWC \citep{hawc_cr}, ICETOP+ICECUBE \citep{icetop_cr}, KASCADE-Grande \citep{kaskade_gr_cr}, KASCADE \citep{kaskade_cr}, NUCLEON \citep{nucleon_cr}, Telescope Array \cite{tel_array_cr}, TIBET-AS \citep{tibet_cr} and TUNKA \citep{tunka_cr} observations. The experiments providing 'proton' measurements are given in the box with solid lines, while for the experiments providing 'all particle' spectra are given in the box with dashed lines. The 'knee' and 'ankle' features are indicated with black arrows on the spectrum.}
\label{cr_spect}
\end{figure}

Since their first discovery, decades of measurements from various experiments have contributed to the reconstruction of a highly detailed energy spectrum for CRs. The spectral energy distribution of CRs, which ranges from a few GeV to over 10$^{20}$~eV energies, is shown in Figure~\ref{cr_spect}. Presently, it is a well-known fact that the CR spectrum consist of a mixture of charged particles, such as protons ($\sim$87$\%$), Helium ($\sim$12$\%$), leptons ($\sim$2$\%$) and heavier ions ($\sim$1$\%$), as discussed in great detail in \citep{elena_rev, elena_rev2, blasi_rev}. While space-based particle detector experiments can in principle detect CRs with energies below 100~TeV energies, ground-based indirect detection techniques, utilizing air showers induced by CRs, are used to detect those with energies above PeV \citep{cr_det_tech}.

The CR spectrum measured from Earth, shown in Figure~\ref{cr_spect}, exhibits a sharp turnover at energies below $\sim$30~GeV, and this spectral feature is related to the magnetized outflowing solar wind generated by our star. This physical phenomenon is known as 'solar modulation' \citep{solar_mod1}, and basically prevents low energy CRs from reaching the Earth. Conversely, the CR spectrum above 30~GeV follows a smooth, single power-law distribution with a spectral index of $\sim$2.7 up to a spectral feature called the "knee", which is observed at energies of $\sim$3~PeV. At this spectral break point, the CR spectrum steepens significantly by changing its spectral index value from 2.7 to 3.1. In principle, all known acceleration mechanisms are rigidity-dependent\footnote{Rigidity is a measure of deflection of a particle of charge Z and momentum p in a magnetic field B, and given as $R={pc}/{Ze}$ (in volts). Different particles will have the same dynamics in a magnetic field if they have the same rigidity.}. Thus, the actual knee feature seen in the all-particle CR spectrum can be considered as the superposition of different CR species exhibiting spectral cutoffs at different energies. This basically means that, if the maximum energy of protons in our Galaxy is assumed to be i.e.~3$-$4~PeV, the maximum energy of heavier nuclei should scale proportionally with the charge of the nucleus in question. For protons, the knee is observed between 700~TeV \citep{knee_below_1pev} and 3$-$4~PeV \citep{crknee_3PeV}, whereas for the heaviest nuclei such as Fe (Z=26), the (corresponding Fe) knee is located around 100~PeV.

The CR spectrum exhibits another characteristic spectral feature at energies below the knee. Around $\sim$300~GeV, there is a noticeable spectral hardening in all nuclear species, with protons and helium exhibiting the most significant effects \citep{300GeV_pam,300GeV_pam2}. The spectral index of protons changes from $\Gamma_{p}$~=~2.89~$\pm$~0.015 below 300~GeV to $\Gamma_{p}$~=~2.67~$\pm$~0.03 above that threshold, while the spectrum of helium nuclei is found to be systematically harder \citep{300GeV_pam3}. Additionally, there are indications of hardening in the spectra of heavier elements. This unique feature in the CR spectrum is connected to the physics of propagation in the Milky Way as reviewed in \citep{elena_rev, sabrina_rev}.

At higher energies around $\sim$10$^{18}$~eV, another characteristic spectral feature known as 'ankle' emerges. At this point, the power-law spectral index reverts its values back to 2.7. Considering that the interstellar medium (ISM) has a typical magnetic field strength of a few $\mu$G\footnote{The exact structure and values of Galactic magnetic fields are poorly constrained. In the vicinity of the solar system, B is $\sim$2 $\mu$G, while average Galactic field strength is 3$-$6 $\mu$G. Assuming an average field strength of 3 $\mu$G, the corresponding energy density of the field is B$^{2}$/8$\pi$ = 4 $\times$ 10$^{-13}$ erg/cm$^{-3}$ $\approx$ 0.25 eV/cm$^{-3}$.}, and that the size of the Galactic halo is a few tens of kpc, CRs with energies ranging from $10^{16}-10^{17}$ eV should be confined within the Galaxy. However, CRs with energies beyond the ankle have a larger gyroradius and cannot be confined within the Galaxy, indicating that they have extra-galactic origin. As a result, the transition between Galactic and extra-galactic CRs is believed to occur between the knee and the ankle energies, although this is still a subject of debate \citep{cr_transition}.

\subsection{Galactic cosmic-rays}
\label{galactic_crs}

As previously mentioned, the CR spectrum detected on Earth reveals a number of characteristic spectral features, particularly the knee feature, the change in composition around the knee, and the spectral hardening observed at $\sim$300~GeV. These features suggest that the majority of CRs are generated and accelerated by a rigidity-dependent acceleration mechanism at CR factories located in our Galaxy, with heavier nuclei capable of achieving higher maximum energies \citep{elena_rev, elena_rev2, blasi_rev}. Consequently, the CR spectrum displaying a notable power-law behavior over several orders of magnitude up to the knee is a clear indication that Galactic sources must accelerate CRs to at least PeV energies and beyond.

The CR sources in the Galaxy inject particles into the ISM, where they can be deflected by turbulent magnetic fields, resulting in confinement of CRs within the Galaxy for up to tens of millions of years through scattering \citep{blasi_rev}. However, it is crucial to understand that the CR spectrum observed from Earth is influenced by several factors, including (i)~the acceleration of CRs in Galactic sources, (ii)~the propagation of CRs within their acceleration sites, (iii)~the escape of CRs from acceleration sites, and (iv)~the propagation of CRs in the ISM until they reach Earth. As a result, CRs continuously bombard Earth's atmosphere in an isotropic manner without carrying information about their origin, making the determination of their origin extremely challenging. Thus, neutral messengers generated by CR interactions, such as $\gamma$ rays and neutrinos, are typically utilized for localizing CR sources. It is worth mentioning that the average energy density of Galactic CRs measured at Earth is around $U_{CR} = 1$ eV/cm$^{3}$, and mainly concentrated around $\sim$1~GeV energies. Assuming that the average confinement time of CRs within the Galaxy is $\sim$10 Myr, the average power needed by CR sources in our Galaxy to sustain the Galactic CR flux at the observed level can be estimated as $P_{CR}$ = (3$-$10)$\times$10$^{40}$ erg/s. Due to the broad scope of Galactic CRs topic, it is not feasible to provide additional specific information within this review. However, there are numerous other captivating reviews available on this topic, i.e. see \citep{amato_casanova_2021,elena_rev,elena_rev2,Sciascio_rev,sabrina_rev,pierre_rev, vink_rev}, offering much more comprehensive details and insights.  

\section{Ground-based $\gamma$-ray astronomy}
\label{vhe_uhe} 

Gamma-ray astronomy is the branch of astronomy that deals with the detection, analysis and interpretation of $\gamma$-ray radiation with photon energies above a few 100~keV. Today, the most energetic photon ever observed, with E$_{\gamma}$ = 1.4~PeV, was detected from the direction of the Cygnus region \citep{LHAASO_Crab}. As it was reviewed in \citep{KNODLSEDER2016663}, emission of nonthermal radiation is a common characteristic shared by all $\gamma$-ray sources. Consequently, $\gamma$-ray astronomy primarily investigates the nonthermal Universe. Unlike thermal radiation generated by the thermal motion of particles in matter, nonthermal radiation can arise from very different origins, such as decay of particles, the interaction of nonthermal particle populations with photons fields and matter, or even from the annihilation of dark matter particles. The significant advantage and uniqueness of $\gamma$-ray observations is that the $\gamma$-ray energy band is devoid of thermal radiation, providing a clear view of the nonthermal physics present in the Universe. It is worth noting that although there is no theoretical limit to the maximum energy of $\gamma$ rays, the pair production effect (see subsection~\ref{gammagammaabs} for a detailed discussion) on interstellar photon fields like cosmic microwave background (CMB) or infrared (IR) dust emission puts a limit on the observable $\gamma$-ray universe.

\begin{figure}[ht!]
\centering
\includegraphics[width=14cm]{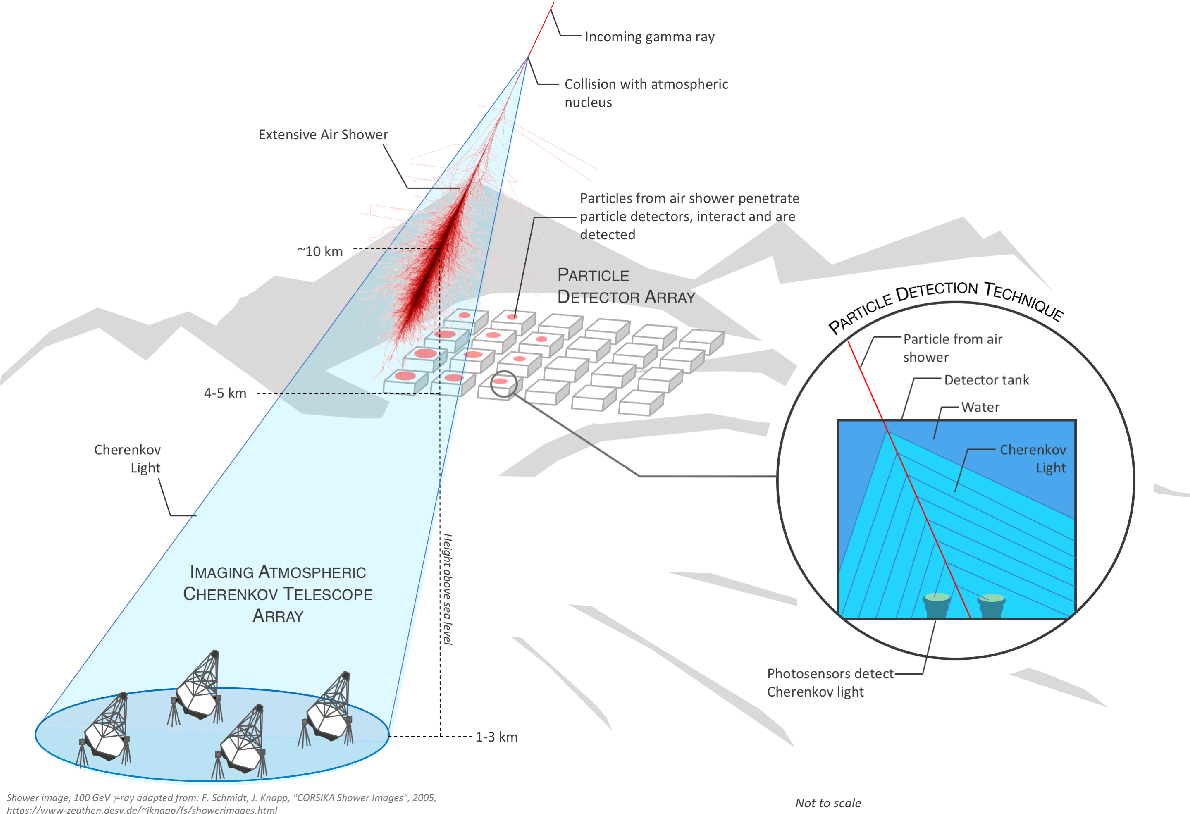}
\caption{Detection techniques for ground-based gamma-ray astronomy. The inset shows the working principle of a water Cherenkov particle detector. Figure and caption are taken from \citep{fig_iact_wcd}.}
\label{iact_wcd}
\end{figure}

When $\gamma$ rays arrive at Earth, they interact with the atmosphere and initiate electromagnetic cascades of particles known as 'extensive air showers (EAS)'. As a result, direct detection of $\gamma$ rays from astrophysical sources is only possible from space, outside the Earth’s atmosphere. Therefore, observations of $\gamma$~rays are conducted either directly from space-based satellite missions, or indirectly from ground-based observatories utilizing EAS. The flux of $\gamma$ rays reaching Earth, defined by the number of $\gamma$-ray photons detected per unit area (i.e.~cm$^{-2}$) per unit time (i.e.~s$^{-1}$), is drastically reduced at higher $\gamma$-ray energies, especially beyond a few hundred GeV. For instance, only a few photons per m$^{2}$ per year are expected above 1~TeV energies \citep{Hinton_2009} even from the brightest $\gamma$-ray sources. As a result, the collection area of space-based satellite missions, which is only a few m$^{2}$, is insufficient for collecting reasonable event statistics above TeV energies that enable reliable analysis of astrophysical data. Instead, ground-based instruments, which detect secondary products resulting from electromagnetic cascades, such as particles reaching the ground or Cherenkov light emitted by these particles, are utilized for $\gamma$-ray astronomy above 0.1~TeV. For instance, Cherenkov telescope systems with their typical effective collection area being in the range of $\sim$10$^{5}$--10$^{6}$~m$^{2}$ allow the collection of reasonable event statistics even after $\mathcal{O}(10)$ hours of observation. The energy threshold of 0.1~TeV marks the lower limit of the VHE $\gamma$-ray regime, while the upper limit is generally quoted as 100~TeV. The UHE $\gamma$-ray regime is defined beyond this upper limit, with energies exceeding 0.1~PeV. Figure~\ref{iact_wcd} illustrates two main approaches that are used to detect $\gamma$-rays from the ground.

\begin{itemize}
\item[1:] Particle detector arrays: Particle detector arrays are typically positioned at high altitudes of around $\sim$4--5 km. The working principle of particle arrays is based on the direct detection of the particles generated in EAS through their water Cherenkov detector tanks (see inset of Figure~\ref{iact_wcd}). As charged EAS particles travel through water, they emit Cherenkov light, which can be detected by photosensors located inside the tanks. Scintillator (electromagnetic) detectors and muon detectors can be incorporated into particle detector arrays, with the former being utilized to the UHE sensitivity and the latter to improve background rejection of the experiment.
\item[2:] Imaging atmospheric Cherenkov telescopes (IACTs): Cherenkov telescopes are typically situated at altitudes ranging from 1 to 3 km. IACTs detect the Cherenkov light generated by EAS particles in the atmosphere. An IACT system typically consists of two or more Cherenkov telescopes, enabling stereoscopic imaging of the same shower from different angles. The stereoscopic imaging technique can achieve an impressive angular resolution better than $\sim$0.1$^{\circ}$ (equivalent to 5$-$6 arc min) depending the number of telescopes.    
\end{itemize}

Particle detector arrays offer few advantages over IACTs, such as a superior duty cycle, wider field of view, and increased energy range up to PeV energies. However, IACTs can achieve much better angular and energy resolution when compared to particle arrays. Table~\ref{table_obser} provides a comparison of key performance characteristics between these two observation techniques, along with examples of currently operating and planned future observatories. For a more comprehensive overview of the current ground-based $\gamma$-ray facilities, including their detailed descriptions and locations, please refer to \cite{alison_ICRC}.

\begin{table*}
\caption{Comparison between key characteristics of IACTs and particle arrays. Examples of currently operating and upcoming IACTs and particle array experiments are also given in a separate rows.}
\label{table_obser}     
\centering                        
\begin{tabular}{c c c }       
\hline\hline                 
\textbf{Instrument} & \textbf{IACTs} & \textbf{Particle Arrays}  \\ 
\textbf{Characteristics} &  &   \\ 

\hline       
Energy range         & [$\sim$30 GeV, $\sim$100 TeV]  & [$\sim$500 GeV, $\sim$1 PeV]  \\
Duty cycle           & $\sim$10$\%$ ($\sim$1000 h/yr)  & $>$95$\%$  \\
Field of view        & $\sim$5$^{\circ}$ ($\sim$ a few msr)    & $\sim$90$^{\circ}$ ($\sim$ sr)  \\
Energy resolution    & $\sim$15$\%$ & $\sim$40$\%$  \\
Angular resolution   & $\sim$0.1$^{\circ}$ & $\sim$0.25$^{\circ}$ \\
Background rejection & $>$99$\%$ & $\sim$90$\%$  \\
\hline  
\hline   
Currently operating & H.E.S.S. \citep{hess_ref}, VERITAS \citep{veritas} & LHAASO \citep{lhaaso_crab_ref}, HAWC \citep{hawc_ref} \\
instruments         & MAGIC \citep{magic_ref}, CTA-LST \citep{cta_lst}& Tibet AS-$\gamma$ \citep{tibet_ref} \\ 
\hline
Planned future      & CTA-South and CTA-North \citep{cta_science} & SWGO \citep{swgo_ref} \\ 
instruments         & ASTRI Mini-array \citep{astri_ref} &  ALPACA \citep{alpaca_ref} \\
\hline  
\hline 
\end{tabular}
\end{table*}

\begin{figure}[ht!]
\centering
\includegraphics[width=17cm]{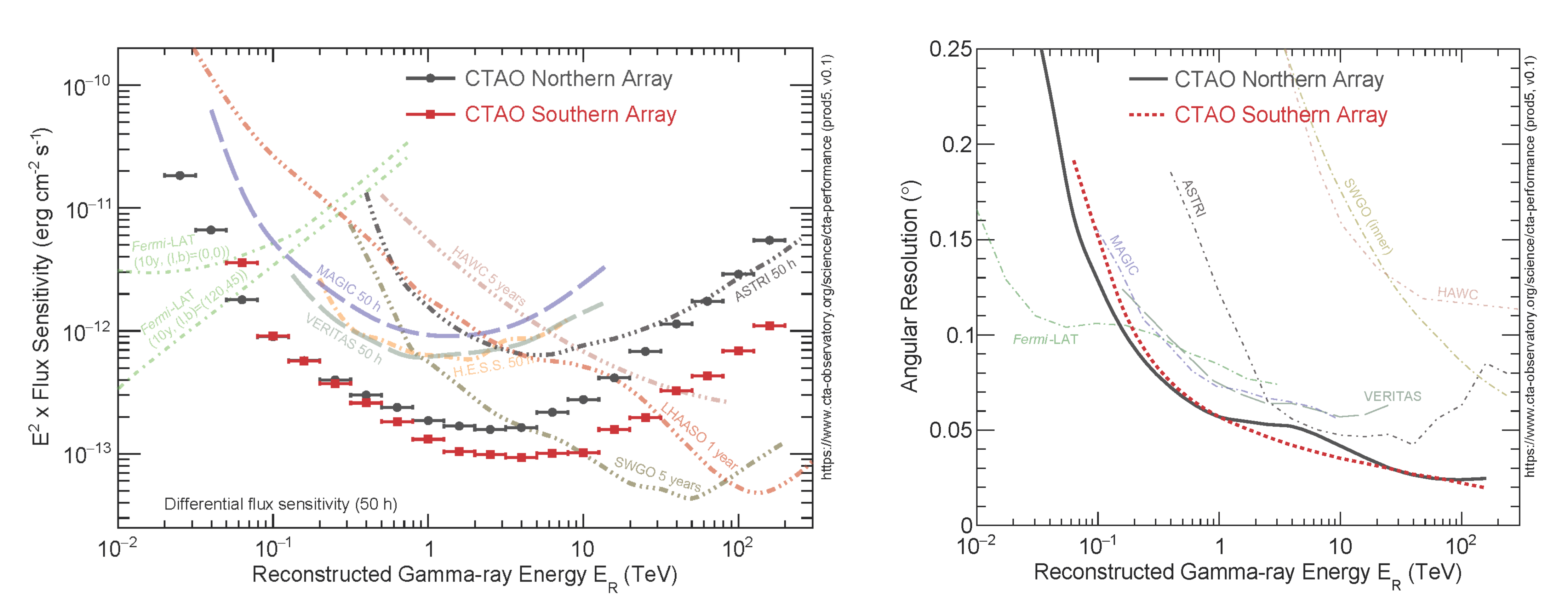}
\caption{Left figure: The 50 h differential sensitivity of future CTA South and North arrays compared to MAGIC, VERITAS and HESS 50 h sensitivity. Fermi (10 years), HAWC (5 years), SWGO (5 years) and LHAASO (1 year) sensitivity curves are also given for comparison. Right figure: comparison of angular resolution, which is expressed as the 68$\%$ containment radius of reconstructed $\gamma$ rays, between CTA arrays and other experiments. The figures are taken from \text{https://www.cta-observatory.org/science/ctao-performance/}.}
\label{sens_cta}
\end{figure}
\begin{figure}[ht!]
\centering
\includegraphics[width=17cm]{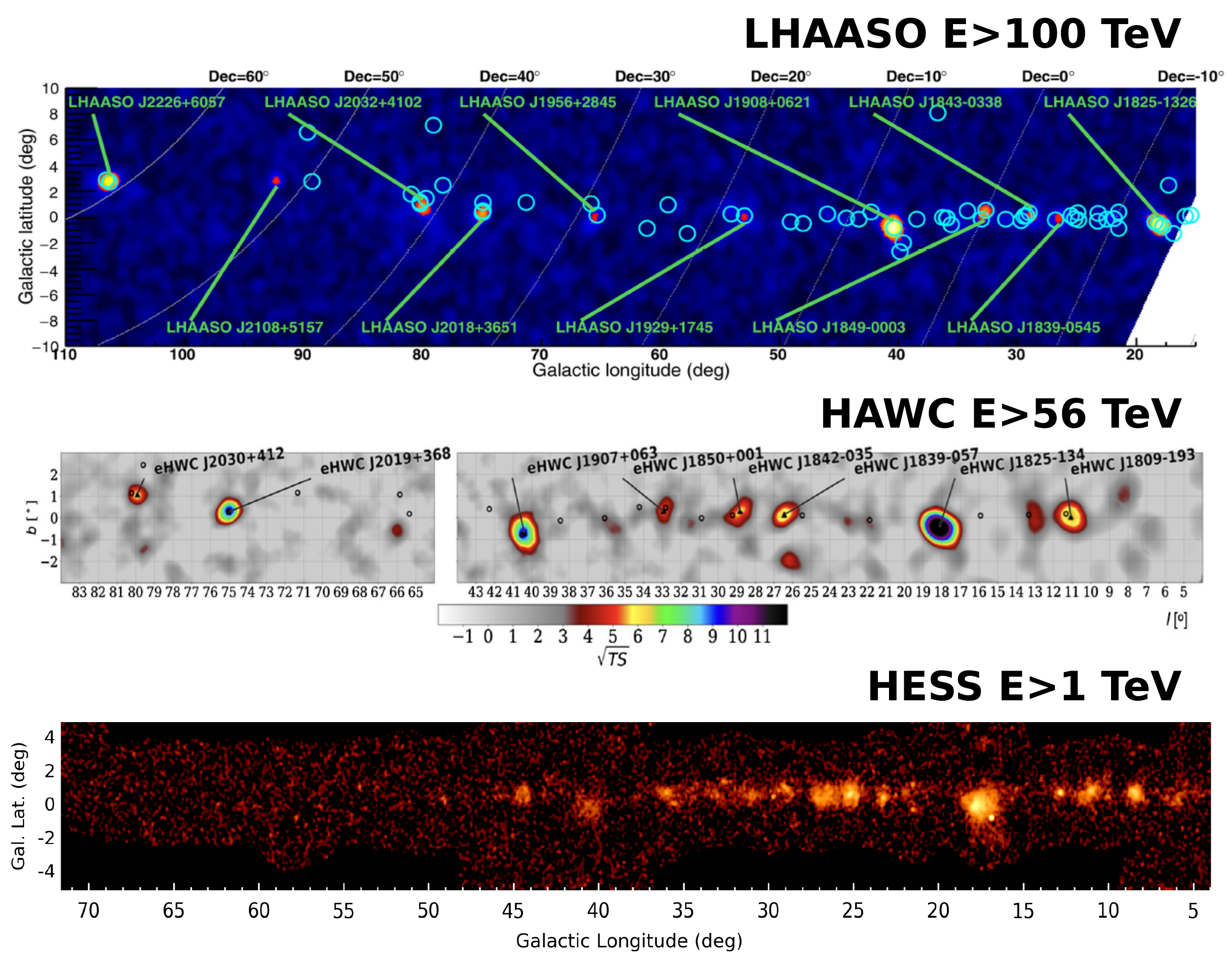}
\caption{Top, middle and bottom show the significance map of a part of Galactic plane above 100~TeV from LHAASO observations \cite{lhaaso2021}, above 56~TeV from HAWC observations \citep{hawc_E56TeV} and above 1~TeV from HESS Galactic plane survey (HGPS) \citep{hgps}.}
\label{skymaps}
\end{figure}

Figure~\ref{sens_cta} shows a comparison of differential flux sensitivity (left) and angular resolution (right) between different ground-based instruments, including both IACT and particle array experiments. One of the most important and notable indication of the figure is the fact that combining IACTs and particle arrays can result in effective synergy for studying and understanding the nature of astrophysical sources, particularly in the energy range of few tens of TeV to PeV energies. As it can be seen from the left figure, IACTs have better flux sensitivity below $\sim$10~TeV energies, whereas particle arrays have much better sensitivity above $\sim$10~TeV. The latter is, in principle, the key factor of locating PeVatrons in the Galaxy, and consequently making particle array experiments 'PeVatron hunters'. On the other hand, IACTs offer superior angular resolution performance throughout the entire energy range due to the stereoscopic imaging technique, providing up to 3$-$4 times improvement over particle arrays above 10~TeV energies (see the right panel of Figure~\ref{sens_cta}). In order to provide a better visualisation, the comparison of significance maps of a part of the Galactic plane, reconstructed by different ground-based experiments, above various energy thresholds is shown in Figure~\ref{skymaps}. The bottom panel (E$>$1~TeV~map) illustrates that the angular resolution of H.E.S.S. telescopes is capable of detecting detailed fine structures and numerous $\gamma$-ray sources. Moreover, given the good angular resolution of HESS telescopes around 0.1$^{\circ}$, even if these $\gamma$-ray sources are located in a very close proximity with an angular separation of less than 0.2$^{\circ}$--0.3$^{\circ}$, they can still be distinguished from each other. However, as the energy threshold increases (see Figure~\ref{skymaps} middle and top panels for E$>56$~TeV and E$>100$~TeV maps, respectively), the number of sources decreases due to both astrophysical nature, i.e.~sources are not powerful enough to produce $\gamma$-ray emission above several 10's of TeV, and degraded angular resolution of observations, i.e.~nearby sources with angular separation less than 0.2$^{\circ}$--0.3$^{\circ}$ will appear as a single source. As a result, observations of E$>$10~TeV regime with particle arrays often suffer from source confusion (see Section~\ref{source_conf} for a very detailed discussions on the source confusion), in case there are two (or more) sources located in a close proximity. To address this issue of source confusion, high angular resolution IACT observations are required. Consequently, it may be possible to pinpoint the actual astrophysical origin of the E$>$100 TeV emission by performing high angular resolution observations of the UHE $\gamma$-ray emission regions.

Furthermore, IACTs offer significantly better energy resolution when compared to particle arrays, enabling detailed spectral studies and the ability to identify and resolve potential spectral structures in source spectra. This feature is particularly valuable and crucial in situations where two different $\gamma$-ray production mechanisms, such as both hadronic and leptonic mechanisms, are simultaneously at work and it is not possible to discriminate them morphologically. For instance, such a source would appear as a point source, while its spectrum is a combination of $\gamma$-ray spectra generated by both hadronic and leptonic mechanisms. High energy resolution observations can be helpful in distinguishing between these two emission mechanisms, especially when the hadronic process dominates at high energies, leading to a hardening of the source spectrum.

On the other hand, particle arrays have two critical advantages: their excellent duty cycle ($>$99$\%$) and large field of view, allowing them to continuously monitor $\sim$15$\%$ of the entire sky. These properties make particle arrays highly suitable for searching for transients, and conducting follow-up observations in response to alerts raised by neutrino, gravitational waves or satellite experiments. 

Finally, a comprehensive understanding of the nature of sources under investigation would require combined $\gamma$-ray observations from IACTs, particle arrays, and even satellite missions like Fermi-LAT \citep{fermi_ref}. Further details on the fundamental principles of ground-based $\gamma$-ray observatories are beyond the scope of this particular review. However, numerous comprehensive and informative review articles are available, i.e.~see \citep{sinnis, Hinton_2009, KNODLSEDER2016663, alison_ICRC, DuVernois2022, Bose_2021}.

\subsection{Galactic diffuse $\gamma$-ray emission at UHEs}
\label{gal_diff}
\begin{figure*}
\centering
\includegraphics[width=17cm]{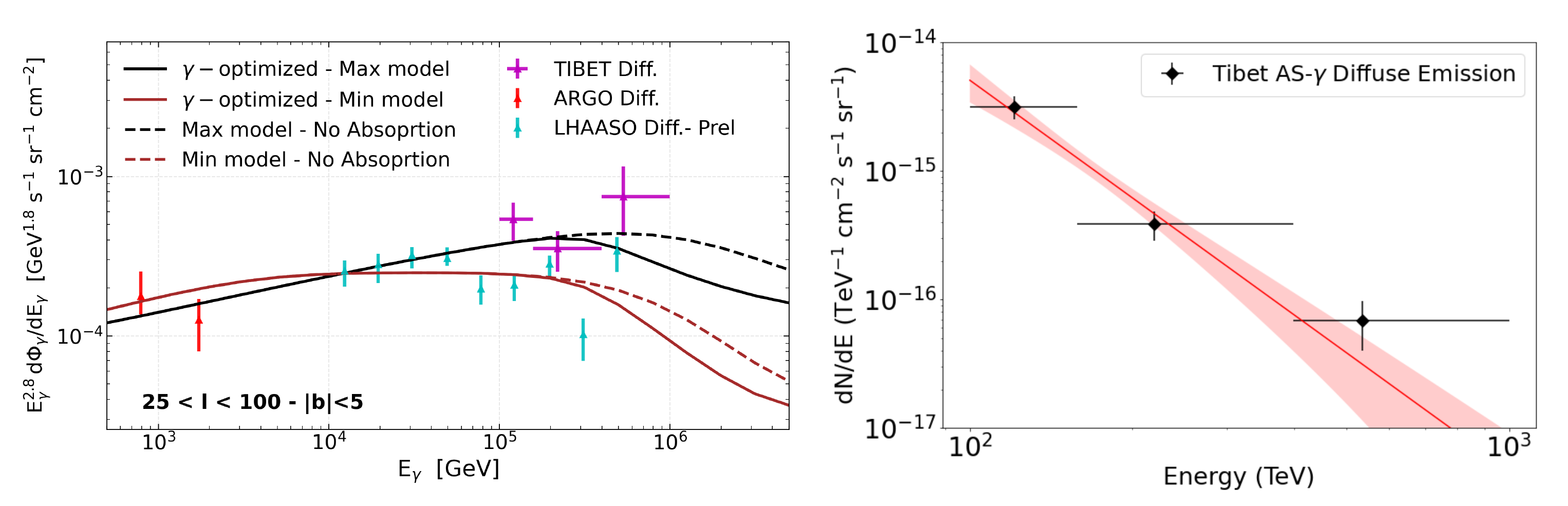}
\caption{Left panel: The $\gamma$-ray spectra computed within the 
conventional (base) and $\gamma$-optimized scenarios are compared 
to ARGO-YBJ \citep{Bartoli_diff_argo}, Tibet AS-$\gamma$ \citep{diff_tibet} and LHAASO \citep{diff_LHAASO} (preliminary) data in the window $|$b$|$ $<5^{\circ}$, 25$^{\circ}$ $<$ \textit{l} $<$ 100$^{\circ}$, showing also the effect of $\gamma$-ray absorption onto the CMB photons for the $\gamma$-optimized scenario (see \citep{diffuse_PeV_Pedro} for more details). The left figure and caption are taken from \citep{diffuse_PeV_Pedro}. Right panel: simple power-law model fit to Tibet AS-$\gamma$ diffuse emission data \citep{diff_tibet} above 100~TeV energies. The black points show the reconstructed $\gamma$-ray flux points with 1$\sigma$ statistical errors, while the red line and shaded region show the best-fit power-law model to the data and 1$\sigma$ error interval, respectively.}
\label{diff_UHE}
\end{figure*}
\begin{table*}[ht!]
\caption{The number of background events above 100~TeV obtained experimentally from LHAASO observations are given, together with the exposure for each source. All values are taken from extended data table 1 of \citep{lhaaso2021}. The number of background events above 100~TeV estimated from Eq.~\ref{bkg_events} for each UHE source is provided. The 'Rel. Diff.' column gives the relative difference between the number of background events.}
\label{table_uhe_diff}     
\centering        
\small
\begin{tabular}{c c c c c}       
\hline\hline                 
\textbf{Source} & \textbf{Exposure} & \textbf{$\#$ of background} & \textbf{$\#$ of background} & \textbf{Rel. diff.}  \\ 
\textbf{name} & \textbf{(h)} & \textbf{events (LHAASO)}  & \textbf{events (GDE model)} & \textbf{($\%$)} \\ 
\hline       
LHAASO~J0534$+$2022 & 2236.4 & 5.5 & 5.7$\pm$2.1 & 3.6 \\
LHAASO~J1825$-$1326 & 1149.3 & 3.2 & 2.9$\pm$1.1 & 9.4 \\
LHAASO~J1839$-$0545 & 1614.5 & 4.2 & 4.1$\pm$1.5 & 2.4 \\
LHAASO~J1843$-$0338 & 1715.4 & 4.3 & 4.4$\pm$1.6 & 2.3 \\
LHAASO~J1849$-$0003 & 1865.3 & 4.8 & 4.8$\pm$1.7 & 0.0 \\
LHAASO~J1908$+$0621 & 2058.0 & 5.1 & 5.3$\pm$1.9 & 3.9 \\
LHAASO~J1929$+$1745 & 2282.6 & 5.8 & 5.8$\pm$2.1 & 0.0 \\
LHAASO~J1956$+$2845 & 2461.5 & 6.1 & 6.3$\pm$2.3 & 3.3 \\
LHAASO~J2018$+$3651 & 2610.7 & 6.3 & 6.7$\pm$2.4 & 6.3 \\
LHAASO~J2032$+$4102 & 2648.2 & 6.7 & 6.8$\pm$2.5 & 1.5 \\
LHAASO~J2108$+$5157 & 2525.8 & 6.4 & 6.5$\pm$2.4 & 1.6 \\
LHAASO~J2226$+$6057 & 2401.3 & 6.2 & 6.1$\pm$2.3 & 1.6 \\
\hline  
\end{tabular}
\end{table*}

The Galactic diffuse $\gamma$-ray emission (GDE) refers to the $\gamma$-ray emissions observed mostly from the direction of the Galactic plane on a large scale. As discussed in \cite{gaisser_CR} in detail, there are three primary components that contribute to the GDE. The first one has hadronic origin and results from the decay of $\pi^{0}$, which is produced from the interactions between CRs in the Galaxy and the gas present in the ISM. The other two components originate from leptons and arise from the inverse Compton (IC) scattering of interstellar radiation fields from cosmic electrons/positrons, and the bremsstrahlung radiation, respectively.

The GDE is a vital tool for studying the production, propagation and interaction of CRs in the Galaxy. While local measurements of CR flux near Earth provide limited information, the GDE allows for direct measurement of CR distribution across the entire Galaxy. As discussed in \citep{diffuse_PeV_Pedro}, the GDE detected from low Galactic longitude regions can provide essential information regarding the propagation of CRs in the inner regions of the Galaxy, while CR transport conditions are expected to provide different picture for the average Galactic plane, resulting in a radially dependent CR spectrum. To study the GDE, a comprehensive survey covering a significant portion of the Galactic plane is necessary. In the initial step of the GDE analysis, the known sources and regions displaying significant emission are masked and removed from the survey data analysis. However, the remaining emission still contains instrument CR background\footnote{CR background of an experiment is basically the CR induced showers which look like $\gamma$-ray induced showers, and therefore can pass $\gamma$-hadron separation.}, necessitating a proper background estimation depending on the experiment. Finally, the GDE can be estimated for the specific Galactic region that the instrument covers. It is worth noting that even after the subtraction of CR background and known sources, the GDE may still include emission from unresolved sources\footnote{Unresolved sources are the source of $\gamma$rays which are not bright enough to be detected significantly (i.e. $2-3\sigma$) in a survey. Some algorithms which can mask emission regions above a certain threshold can be applied to survey data in order to minimize contribution coming from unresolved sources.}. For instance, the process of estimating GDE at TeV energies is discussed in detail in \citep{hess_GDE,diff_TEV}. Investigating the GDE at UHEs is particularly crucial, since it is expected that the primary contribution at these extreme energies will come from the decay of $\pi^{0}$ resulting from interactions between Galactic CRs with energies of 1~PeV (and beyond) and the interstellar gas. The ARGO-YBJ \citep{Bartoli_diff_argo}, LHAASO \citep{diff_LHAASO}, and Tibet AS-$\gamma$ \citep{diff_tibet} experiments have conducted GDE studies at energies above 100~TeV. The findings of these studies have been combined and thoroughly discussed in \citep{diffuse_PeV_Pedro} as presented in Figure~\ref{diff_UHE} (left).

In this subsection, an initial benchmark estimate of the diffuse UHE $\gamma$-ray emission is provided by fitting a simple power-law to the Tibet AS-$\gamma$ flux data published in \citep{diff_tibet}. The obtained flux model is then employed to estimate the background event counts of UHE LHAASO sources. The best-fit power-law model to UHE diffuse flux data is shown in Figure~\ref{diff_UHE} (right), and the best-fit parameters are estimated as $\Phi_{0}$(300~TeV) = (1.8135 $\pm$ 0.6175)$\times$10$^{-16}$ cm$^{-2}$s$^{-1}$sr$^{-1}$TeV$^{-1}$, and $\Gamma$ = 3.03 $\pm$ 0.54, leading to integral diffuse $\gamma$-ray flux of F$_{\gamma, \text{diff}}$($>$100~TeV) = (2.48 $\pm$ 0.91)$\times$10$^{-13}$ cm$^{-2}$s$^{-1}$sr$^{-1}$. With some basic assumptions, this model can be used to forecast the number of background events detected by particle array experiments. The detected number of diffuse $\gamma$-ray events above 100~TeV can be estimated as 
\begin{equation}
\label{bkg_events} 
N_{\text{bkg,diff}} (E > 100~\text{TeV}) = \text{F}_{\gamma, \text{diff}}(>100~\text{TeV}) \times \text{A}_{\text{Eff}} \times \text{Area}_{\text{Off}} \times \text{T}_{\text{Obs}},
\end{equation}\\
where F$_{\gamma, \text{diff}}$($>$100~TeV) is the diffuse integral flux above 100~TeV obtained from the GDE model, $\text{A}_{\text{Eff}}$ is the effective area of the instrument (above 100~TeV), $\text{Area}_{\text{Off}}$ is the area of the background region (in steradian) where the background counts are estimated from, and $\text{T}_{\text{Obs}}$ is the observation time. Table~\ref{table_uhe_diff} provides a comparison between the number of background events observed experimentally from LHAASO observations of 12 Galactic UHE sources and the number of background events estimated from Eq.~\ref{bkg_events}. For the estimation, the LHAASO effective area above 100~TeV is assumed to be constant at a value of 3$\times$10$^{5}$ m$^{2}$ for zenith angles between 10$^{\circ}$ and 40$^{\circ}$, as shown in Figure~\ref{aeff} (left). The $\text{Area}_{\text{Off}}$ is fixed at 1 deg$^{2}$ for all sources\footnote{As it was mentioned in \citep{lhaaso2021}, background events are integrated over the angular range, which corresponds to 90$\%$ of the probability of the Gaussian templates used for modelling. Given that the 68$\%$ containment radii are 0.49$^{\circ}$, 0.45$^{\circ}$ and 0.62$^{\circ}$ for LHAASO~J2226$+$6057, LHAASO~J1908$+$0621 and LHAASO~J1825$-$1326, respectively, the choice of 90$\%$ containment radius of 1.0$^{\circ}$ off region is a reasonable assumption.}, and $\text{T}_{\text{Obs}}$ is taken from each source independently. The table indicates that with the given basic assumptions, the number of background events can be predicted with an accuracy of better than 10$\%$, with a mean relative difference value of $\sim$3$\%$. Therefore, the diffuse flux above 100~TeV obtained in this subsection using Tibet AS-$\gamma$ observations can provide a rough benchmark estimate for the Galactic window of ($|$b$|$~$<5^{\circ}$) and (25$^{\circ}$~$<$~\textit{l}~$<$~100$^{\circ}$). It is worth noting that LHAASO observations above 400~TeV are background-free \citep{lhaaso2021}. Taking into account the integral flux only between 100~TeV and 400~TeV for the estimation of E$>$100~TeV diffuse background events does not significantly alter the numbers given in Table~\ref{bkg_events}, leading to 5$\%$ smaller number of events on average. The forthcoming results from the HAWC experiment \citep{diff_HAWC}, along with future observations from the CTA, SWGO, and ALPACA experiments will help improve our understanding of the UHE diffuse $\gamma$-ray emission, and provide insights into the propagation and interaction of CRs in the Galaxy.

\subsection{Detection of PeV photons with ground-based observatories}
\label{pev_photons}

\begin{figure*}
\centering
\includegraphics[width=17cm]{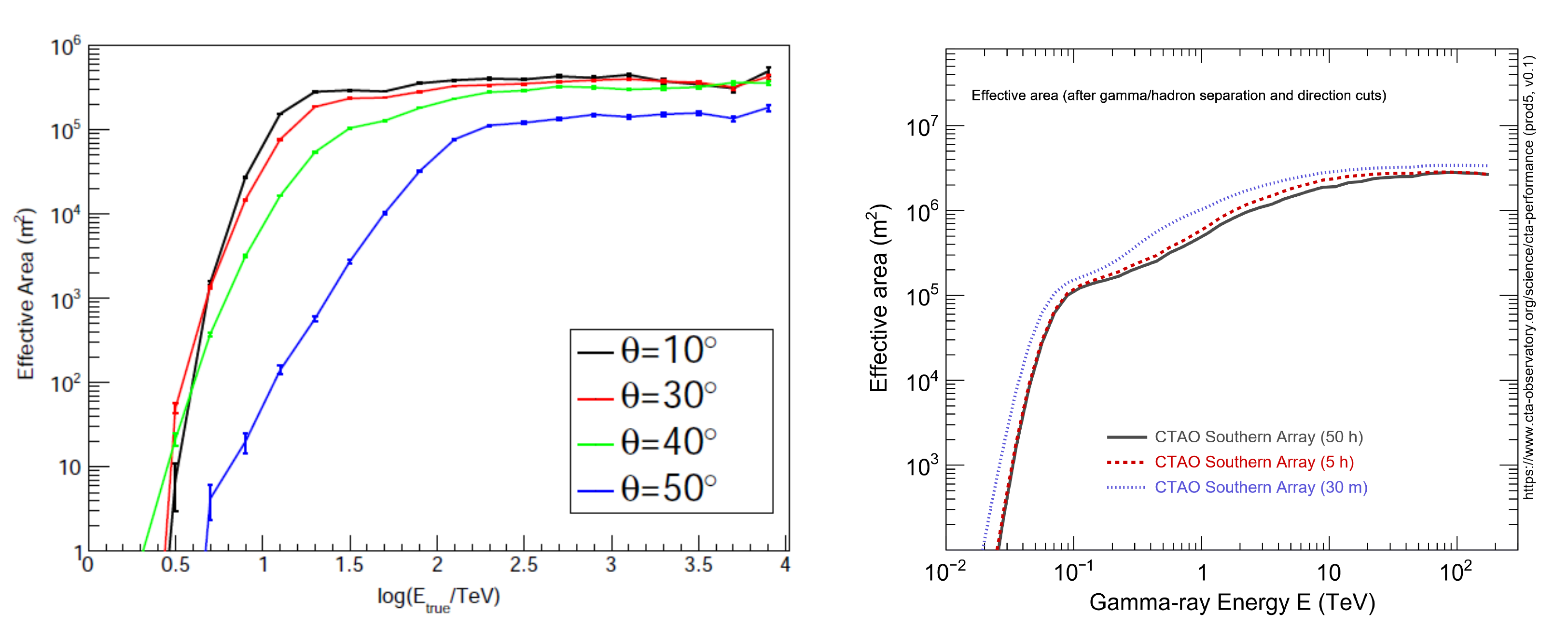}
\caption{Left panel: The effective area of the KM2A for gamma-ray showers at four zenith angles after applying the data quality and gamma-ray/background discrimination cuts. The left figure and caption are taken from \cite{lhaaso_crab_ref}. Right panel: The effective area for gamma rays from a point-like source is shown below as a function of true energy for gamma/hadron cuts optimised for 0.5-, 5-, and 50-h observations with cuts in the reconstructed event direction. The right figure is taken from \text{https://www.cta-observatory.org/science/ctao-performance/.}}
\label{aeff}
\end{figure*}
The detection of PeV photons from the direction of the Crab Nebula and the Cygnus region by the LHAASO collaboration \citep{LHAASO_Crab}, using particle arrays consisting of the square kilometer array (KM2A) with surface counters and subsurface muon detectors, marks the beginning of a new era in $\gamma$-ray astronomy. Although only a limited number of PeV photons have been detected so far, researchers are actively working towards the development of more advanced particle arrays to increase the photon statistics at these extreme energies. The main challenge in this regard is related to the extremely low photon flux at these energies, necessitating particle arrays spread across very large areas to collect sufficient event statistics. However, the key advantage of conducting astronomy at these extremely high energies (E$>$400~TeV) is that observations are free from background \citep{lhaaso2021}, and thus $\gamma$/hadron separation requires no special attention.

\begin{figure}[ht!]
\centering
\includegraphics[width=17cm]{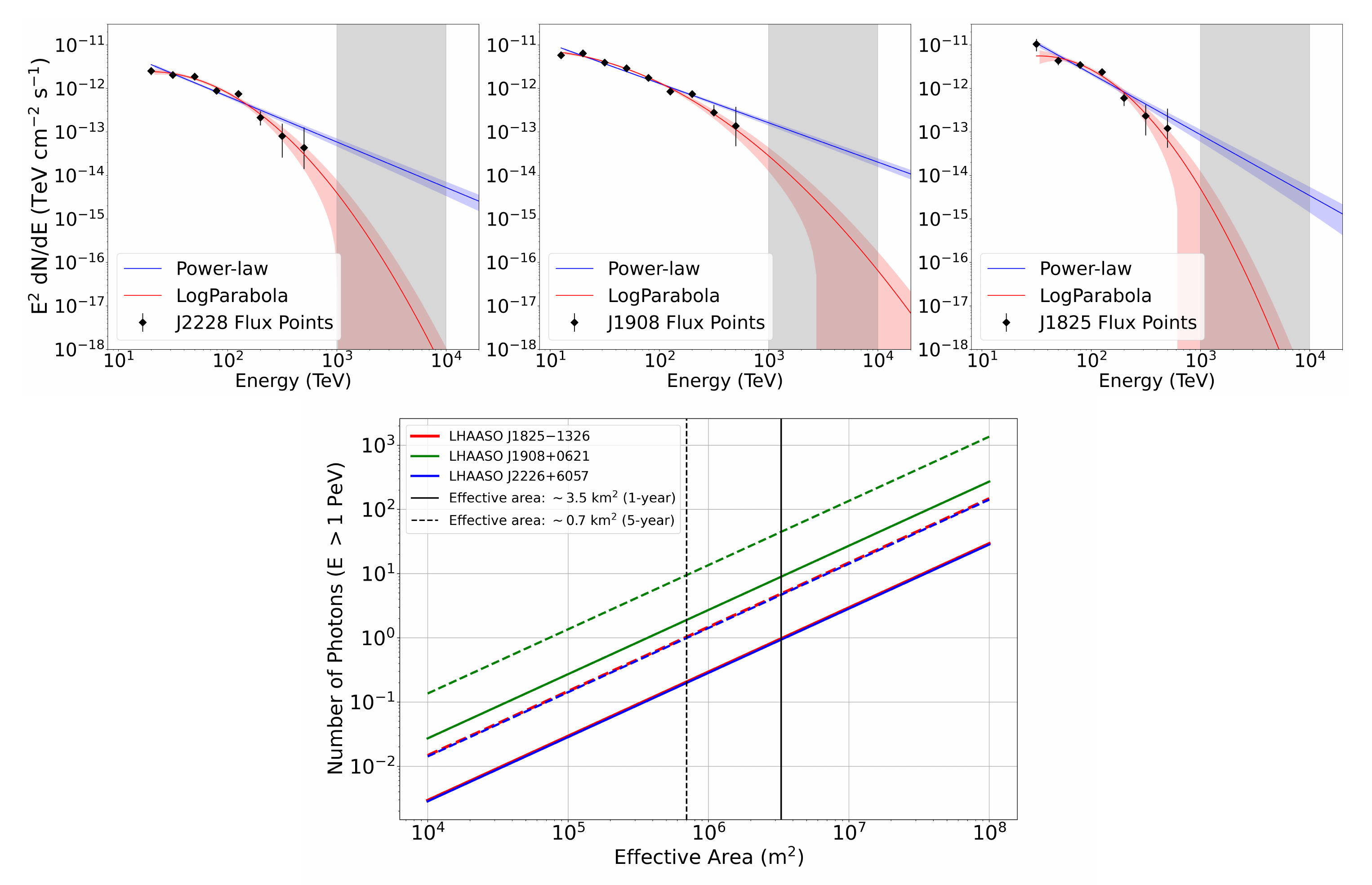}
\caption{Top panel: The reconstructed $\gamma$-ray spectra of LHAASO~J2226$+$6057, LHAASO~J1908$+$0621 and LHAASO~J1825$-$1326 obtained from LHAASO observations are shown in left, middle, and right figures, respectively. The reconstructed flux point with 1$\sigma$ statistical errors are shown with black points, while the best-fit log-parabola and power-law models are shown with red and blue lines, respectively. The shaded regions show 1$\sigma$ error intervals of the models. The models are extrapolated up to above 10~PeV energies. The gray shaded regions cover 1$-$10~PeV energy range. Bottom figure: Estimated number of detected PeV photon from these three LHAASO UHE sources as a function of effective areas, for 1-year (solid-lines) and 5-year (dashed-lines) of observations. The solid and dashed vertical lines mark the specific effective area value needed for detection of at least 1~PeV photon from all of these sources after 1-year and 5-year of observations, respectively.}
\label{pev_detect}
\end{figure}

This subsection presents a study that aims to estimate the minimum required effective area of particle arrays to detect PeV photons. Effective area of an instrument varies with energy and zenith angle, and is generally larger than the physical area of the instrument. Figure~\ref{aeff} left and right panels show examples of effective areas from LHAASO KM2A and future CTA experiment, respectively. To estimate minimum required effective areas, three UHE sources (LHAASO~J2226$+$6057,~LHAASO~J1908$+$0621 and LHAASO~J1825$-$1326) that exhibit significant E$>$100~TeV emission are selected as reference sources \cite{lhaaso2021}. Figure~\ref{pev_detect} (top panel) shows the best-fit log-parabola\footnote{The best-fit log-parabola spectral models, [$E$/(10~TeV)]$^{-a-blog[E/(10~\text{TeV})]}$, provided in \cite{lhaaso2021} as a = 1.56 and b = 0.88 for LHAASO~J2226$+$6057, a = 2.27 and b = 0.46 for LHAASO~J1908$+$0621, and a = 0.92 and b = 1.19 for LHAASO~J1825$-$1326} (red) and power-law (blue) spectral models obtained from the analysis of these sources, along with the reconstructed flux points from LHAASO observations. The number of PeV photons from these sources is estimated using the integral flux between E=[1.0, 10.0]~PeV (gray shaded regions in Figure~\ref{pev_detect}, top panel) obtained by extrapolating the provided best-fit log-parabola models to high energies, assuming two different exposure times of 1-year and 5-year for a set of effective area values.

The outcomes of the study are illustrated in Figure~\ref{pev_detect} (bottom panel), where each source is represented by a line with a different color, and the solid and dashed lines demonstrate the evolution of effective areas for 1-year and 5-year of observation, respectively. It is evident from the figure that, to detect at least one PeV photon from all three sources within 1-year of observation time, a minimum of $\sim$3.5~km$^{2}$ effective area is required, whereas the requirement is reduced to $\sim$0.7~km$^{2}$ when considering 5-year observations. From Figure~\ref{pev_detect} (top panel), it is evident that LHAASO~J1908$+$0621 is the brightest source in the energy range of E=[1.0, 10.0]~PeV. Assuming an effective area of 3.5 km$^{2}$, it is expected to detect around $\mathcal{O}(10)$~PeV photons from this source after 1-year of observation, providing reasonable statistics for spectral studies. In conclusion, future ground-based particle arrays would require effective areas A$_{\text{Eff}}$ $\geq\sim4-5$ km$^{2}$ to perform detailed UHE spectroscopy that goes beyond PeV energies.

\section{Galactic PeVatrons}
\label{pev_def_main} 

\begin{table*}
\caption{List of Galactic sources currently known to produce $\gamma$-ray emission above 100~TeV, together with their possible nature. PSR: Pulsar, SNR: Supernove Remnant, YMC: Young Massive Clusters, HII region: Interstellar matter consisting of ionized hydrogen atoms, Dark: Having no known counterpart. The table is adopted from \citep{alison_ICRC}. There are no photons with E$>$100~TeV detected from the direction of Westerlund~1 \citep{westerlund1}, Galactic~center \citep{hessGCNat} and HESS~J1641$-$463 \citep{j1641_463} but the source spectra are expected to go beyond 100~TeV energies.}
\small
\label{table_e100}     
\centering                        
\begin{tabular}{c c c c}       
\hline\hline                 
\textbf{Source } & \textbf{Location} & \textbf{E$>$100 TeV} & \textbf{Possible origin} \\ 
\textbf{name } & \textbf{(\textit{l, b})} &  \textbf{Detection instrument} &\\ 
\hline       
Crab Nebula         & (184.557, $-$5.784) & HAWC, MAGIC, LHAASO, Tibet-AS$\gamma$ & PSR \\
HESS~J1702$-$420    & (344.304, $-$0.184  & H.E.S.S. & Dark \\
eHWC~J1825$-$134    & (18.116, $-$0.46)   & HAWC, LHAASO & PSR \\
LHAASO~J1839$-$0545 & (26.49, $-$0.04)    & LHAASO & PSR\\
LHAASO~J1843$-$0338 & (28.722, 0.21)      & LHAASO & SNR \\
LHAASO~J1849$-$0003 & (32.655, 0.43)      & LHAASO & PSR, YMC \\
eHWC~J1907$+$06     & (40.401, $-$0.70)   & HAWC, LHAASO & PSR, SNR \\
LHAASO~J1929$+$1745 & (52.94, 0.04)       & LHAASO & PSR, SNR \\
LHAASO~J1956$+$2845 & (65.58, 0.10)       & LHAASO & PSR, SNR \\
eHWC~J2019$+$368    & (75.017, 0.283)     & HAWC, LHAASO & PSR, HII/YMC \\
LHAASO~J2032$+$4102 & (79.89, 0.79)       & LHAASO & YMC, PSR, SNR \\
LHAASO~J2108$+$5157 & (92.28, 2.87)       & LHAASO & Dark \\
TeV~J2227$+$609     & (106.259, 2.73)     & Tibet-AS$\gamma$, LHAASO & PSR, SNR \\
\hline 
Westerlund 1        & (339.55, $-$0.35)   & H.E.S.S. & YMC \\
Galactic center     & (359.94, $-$0.04)   & H.E.S.S. & SMBH \\
HESS~J1641$-$463    & (338.52, 0.09)      & H.E.S.S. & Dark \\
\hline  
\hline 
\end{tabular}
\end{table*}

Before the remarkable results from LHAASO \citep{lhaaso2021} in 2021 and the start of UHE $\gamma$-ray astronomy era, the term PeVatron was exclusively used to refer to hadronic CR accelerators. This was motivated by the quest to identify the sources of CRs, particularly the factories in the Galaxy that can accelerate hadrons to PeV energies. As a result of this definition, researchers focused on searching for CR sources that exhibit hard hadronic spectra (i.e.~$\Gamma_{p}$~$\sim$~2.0) without showing any spectral cutoff up to PeV energies and beyond. This definition has been historically used to investigate the origin of Galactic CRs. The energy of $\gamma$ rays resulting from pp-interactions is typically around 10 times lower than the energy of their parent protons \citep{kernel2006}. Hence, it was believed that the detection of $\gamma$-ray events with energies E$>$100~TeV would be a direct indication of Galactic PeVatrons. Furthermore, it was believed that the ambiguity between the hadronic and leptonic origins of the observed emissions would be naturally resolved, since electrons are dramatically inefficient in producing $\gamma$ rays at energies above 50~TeV due to the Klein-Nishina (KN) suppression \citep{aharonian2004Book}. As a result, it was anticipated that any $\gamma$-ray facility capable of operating efficiently above 100~TeV would solve the enigma of the origin of Galactic CRs, which has puzzled scientists for over a century.

The groundbreaking results of the LHAASO collaboration have completely revolutionized our understanding of PeVatrons. The collaboration has reported significant\footnote{The detection significance threshold of UHE sources is set to 7$\sigma$ level for these results.} UHE $\gamma$-ray emission of E$>$100~TeV from 12 Galactic sources located in the Northern sky, with photon energies ranging from 0.1~PeV to 1.4~PeV \citep{lhaaso2021}. Despite the presence of potential counterparts in their vicinity, such as SNRs and SFRs, several objects detected above 100~TeV are plausibly associated to pulsars and PWNe, where leptonic processes are expected to dominate the observed $\gamma$-ray emission. Currently, there are several Galactic source regions where UHE $\gamma$-ray emission is observed (or highly anticipated), and almost all of them have been detected by the LHAASO experiment.~Table~\ref{table_e100} provides a list of these source regions, as well as their possible origins. Moreover, for the first time in history, PeV photons have been detected from the direction of the Crab Nebula \citep{LHAASO_Crab}, which is a very well-known leptonic accelerator. These remarkable experimental results have given rise to a new concept referred to as  'leptonic PeVatron' in the literature, which further complicates the mystery of Galactic PeVatrons.

\subsection{Definition of Galactic PeVatrons}
\label{pev_def}
At present, there exists an ambiguity in the literature regarding the definition of 'PeVatron'. Nonetheless, it is clear that a 'PeVatron' refers to an astrophysical source that can accelerate 'individual' particles, either leptons or hadrons, to energies of at least a PeV and beyond. The ambiguity arises from the historical context of the term as discussed above. Based on the recent observational results, it is possible to provide general definitions for PeVatrons, which can be classified into three subcategories:

\begin{itemize}
  \item Leptonic PeVatron: an astrophysical object capable of accelerating 'individual' leptons to at least 1 PeV energies and beyond. 
  \item Hadronic PeVatron: an astrophysical object capable of accelerating 'individual' hadrons to at least 1 PeV energies and beyond.
  \item CR knee PeVatron: an astrophysical object capable of accelerating 'bulk of hadrons' to at least knee energies seen in the CR spectrum and beyond.
\end{itemize}

The categorization of PeVatrons as above has two crucial implications. Firstly, regardless of their strength, leptonic PeVatrons have negligible impact on the observed CR spectrum as the composition of leptons is only $\sim$2$\%$. Thus, they cannot be considered as the origin of Galactic CRs. Secondly, a source that is a hadronic PeVatron capable of producing detectable E$>$100 TeV $\gamma$-ray emissions does not necessarily have to be the source of Galactic CRs with energies around the knee of the CR spectrum, since the all-particle CR spectrum knee is well above 1~PeV. Although it is plausible that individual hadrons having PeV energies exist in the source region, the abundance of these hadrons with PeV energies may not be sufficient to make a significant contribution to the CR spectrum at knee energies. To make this distinction clear, a new category called 'CR knee PeVatron' has been introduced, which represents the class of hadronic sources capable of accelerating CRs well above the knee energy. In this case, the knee energy should be at least around $\sim$700~TeV if only protons and helium are considered \cite{knee_below_1pev}. As a result, the hadronic spectrum of CR knee PeVatrons must extend well beyond knee energies without showing any clear indication of a spectral cutoff, i.e. appearing as a power-law spectrum. At that point, it is important to stress on the distinction between the 'cutoff energy' of a hadronic spectrum and 'the maximum energy of individual particles' (hadrons) within that spectrum. For instance, even though a hadronic spectrum may have a spectral cutoff at $E_\mathrm{cut,\,p}$~=~500~TeV, it cannot contribute significantly to the CR spectrum at knee energies of 3~PeV, as it cuts off well below 3~PeV. However, there may still be PeV particles present in this spectrum due to the falling edge around the cutoff energies. This analogy can be compared to a VHE $\gamma$-ray source that exhibits a significant spectral cutoff at 15~TeV, but still having $\gamma$-ray events well above 15~TeV, i.e. 30$-$40 TeV. In this specific scenario, the source is considered as a hadronic PeVatron, but not a CR knee PeVatron.

\subsection{An example of a CR knee PeVatron}
\label{CRKneeEx}
\begin{figure}[ht!]
\centering
\includegraphics[width=17cm]{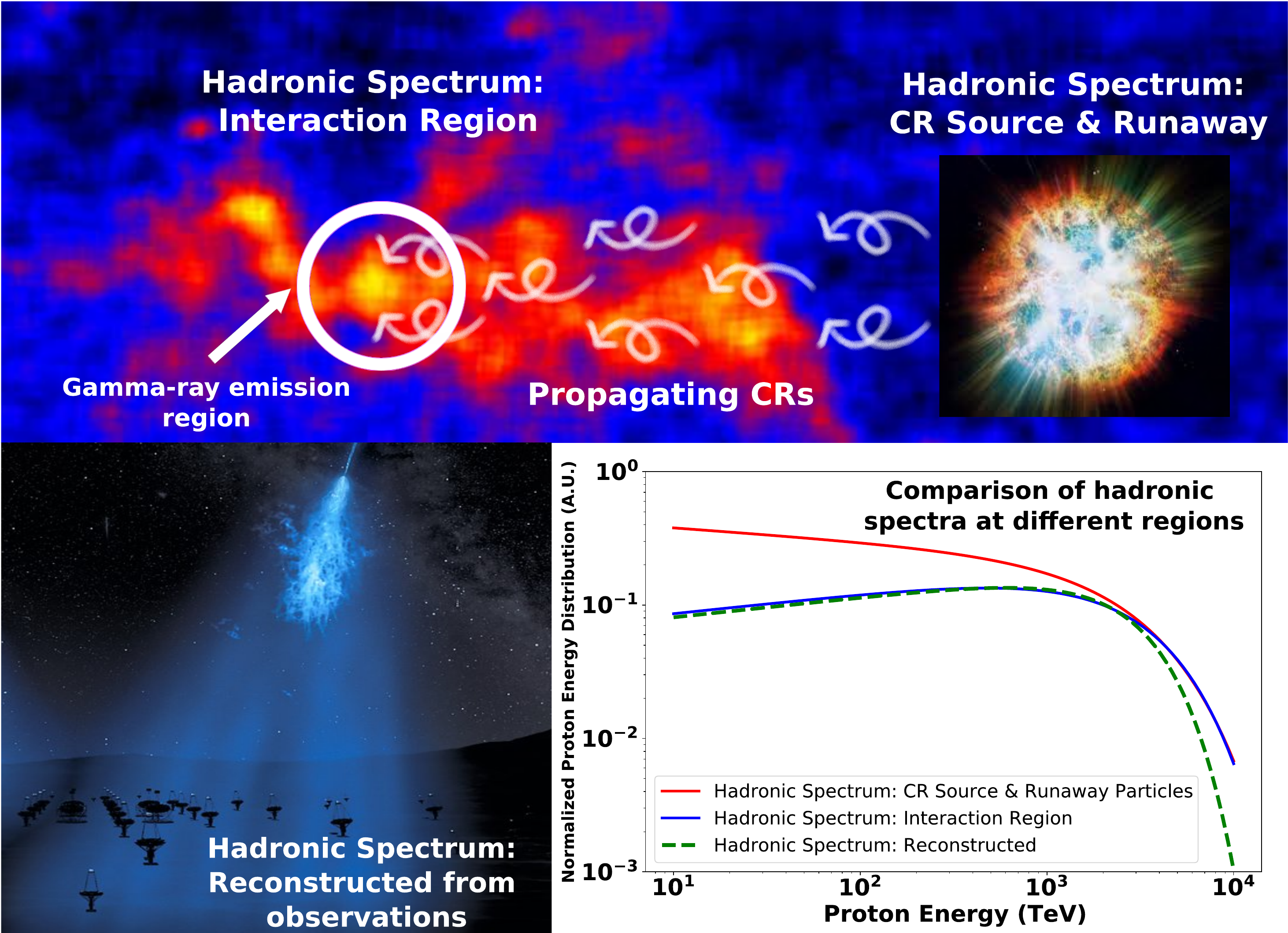}
\caption{Top panel: artistic illustration of CRs accelerated by a CR knee PeVatron source (shown at the right hand side as an exploding astrophysical object), first propagating in the interstellar medium as marked by white arrows, and then interacting with target material, a molecular cloud, giving rise to UHE $\gamma$-ray emission from a particular region of the cloud. Bottom panel: (left) demonstration of $\gamma$ rays detected by a ground-based observatory; (right) comparison between the hadronic spectra at the source region (red solid-line), at the region of interaction (blue solid-line) and reconstructed using the data from ground-based instruments (green dashed-line).}
\label{pev_illust}
\end{figure}
The artistic illustration shown in Figure~\ref{pev_illust} depicts how CRs are accelerated by a 'CR knee PeVatron' source with a true proton cutoff energy of 3~PeV, and then interact with dense molecular clouds in its vicinity, leading to the emission of UHE $\gamma$ rays. The red spectrum in the lower right panel of Figure~\ref{pev_illust} represents the runaway CRs from the source region. It is worth noting that the spectrum of runaway CRs, and the CRs that are actually accelerated at the source may differ depending on environmental conditions, i.e.~low energy CRs may be trapped in the source region due to strong magnetic fields, causing the runaway spectrum to be harder. The CRs that run away from their source begin to travel through the ISM, where the density of gas may vary depending on location. Eventually, these runaway CRs may reach a dense gas region where pp interaction takes place, resulting in the efficient production of $\gamma$ rays. It is important to note that while a dense gas region illuminated by CRs is evidence of PeVatron activity, the region does not necessarily host the PeVatron source itself, as illustrated in Figure~\ref{pev_illust}. In other words, the PeVatron source itself and the region where PeVatron activity is detected may be located far away from each other. The spectrum shown with blue solid line in the inset of Figure~\ref{pev_illust} represents an illustration of the CR spectrum at the interaction region. During the propagation from the source to the interaction region, some of the runaway CRs, particularly those with low energy, may lose their energy along the way. As a result, the modified spectrum at the interaction region (blue) can be harder than the initial spectrum at the source (red), and this modification depends strongly on the distance between the source and the interaction region, as well as the ISM properties, i.e.~gas density profile along the propagation path.

The interaction region (indicated by a white circle in Figure~\ref{pev_illust}) produces significant $\gamma$-ray emissions that can be detected by current $\gamma$-ray observatories. Different analysis techniques are then used to reconstruct the underlying hadronic spectrum based on the observed $\gamma$-ray spectrum. However, another important factor must be taken into consideration if relevant. It is well-known that UHE $\gamma$ rays with energies above 200$-$300~TeV are attenuated due to $\gamma\gamma$ absorption effects, caused by interstellar radiation fields, particularly the cosmic microwave background (CMB) radiation \citep{gamma_abs} (a detailed discussion is provided in subsection~\ref{gammagammaabs}). As a result, the total number of UHE photons detected from Earth, particularly with energies above 200$-$300~TeV, will be reduced depending on the distance from the source, leading to decreased flux at high energies, which may even appear as an artificial spectral cutoff effect. An example of the reconstructed hadronic spectrum based on ground-based $\gamma$-ray observations is shown with the dashed green line in the inset of Figure~\ref{pev_illust}. Therefore, if the distance to the $\gamma$-ray source is far enough for such absorption effects to take place, the reconstructed spectrum should be corrected. On the other hand, if the distance is close enough, these effects can be disregarded.

The $\gamma$-ray emission region (marked with a white circle in Figure~\ref{pev_illust}) used to reconstruct the underlying hadronic spectrum is a crucial factor when searching for the origin of Galactic CRs. The size of the emission region is determined by observational (e.g., angular resolution of the experiment) and physical (e.g., actual size of the interaction region) conditions, and can be large in many cases, leading to contamination effects coming from nearby $\gamma$-ray sources in the region. Therefore, it is crucial to locate the actual hadronic emission region accurately by utilizing sophisticated analysis techniques and high angular resolution observations to resolve any source confusion. Furthermore, the search for the origin of CRs should not always focus on individual sources but instead take into account the entire CR spectrum, which is actually the cumulative effect of acceleration in various types of astrophysical sources. From an experimental point of view, the following critical conditions must be met for a robust solution to the question of the origin of Galactic CRs:

\begin{itemize}
  \item Sources of the origin of Galactic CRs must accelerate hadrons above knee energies, and the hadronic spectrum must go well beyond knee energies without showing a clear indication of a spectral cutoff.
  \item There must be enough dense target material for the accelerated CRs to interact and produce secondary $\gamma$-ray emission, and the effects of CR propagation between the source and the interaction region should be taken into account if necessary.
  \item Significant E$>$100 TeV $\gamma$-ray emission from the source is mandatory to establish its PeVatron nature. Additionally, broadband $\gamma$-ray data from HE to UHE supported by multiwavelength observations are necessary to prove the hadronic origin of the emission, and determine the shape of the underlying hadronic spectrum.
  \item $\gamma$-$\gamma$ absorption effects, especially for the sources emitting $\gamma$ rays well above 100~TeV, should be considered. 
  \item Observations with high angular resolution are needed to accurately locate the actual emission region and resolve any source confusion.
\end{itemize}

\subsection{A special case: the Crab Nebula}

The Crab Nebula is a well-studied astronomical object in the Milky Way, and powered by the energetic Crab pulsar with the spin down luminosity of $\dot{E}$ = $\sim$5$\times$10$^{38}$ erg/s and the characteristic age of $\tau_{c}$=1260 years \citep{crab_rev,crab_rev2}. Significant fraction of the rotational energy from the pulsar is transformed into the generation of a confined relativistic magnetized wind, which is commonly referred to as pulsar wind nebula (PWN). The Crab Nebula is considered the prototypical example of PWNe sources, and bright enough to be detectable across a very wide energy range, spanning over 20 decades from low radio at the tens of MHz range to recently detected UHE $\gamma$-rays. The $\gamma$-ray emission from the nebula has dominantly leptonic origin and produced by the IC radiation of electrons \cite{crab1996_felix, blumenthal}, with the presence of far-infrared (FIR), CMB, and synchrotron photons acting efficiently as target photon fields for IC. The TeV emission from the nebula can be well-explained by upscattering of synchrotron photon fields through the synchrotron self-Compton (SSC) process, while at UHEs, the SSC process is suppressed due to KN suppression effects. Instead, IC scattering of CMB photons becomes the dominant process at these extremely high energies.

\begin{figure}[ht!]
\centering
\includegraphics[width=17cm]{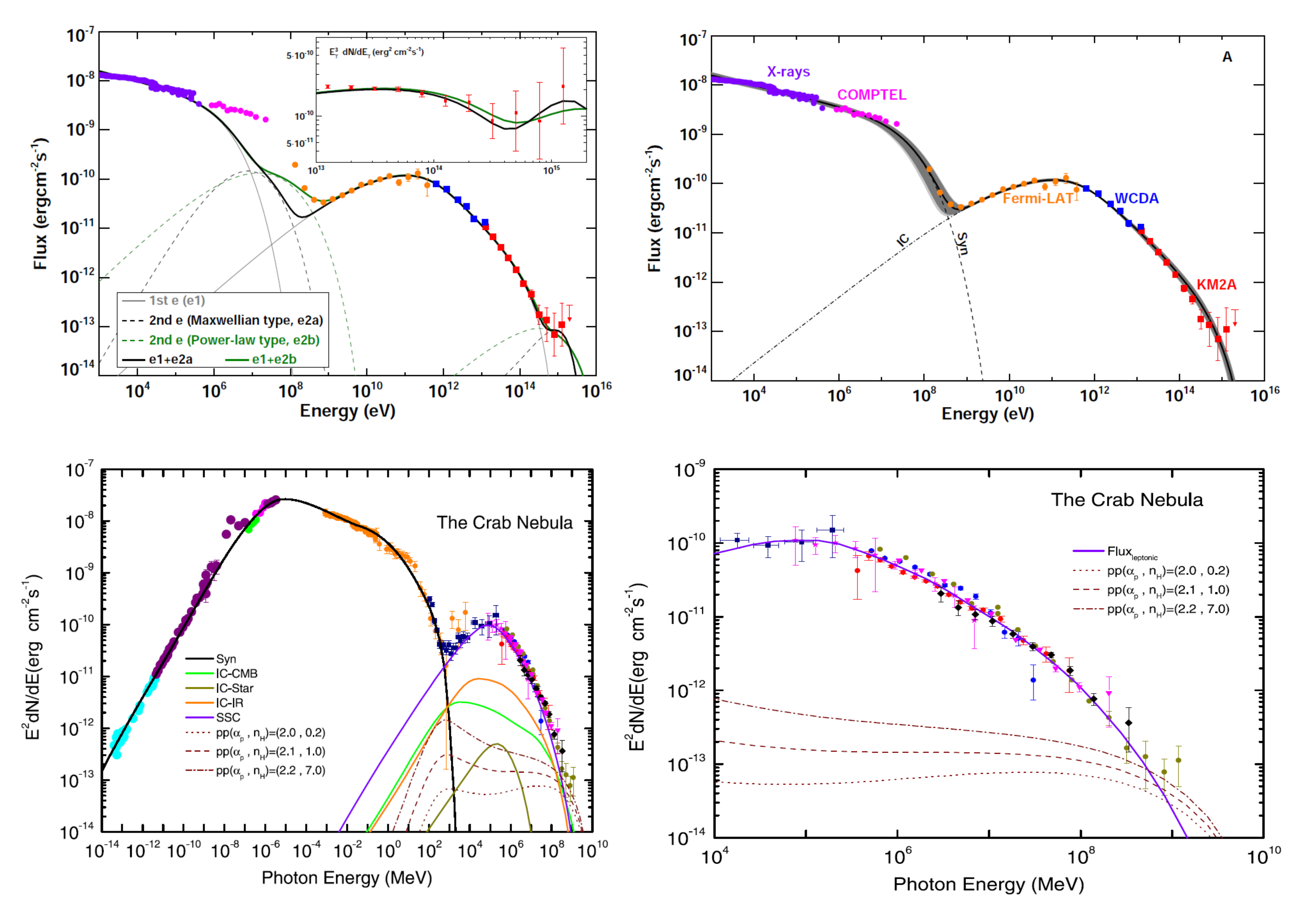}
\caption{Top panels: (left) a two-zone scenario with two electron populations. The first electron component (e1) is the same as the one in the right figure, except for the smaller cutoff energy E$_{0}$=450~TeV. Two type of distributions are assumed for the 2nd electron population. (a) Maxwellian-type spectrum of the temperature E$_{2}$=450~TeV (e2a), and (b) power-law spectrum with a slope fixed at 1.5, an exponential cutoff at E$_{2}$=2~PeV (e2b). The solid grey curve shows the radiation of the first component. The dashed black and green curves represent respectively the radiations of the second component in two spectral forms. The thick solid black curve shows the sum of e1 and e2a, while the thick solid green curve shows sum of e1 and e2b. (Right) The black curves represent the fluxes of the synchrotron and IC components of radiation of an electron population calculated within the one-zone model. The electron spectrum above 1~TeV is assumed to be a power-law function terminated by an superexponential cutoff. The dark-grey and light-grey shaded regions show the 1$\sigma$ and 3$\sigma$ error bands, respectively. The purple and the magenta circles show the X-ray and the MeV emission of the Crab Nebula. The orange circles represent the Crab observations by Fermi-LAT in the nonflare state. The blue and red squares represent LHAASO WCDA and KM2A measurements. The top panels and captions are taken from \citep{LHAASO_Crab}. Bottom panels: (Left) The SEDs of the Crab Nebula are plotted where both leptonic and hadronic contributions are considered. For a direct comparison, observed multiband data are also displayed. (Right) A zoomed-in plot of bottom left figure where the range of photon energy is from 10$^{4}$~MeV to 10~PeV. The solid line stands for the summed leptonic contributions while the dashed (dotted and dotted–dashed) dark red lines are proton spectra with different spectral indices. The bottom figures and captions are taken from \citep{Peng_2022}.}
\label{crab_spectra}
\end{figure}

For the first time in history, the LHAASO collaboration has detected PeV photons coming from the direction of the Crab Nebula \citep{LHAASO_Crab}, marking a significant breakthrough in PWN research. This groundbreaking experimental result has opened up a new era for PWN studies. The published UHE spectrum of the Crab Nebula by the LHAASO collaboration is shown in Figure~\ref{crab_spectra} along with spectra derived from observations in other energy bands. The most remarkable feature seen in the nebula spectrum is the potential hardening observed at the high energy end, which appears around PeV energies. As discussed in detail in \citep{LHAASO_Crab}, while a one-component electron population model can adequately explain the Crab Nebula's spectral energy distribution (SED) from X-rays to multi-MeV energies with synchrotron radiation, and from TeV to PeV energies with IC radiation, as seen in the top right panel of Figure~\ref{crab_spectra}, it fails to explain the spectral hardening observed at the high energy end. Since it is difficult to explain hardening of the electron spectrum within reasonable assumptions, the hardening feature is instead interpreted by the authors as the presence of a second spectral component hidden under the IC component emerging at high energies. The authors of \citep{LHAASO_Crab} discussed two main scenarios to explain this hardening feature: the first scenario involves a second electron component that could extend the spectrum up to $\sim$4~PeV energies, with two different possible types of secondary electron spectrum components, Maxwellian and power-law, being used to explain the SED (as seen in the top left panel of Figure~\ref{crab_spectra}). The second scenario assumes the existence of an underlying proton spectrum that contributes solely to PeV energies, as demonstrated in the bottom left and right panels of Figure~\ref{crab_spectra}. Regardless of whether the UHE photons are produced through hadronic or leptonic processes, the presence of PeV particles at the source region is evident for both cases, making the Crab Nebula the first "robust" PeVatron source identified in the Galaxy. Furthermore, if the hardening feature observed at the high energy end of the Crab spectrum originates from an underlying proton spectrum, this would classify the Crab Nebula as a CR knee PeVatron, since the expected cutoff in the proton spectrum is predicted to be between 10~PeV and 100~PeV \cite{pev_crab_hadron} in this case.

\subsection{A word on leptonic PeVatrons}

The term "leptonic PeVatron" has become increasingly prevalent in scientific literature, particularly after the groundbreaking results from LHAASO, as it discusses the possibility that several UHE sources are associated with leptonic accelerators \citep{lhaaso2021}. As defined before in subsection ~\ref{pev_def}, a leptonic accelerator must contain PeV leptons in order to be classified as a PeVatron. As it has just been discussed, the Crab Nebula is a prime example of leptonic PeVatrons and actually it has been known to contain PeV leptons for at least a decade based on its synchrotron spectrum in the GeV band \citep{crab_rev2}.

It is widely known that PWNe are the most efficient electron accelerators in the Milky Way. These nebulae are powered by energetic pulsars that inject ultrarelativistic particles into their magnetosphere, forming an expanding wind that extends up to the termination shock. The VHE and UHE $\gamma$ rays emitted by these objects are primarily produced by electrons via IC scattering of low energy photon fields, especially the 2.7~K CMB radiation. More precisely, as it was reviewed in \citep{Breuhaus_2021} in detail, an injection electron spectrum accelerated by the leptonic accelerator that follows a power-law distribution with an index of $\alpha_{e, inj}$ results in a softer equilibrium electron spectrum\footnote{Equilibrium electron spectrum is the case where energy losses due to by synchrotron and IC emission are balanced by the injection and acceleration of new electrons from the CR source.}, with a power-law index of $\alpha_{e,eq}$~=~($\alpha_{e,inj}$~+~1). The resulting $\gamma$-ray emission from this equilibrium electron population then follows a power-law distribution with a photon index of $\Gamma_{\gamma,Th}$~=~($\alpha_{e, inj}$~+~2)/2 in the Thomson regime, and steepens to $\Gamma_{\gamma,KN}$~=~($\alpha_{e,inj}$~+~2) in the KN regime for energies above $E_{KN}$ $\sim$ m$_{e}^{2}$c$^{4}$/E$_{Rad}$, where E$_{Rad}$ is the target photon energy. At energies above several TeV, the scattering process happens in the deep KN regime, where electrons can lose almost all of their energy even in a single scattering event, making the IC process extremely inefficient. Consequently, the highest energy of photons produced $\sim$corresponds to the maximum energy of the accelerated electrons. According to \cite{LHAASO_Crab}, the relationship between the maximum energies of $\gamma$-rays and electrons are given by equations
\begin{equation}
\label{e_gamma} 
E_{\gamma,max} = 0.37 (E_{e,max} / 1~\text{PeV})^{1.3} \text{ PeV},
\end{equation}
\begin{equation}
\label{e_elec} 
E_{e,max} = 2.15 (E_{\gamma,max} / 1~\text{PeV})^{0.77} \text{ PeV},
\end{equation}
which can provide accuracies better than 10$\%$ above 30 TeV energies.

Extensive research has been conducted in \citep{pevatron_pulsars} on the relationship between pulsars and PeVatrons, providing valuable relation on the maximum energy of accelerated particles by such objects as
\begin{equation}
\label{e_par} 
E_{\text{par,max}} = 2~\eta_{e}~\eta_{B}^{1/2}~\dot{E}^{1/2} \text{ PeV},
\end{equation}
where $\eta_{e}$ is the ratio between the electric field strength and magnetic field strength, $\eta_{B}$ is the magnetic energy density expressed as a fraction of the pulsar wind energy flux, and $\dot{E}$ is the spin-down luminosity of the pulsar. As it was pointed out by the authors, Eq.~\ref{e_par} applies to all kinds of particles that can be accelerated in pulsars, irrespective of whether they are electrons, positrons, or hadrons. For instance, assuming a magnetic field strength of 112~$\mu$G \citep{LHAASO_Crab} and using this value in Eq.~\ref{e_par}, an efficiency of $\eta_{e}$~=~0.16 is required for the Crab Nebula. Although such an high efficiency of $\eta_{e}$~=~0.16 sounds particularly impressive when compared to other powerful Galactic CR accelerators like SNRs, which typically have efficiencies just around $\eta_{e}$ of $\sim$10$^{-3}$ \citep{non_dsa_hard}, this is actually not surprising given that acceleration in PWNe occurs in a relativistic environment where the electric field can be comparable to the magnetic field.

\begin{figure}[ht!]
\centering
\includegraphics[width=14cm]{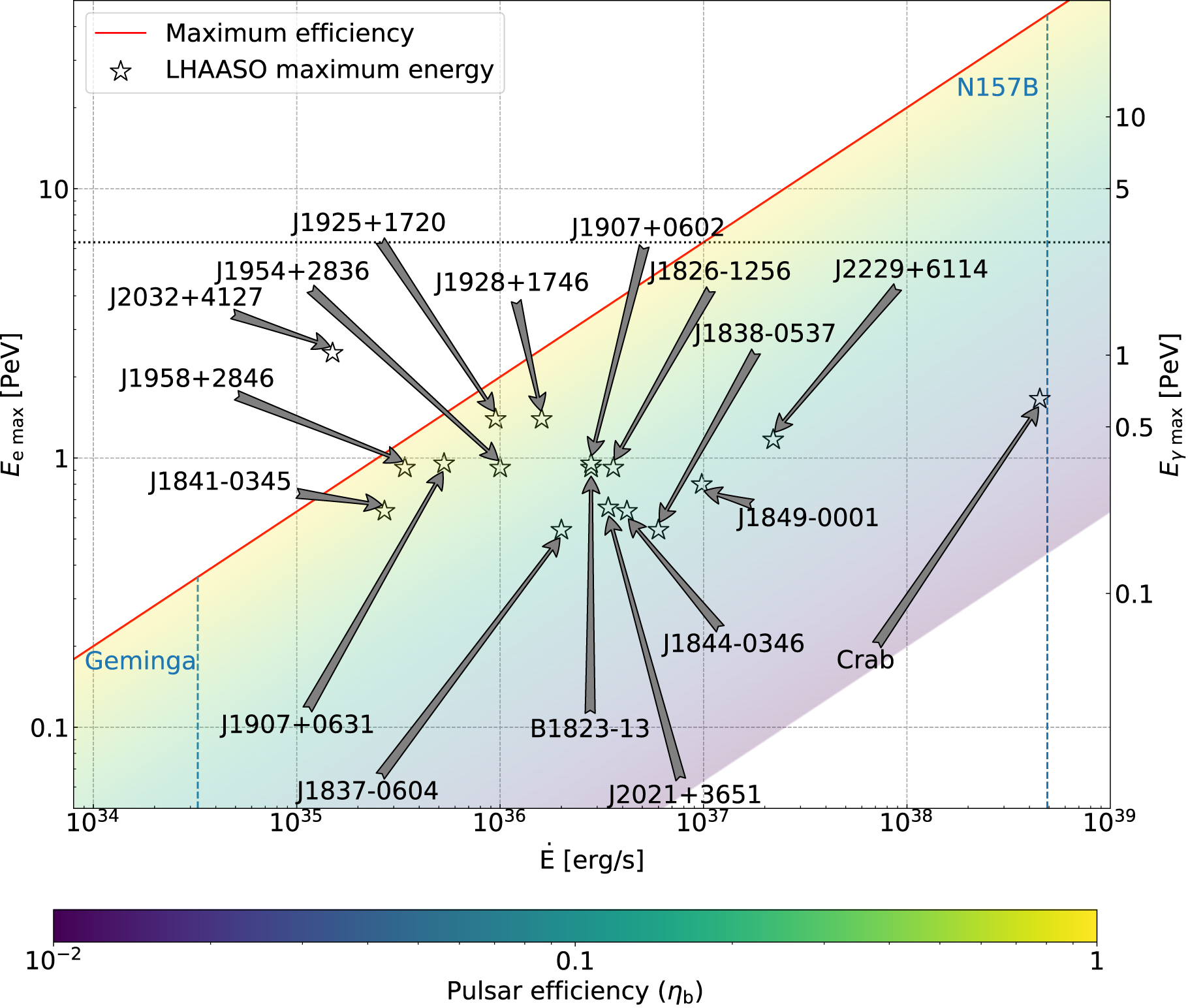}
\caption{Maximum electron energy derived from the LHAASO spectra versus spin-down power of the colocated pulsars. The right Y-axis shows the corresponding gamma-ray energy. The colored area shows the values for $\eta_{e}$ and $\eta_{B}$ ranging from 0.01 to 1, with the red line indicating the limiting value corresponding to maximally efficient acceleration $\eta_{e}$ = 1 and $\eta_{B}$ = 1. The dotted black line marks the upper limit to the maximum energy for young pulsars with large magnetic field of 100 $\mu$G. The blue dashed horizontal lines show the predicted values for PWNe associated to Geminga and N157B. The figure and caption are taken from \citep{pevatron_pulsars}.}
\label{pwn_photons}
\end{figure}

Figure \ref{pwn_photons} illustrates the highest achievable energy of electrons that can be accelerated by pulsars, as well as the corresponding maximum energy of $\gamma$-rays that can be produced through IC scattering of the CMB. The selected pulsars in the vicinity of LHAASO sources are also shown to investigate whether these pulsars have the potential to power the UHE LHAASO sources. The potential associations of these pulsars in the $E_{e,max}$--$\dot{E}$ plane are presented, along with the known $\dot{E}$ values, and then compared to theoretical predictions based on Eq.~\ref{e_par} using reasonable values of $\eta_{e}$ and $\eta_{B}$ ranging between 0.01 and 1. It is evident from the figure that powerful pulsars ($\dot{E}$ $\geq$ 10$^{36}$ erg/s) with reasonable efficiencies can indeed produce UHE $\gamma$ rays well above 100~TeV energies. 

\begin{figure}[ht!]
\centering
\includegraphics[width=18cm]{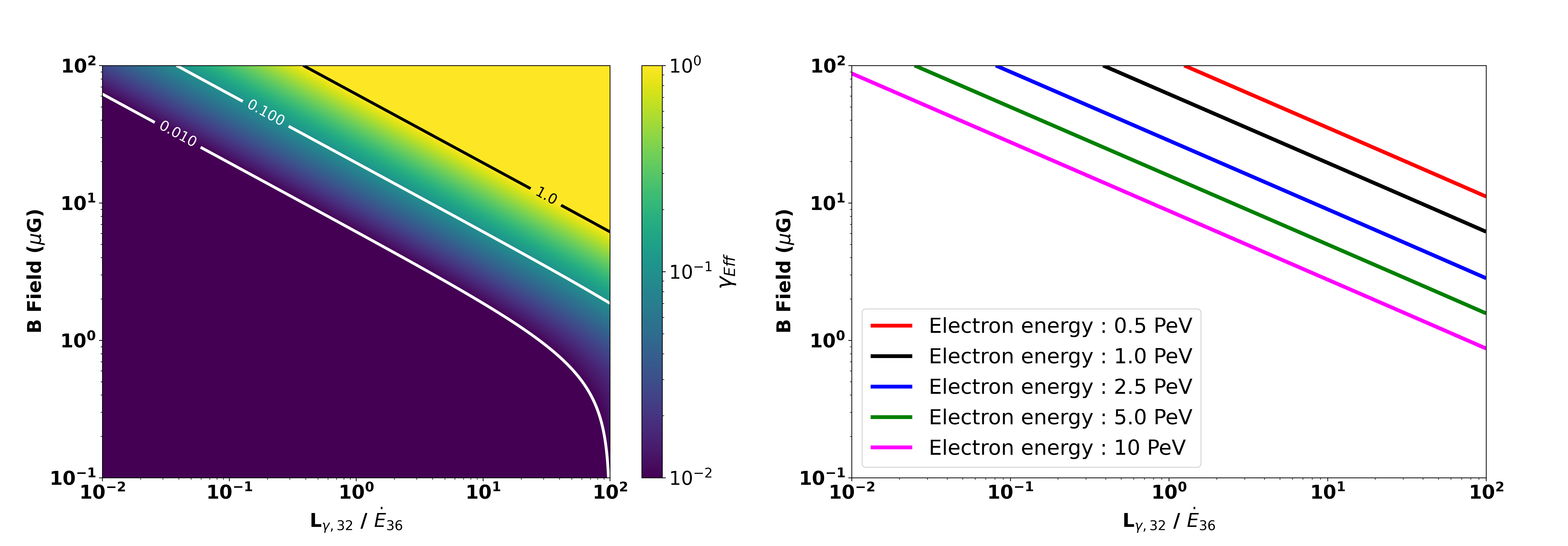}
\caption{Left figure: gamma-ray efficiencies given in 2D B -- ($L_{\gamma,32}$/$\dot{E}_{36}$) phase space for an electron energy of 1~PeV. The z-axis shows the $\gamma_{Eff}$ values obtained from Eq.~\ref{gamma_eff}. The black line indicates $\gamma_{Eff}$=1 values throughout the phase space, therefore provides upper limit boundaries. In addition, $\gamma_{Eff}$=0.1 and $\gamma_{Eff}$=0.01 contours are also shown with white lines. Right figure: comparison of $\gamma_{Eff}$ boundaries for a set of different electron energies. The black line corresponds to the electron energy of 1~PeV shown in the left plot, while $E_{e}$ = 0.5, 2.5, 5.0 and 10.0~PeV $\gamma_{Eff}$ boundaries are shown with red, blue, green and magenta lines, respectively.}
\label{gamma_eff_plot}
\end{figure}

Another limiting factor was pointed out by the authors and can be used to evaluate pulsars as UHE sources is based on the maximum allowed conversion efficiency from pulsar rotational power to $\gamma$ rays \citep{pevatron_pulsars}. This constraint can be determined by comparing the IC luminosity of UHE sources with the pulsar spin-down luminosity. The $\gamma$-ray luminosity of UHE sources, denoted as L$_{\gamma}$ = 4$\pi$d$^{2}$F$_{\gamma}$, can be obtained through related $\gamma$-ray observations, where d represents distance to the source. As it was derived in \citep{pevatron_pulsars}, the $\gamma$-ray efficiency $\gamma_{Eff}$, which is the fraction of $\dot{E}$ converted to $\gamma$-ray emitting electrons, can be written as 
\begin{equation}
\label{gamma_eff} 
\gamma_{Eff} = 10^{-4} \frac{L_{\gamma,32}}{\dot{E}_{36}}(1+260 \text{ }E^{1.7}_{e,15}\text{ }B^{2}_{-5}),
\end{equation}
where $L_{\gamma,32}$ is the $\gamma$-ray luminosity of the UHE source (in units of 10$^{32}$ erg/s), $\dot{E}_{36}$ is the spin-down luminosity of the pulsar (in units of 10$^{36}$ erg/s), $E_{e,15}$ is the electron energy (in units of 10$^{15}$ eV, or equivalently PeV), and B$_{-5}$ is the magnetic field strength (in units of 10$^{-5}$ G, or in 10$\mu$G). The maximum allowed value of $\gamma_{Eff}$ cannot exceed 1, consequently, boundaries can be determined using Eq.~\ref{gamma_eff}. The black line in Figure~\ref{gamma_eff_plot} (left) shows the $\gamma_{Eff}$=1 boundary for a given B--($L_{\gamma,32}$/$\dot{E}_{36}$) phase space for the fixed electron energy of 1~PeV; therefore, the yellow region in the plot represents the forbidden regions of the phase space. Similarly, Figure~\ref{gamma_eff_plot} (right) shows how $\gamma_{Eff}$=1 boundary changes within the same phase space depending on the electron energies, given with different colored lines.

\begin{figure}[ht!]
\centering
\includegraphics[width=17cm]{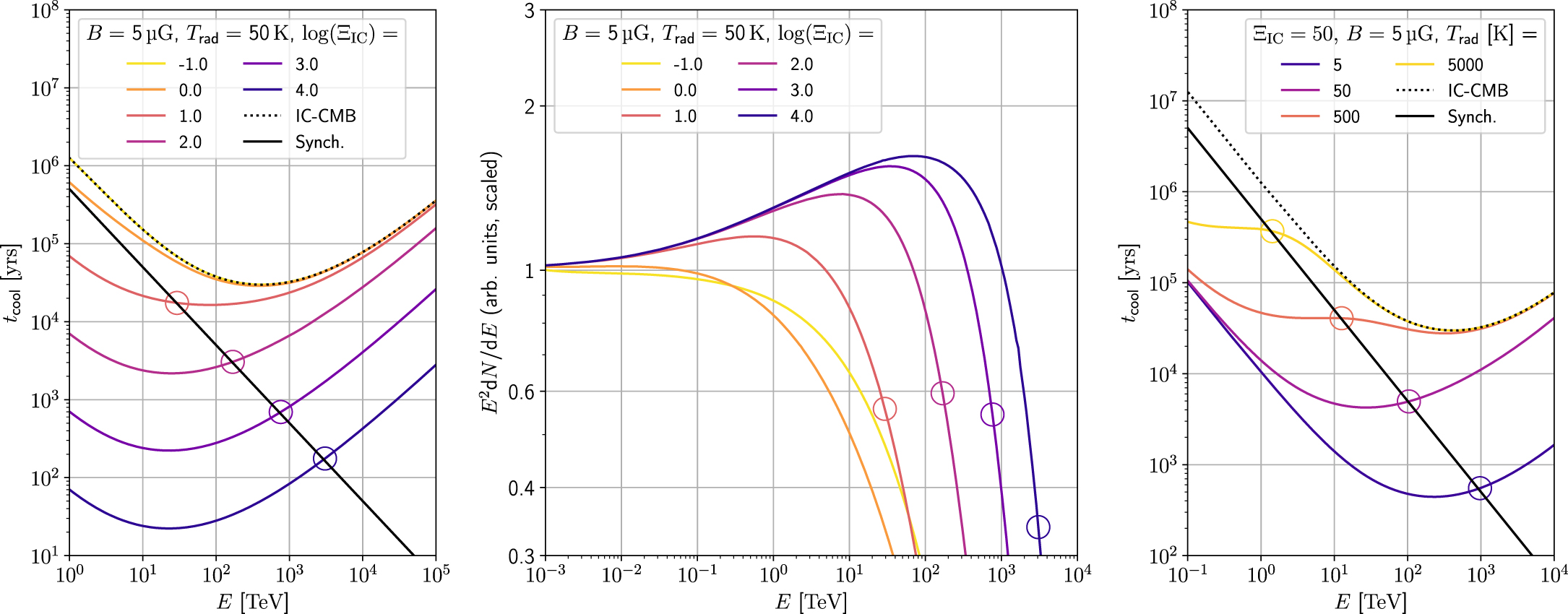}
\caption{Electron cooling time-scales t$_{cool}$ (left) and the steady-state $\gamma$-ray (middle) spectra assuming different energy densities of a 50 K
radiation field including the CMB, for a representative fixed magnetic field of 5 $\mu$G. The $\gamma$-ray spectra are those arising in equilibrium from
continuous injection of an E$^{-2}$ spectrum with exponential cutoff at 10 PeV. The right panel shows t$_{cool}$ for fixed $\Xi_{IC}$ (or $U_{Rad}$/$U_{B}$) = 50, but different temperatures. The circles on all panels indicate the transition energy E$_{X}$ for each radiation field density. The figure and caption are taken from \citep{Breuhaus_2021}.}
\label{hard_ic}
\end{figure}

Despite the fact that powerful pulsars can produce UHE $\gamma$ rays, one of the biggest challenge posed by observations is explaining the hard $\gamma$-ray spectra up to these extreme energies. It has been demonstrated and discussed in detail in \citep{Breuhaus_2021} that hard IC spectra at UHE energies can be possible when IC losses dominate over synchrotron losses. When high energy density radiation fields are present, and the condition of $U_{Rad}$/$U_{B}$~$\gg$~1 is satisfied, the equilibrium electron spectrum becomes harder due to the energy dependence of the KN cross-section. Although the resulting IC $\gamma$-ray spectrum from the hardening electron spectrum is less pronounced at these high energies due to KN suppression, it may still be possible to maintain this hard $\gamma$-ray spectra well beyond 100~TeV energies, particularly in astrophysical environments where radiation energy density is high. Figure~\ref{hard_ic} (left and middle) illustrates the relationship between E$_{X}$, which is the electron energy at which the synchrotron cooling time equals that of IC, and the produced $\gamma$-ray emission for a given $\Xi_{IC}$~=~$U_{Rad}$/$U_{B}$ value. It can be seen from these figures that the resulting $\gamma$-ray spectra can be significantly hardened with respect to the IC CMB case (black dotted line) and produce a stable hard UHE spectrum when $U_{Rad}$ is an order of magnitude higher than $U_{B}$, i.e. $\Xi_{IC}$ = 10$^{4}$ (see dark magenta lines in left and middle plots of Figure~\ref{hard_ic}). It is worth pointing out that the IC CMB case (dashed black line) produces a quite similar spectrum to the case of $\Xi_{IC}$ = 0.1 and radiation field temperature T$_{rad}$ = 50 K (yellow line), especially at high energies. Furthermore, the effect of the energy density of the radiation field on cooling times is illustrated in the right panel of Figure~\ref{hard_ic}. As depicted in the figure, the hardest IC spectrum is produced when T$_{rad}$ is 5~K. The radiation temperature range of 5$-$50~K is associated with IR fields. This means, in environments where strong IR fields are present, it is feasible that leptonic accelerators can produce and maintain hard IC spectrum which can extend up to and even beyond UHE energies. These kind of astrophysical environments can be found within our Galaxy, such as a young, powerful pulsar and a massive star are comprised in a binary system. In such systems, intense IR photon field produced by the star can act as a target photon field for IC scattering from the electron population provided by the pulsar, potentially satisfying the conditions discussed above. Therefore, these predictions indicate that binary systems are a potentially promising source class for leptonic PeVatrons. Additionally, similar astrophysical environments can be present and conditions can be met in Galactic spiral arms and particularly in dense regions where star formation activity is ongoing \citep{Breuhaus_2021}. Nevertheless, it should be emphasized that meeting these conditions requires extremely precise fine tuning and a very large system size, otherwise, accelerating particles to such high energies becomes exceedingly difficult. \citep{hillas}.

\section{Spectral signatures of hadronic PeVatrons}
\label{sig_pev} 

The quest for the origin of Galactic CRs essentially involves robust identification of hadronic Galactic PeVatrons by means of neutral messengers. This indicates that the parent hadronic spectrum should be reconstructed primarily based on $\gamma$-ray observations. The production of $\gamma$ rays from pp interactions is therefore crucial for understanding hadronic PeVatrons. There are two primary channels for $\gamma$-ray production in pp interactions, resulting in the subsequent decay of light mesons, specifically the neutral pion ($\pi_{0}$ $\rightarrow$ 2$\gamma$) and eta meson ($\eta$ $\rightarrow$ 2$\gamma$). Although other heavy particles are also produced in pp interactions, they generally tend to decay quickly, producing neutral pions. As a result, the decay of $\pi_{0}$ is the primary channel responsible for the observed $\gamma$ rays. Apart from $\gamma$ rays, neutrinos are also stable and neutral products of pp interactions. They are generated by the primary decay mode of charged mesons, specifically originating from the $\pi^{+}$~$\rightarrow$~$\mu^{+}$~+~$\nu_{\mu}$ and $\pi^{-}$~$\rightarrow$~$\mu^{-}$~+~$\bar{\nu}_{\mu}$ channels. While future observations of neutrinos can potentially play a crucial role and provide valuable complementary information for identifying Galactic PeVatrons, current neutrino spectroscopy is exceedingly challenging due to low neutrino statistics. Detailed discussions on the $\gamma$-ray and neutrino spectra produced by pp interactions can be found in \citep{kernel2006,kafexhiu2014,celli2020}.

In summary, it is evident that hadronic PeVatrons exhibit recognizable spectral signatures that can be determined through $\gamma$-ray (and possibly neutrino) observations. This section discusses such signatures within the framework of observational $\gamma$-ray astronomy.

\subsection{Connections between $\gamma$-ray and hadronic spectral parameters}
\label{signatures_conn}

The investigation on the interrelation between proton and $\gamma$-ray spectra, as presented in \citep{kernel2006,kafexhiu2014}, serves as a milestone in reconstructing parent proton spectra based on $\gamma$-ray observations. In these studies, the primary approach involves approximating the $\pi^{0}$ spectrum initially from pp interactions, and then estimating the resulting $\gamma$-ray spectrum from the decay of this $\pi^{0}$ spectrum. In practice, the {\tt naima} python package \citep{naima}, which utilizes the parametrization developed in \citep{kafexhiu2014}, is commonly used in $\gamma$-ray astronomy to reconstruct proton spectra from observational data.

A generalized expression for rigidity-dependent proton spectrum in the form of exponential cutoff power-law can be written as  
\begin{equation}
\label{e1} 
J_{p}(E_{p}) = \frac{dN_{p}}{dE_{p} dV} = J_{0,p}~\left(\frac{E_{p}}{E_{0}}\right)^{-\Gamma_{p}}\exp\left[-\left(\frac{E_\mathrm{p}}{\mathrm{Z}E_\mathrm{cut,\,p}}\right)^{\beta_\mathrm{p}}\right],
\end{equation}
where $J_{0,p}$ is the normalization at reference energy of $E_{0}$ and can be obtained from the normalization condition given in \citep{celli2020} as 
\begin{equation}
\label{e2} 
\int_{100~\text{GeV}}^{\infty} E_{p} J_{p}(E_{p}) \,dE_{p} \ = 1~\text{erg~cm}^{-3},
\end{equation}
while the other parameters $\Gamma_{p}$ is the spectral index, $E_\mathrm{cut,\,p}$ is the spectral cutoff energy and $\beta_{p}$ is the sharpness parameter describing the decay rate of exponential cutoff. The case of $\beta$ = 1.0 is called 'normal exponential', while the cases $\beta<$1 and $\beta>$1 are called 'sub-exponential' and 'super-exponential', respectively. In Eq.~\ref{e1}, the parameter Z represents the mass number, which is set to Z=1 for the case of protons. A fascinating implication of this general expression is that the apparent spectral cutoff of the hadronic spectrum increases linearly with the mass number, dependent on $E_\mathrm{cut,\,p}$, as already discussed in Section~\ref{sec_cr}. Consequently, it is possible to characterize the hadronic spectrum of diverse astrophysical objects just by adjusting these spectral parameters in different ways.

The generalized expression of resulting $\gamma$-ray spectrum from pp interactions, assuming a parent proton spectrum given in Eq.~\ref{e1}, will also have a similar exponential cutoff power-law form given as
\begin{equation}
\label{e3} 
\Phi(E_{\gamma}) = \Phi_{0,\gamma}~\left(\frac{E_{\gamma}}{E_{0,\gamma}}\right)^{-\Gamma_{\gamma}}~\exp\left[-\left(\frac{E_\mathrm{\gamma}}{E_\mathrm{cut,\,\gamma}}\right)^{\beta_\mathrm{\gamma}}\right],
\end{equation}
where $\Phi_{0,\gamma}$ is the normalization at reference $\gamma$-ray energy of $E_{0,\gamma}$, $\Gamma_{\gamma}$ is the $\gamma$-ray spectral index, $E_\mathrm{cut,\,\gamma}$ is the cutoff energy of resulting $\gamma$-ray spectrum and $\beta_{\gamma}$ is again the sharpness parameter describing the rate of exponential decay. The relationship between the spectral parameters of $J_{p}$ and $\Phi$ is particular crucial in $\gamma$-ray astronomy, as $J_{p}$ needs to be reconstructed from $\Phi$, which is derived from $\gamma$-ray observations of astrophysical sources. These connections have been extensively investigated in \citep{kernel2006,kafexhiu2014}, and direct correlations between these parameters are outlined in \citep{celli2020} for practical purposes as
\begin{equation}
\label{e4} 
\Gamma_{\gamma} = 0.94\Gamma_{p} - 0.15,
\end{equation}
\begin{equation}
\label{e5} 
\beta_{\gamma} = \frac{\beta_{p}}{(-0.7\Gamma_{p}+2.4)\beta_{p}+0.1\Gamma_{p}+0.7},
\end{equation}
\begin{equation}
\label{e6} 
\text{log}_{10}\left(\frac{E_\mathrm{cut,\,\gamma}}{E_\mathrm{cut,\,p}}\right) = (-1.15\Gamma_{p}+3.30)\beta_{\gamma}+(1.33\Gamma_{p}-4.61).
\end{equation}

\begin{figure*}[ht!]
\centering
\includegraphics[width=17cm]{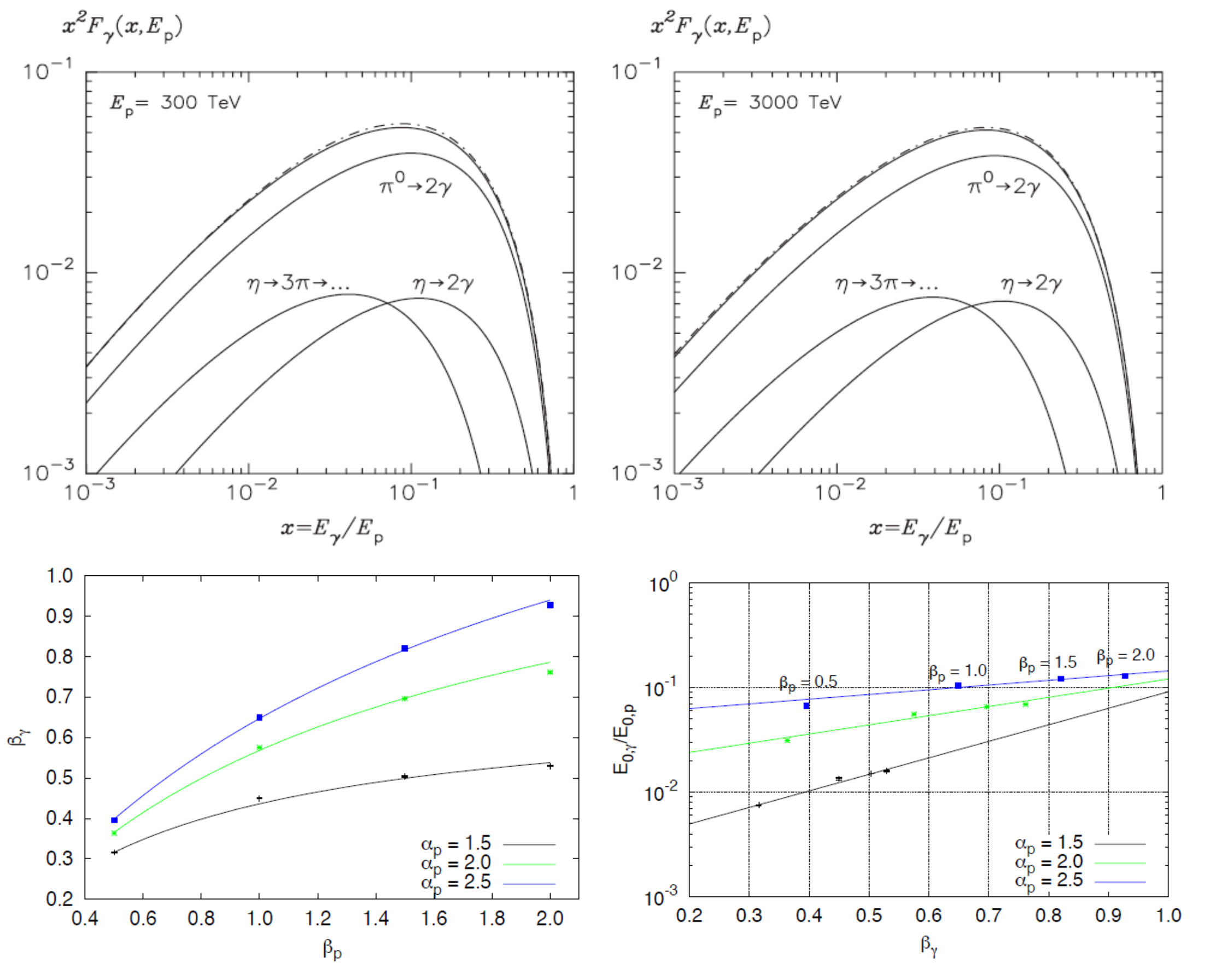}
\caption{Top panel: energy spectra of gamma rays from pp interactions corresponding to 300~TeV (left) and 3~PeV (right) energies of incident protons. Solid lines correspond to numerical calculations based on simulations of production of $\pi$ and $\eta$ mesons using
the SIBYLL code, while dotted-dashed line shows analytical presentation given by equation 58 of \citep{kernel2006}. The partial contributions from $\pi^{0}$ and $\eta$ meson decays are also shown. Top figures and caption are taken from \citep{kernel2006}. Bottom panel: The left figure shows the correlation between values of $\beta_{\gamma}$, obtained in the fitting procedure of secondary particle spectra, and the values of
$\beta_{p}$ assumed for protons. Dots indicate simulated spectral values. The cut-off energy of protons is here fixed to $E_\mathrm{cut,\,p}$ = 1~PeV. The lines refer to the analytical parametrization in the form of Eq.~\ref{e5}. The right figure shows ratios $E_\mathrm{cut,\,p}$/$E_\mathrm{cut,\,\gamma}$ as a function of $\beta_{\gamma}$ calculated for $E_\mathrm{cut,\,p}$ = 1~PeV and three different values of proton index $\alpha_{p}$ = 1.5, 2.0, 2.5, which is referred as $\Gamma_{p}$ in Eq.~\ref{e6}. Bottom figures and caption are taken from \citep{celli2020}.}
\label{kernel_param}
\end{figure*}

The authors of \citep{celli2020} emphasized that the simple relationships discussed here are only approximations, but they can still provide an accuracy of better than 20$\%$ for a wide range of $E_\mathrm{cut,\,p}$ values, between 100~TeV and 10~PeV, making them highly valuable in spectral PeVatron studies. For example, one common assumption used in $\gamma$-ray astronomy is that the ratio between the cutoff energies of $\gamma$-rays and protons ($E_\mathrm{cut,\,\gamma}$/$E_\mathrm{cut,\,p}$) is $\sim$0.1, which was provided in \citep{kernel2006}. This means, if a proton spectrum has a 1~PeV cutoff energy, the resulting $\gamma$-ray spectrum from pp interactions will have a cutoff energy of around 100~TeV. The top panel of Figure~\ref{kernel_param} shows the $\gamma$-ray spectra produced by different pp interaction channels for incident proton energies of 300~TeV (left) and 3~PeV (right), taken from \citep{kernel2006}. It can be seen from the figure that the $\pi^{0}$ $\rightarrow$ 2$\gamma$ channel reaches its peak $\gamma$-ray emission when $E_\mathrm{cut,\,\gamma}$/$E_\mathrm{cut,\,p}$~=~$\sim$0.1. However, it is important to note that this ratio is not a fixed number, but rather a distribution, i.e.~can go up to 0.5 or down to 0.01. The bottom panel of Figure~\ref{kernel_param} focuses on the relationship between $\beta_{p}$ and $\beta_{\gamma}$ on the left, and the connection between $E_\mathrm{cut,\,\gamma}$/$E_\mathrm{cut,\,p}$ and $\beta_{\gamma}$ on the right, taken from \citep{celli2020}. For example, the bottom right figure clearly shows that the commonly assumed ratio of $E_\mathrm{cut,\,\gamma}$/$E_\mathrm{cut,\,p}$ = 0.1 can be obtained for soft sources with $\Gamma_{p}$ = 2.5 and $\beta_{p}$=1, while as it can be seen from the bottom left plot, it will be impossible for a hard source with $\Gamma_{p}$ = 1.5 to reach such a ratio of $\sim$0.1, since $\beta_{\gamma}$ will never get close to unity even if $\beta_{p}$ increases.

\begin{figure}[ht!]
\centering
\includegraphics[width=17cm]{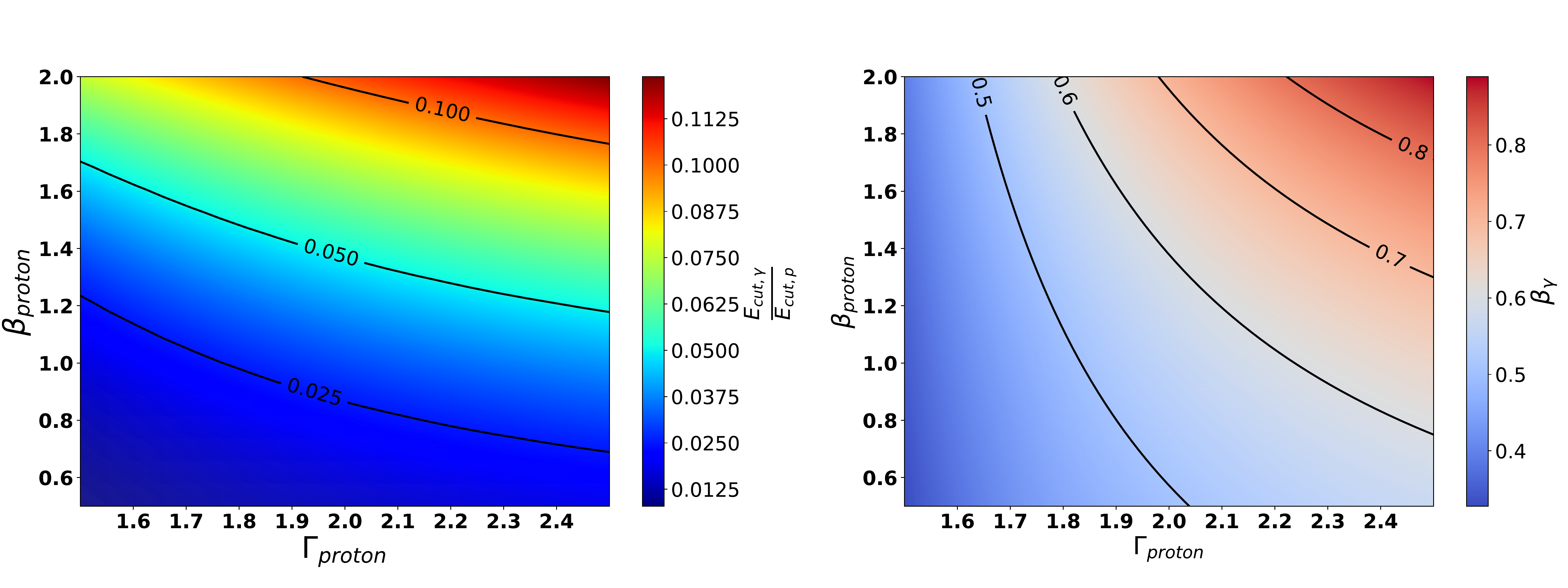}
\caption{Connections between $\gamma$-ray and hadronic spectral parameters. Both plots are defined in a 2D proton spectral parameter phase space defined by ($\Gamma_{p}$, $\beta_{p}$), where $\Gamma_{p}$ and $\beta_{p}$ are the proton spectral index and exponential decay rate of proton spectrum, respectively. The left figure shows the $E_\mathrm{cut,\,\gamma}$/$E_\mathrm{cut,\,p}$ ratio within the phase space, together with the 0.1, 0.05, and 0.025 contour lines in black, which corresponds to inverse ratio of $E_\mathrm{cut,\,p}$ / $E_\mathrm{cut,\,\gamma}$ of 10, 20, and 40, respectively. The right figure shows the exponential decay rate ($\beta_{\gamma}$) of the $\gamma$-ray spectrum resulting from pp interactions, together with the contour lines between $\beta_{\gamma}$ = 0.5 and $\beta_{\gamma}$=0.8 given in black for a step of 0.1.}
\label{cutoff_relation}
\end{figure}

By combining Eqs.~\ref{e5} and \ref{e6}, it is possible to directly derive the $E_\mathrm{cut,\,\gamma}$/$E_\mathrm{cut,\,p}$ relation from proton spectral parameters in a 2D phase space defined by ($\Gamma_{p}$, $\beta_{p}$). Figure~\ref{cutoff_relation} demonstrates how the $E_\mathrm{cut,\,\gamma}$/$E_\mathrm{cut,\,p}$ ratio and $\beta_{\gamma}$ change in a 2D proton spectrum phase space, as seen in the left and right figures, respectively. The left figure reveals that an $E_\mathrm{cut,\,\gamma}$/$E_\mathrm{cut,\,p}$ ratio of 0.1 can only be achieved when $\beta_{p}$ is $\sim$2.0, meaning that the proton spectrum is cutting super-exponentially. Conversely, for proton spectra with normal exponents ($\beta_{p}$=1.0), a wide range of proton spectral index values ($\Gamma_{p}$) can result in a ratio of $\sim$0.025, meaning the observed $\gamma$-ray cutoff would be $\sim$40 times lower than that of the proton spectrum, instead of $\sim$10 times. The right figure shows that even for very soft and steeply cutting exponential proton spectra, a $\beta_{\gamma}$ value of $\sim$1.0 cannot be attained within this wide proton phase space, which indicates that super or even normal exponential cutoff $\gamma$-ray spectra are not expected from pp interactions, instead mostly sub-exponential $\gamma$-ray spectra are expected.

It is worth discussing the interpretation of the $\alpha_{p}$, $\beta_{p}$ and $\beta_{\gamma}$ parameters at this point. Firstly, it should be noted that the $\beta_{\gamma}$ parameter, which can be obtained experimentally from $\gamma$-ray observations, is a combination of $\alpha_{p}$ and $\beta_{p}$ parameters, as approximated in Eq.~\ref{e5}. From the experimental point, robust statistical determination of the precise $\beta_{\gamma}$ value is extremely challenging due to low energy resolutions, $\%$15 and $\%$40, of current IACTs and particle arrays, respectively, but can be well possible with the planned future experiments reaching $\%$7 at energies above 1~TeV (i.e.~CTA \citep{cta_science}). The spectral index parameter of the proton spectrum, $\alpha_{p}$, is typically 2.0 for the diffusive shock acceleration (DSA) in the test-particle limit. However, in more realistic scenarios, the value of $\alpha_{p}$ may differ depending on the properties of the acceleration region, with $\alpha_{p}$ being as hard as 1.5 in some cases \citep{non_dsa_hard}. On the other hand, the $\beta_{p}$ parameter is responsible for determining the high energy efficiency of acceleration, and is therefore heavily dependent on the limiting factor of CR acceleration at the source region. If one assumes that DSA is the primary mechanism for CR acceleration at the source region, which is often the case for sources such as SNRs or YMCs, then the strongest limiting factor would typically be the size of the accelerator relative to the diffusion distance of the most energetic particles. In such cases, the parameter $\beta_{p}$ is generally set to 1.0, especially when the diffusion is close to the Bohm regime, while subexponential models ($\beta_{p}$ $<$ 1.0) can be plausible for different diffusion models used in astrophysics \citep{expcutoff_ref}. On the other hand, superexponential models ($\beta_{p}$ $>$ 1.0) can be employed to explain some acceleration mechanisms at SNRs, in which the achievable maximum energy is restricted by magnetic field growth rather than the system size \citep{Bell04,Schure13,Cristofari20}. While the value of $\beta_{p}$ can vary depending on the characteristics of the CR acceleration at the source region, the resulting $\beta_{\gamma}$ of the $\gamma$-ray spectrum is generally less than 1.0 for a significant portion of the parameter space, as shown in Figure~\ref{cutoff_relation} (right).

\subsection{Analysis methods to search for PeVatrons}

The first step in the search for hadronic PeVatrons is to examine the spectral characteristics of the observed $\gamma$-ray spectrum. A significant advancement in our understanding of the origin of Galactic CRs was the investigation of the Galactic center using H.E.S.S. observations \cite{hessGCNat}. According to the study, a strong PeVatron was discovered in the center of the Milky Way, and it was linked to the radio source Sgr A*. This PeVatron identification was made based on the 95$\%$ confidence level (CL) lower limit on the cutoff energy of the parent proton spectrum, which was found to be 400~TeV. Following this pioneering study, quoting the 95$\%$ CL lower limit on the proton spectral cutoff and interpreting a source's PeVatron nature based on the lower limit values became a kind of standard in PeVatron searches. However, the latest breakthrough LHAASO results reported the detection significance of $\gamma$-ray emission above 100~TeV \citep{lhaaso2021} instead of the 95$\%$ CL on the proton spectral cutoff energy.

\begin{figure}[ht!]
\centering
\includegraphics[width=15cm]{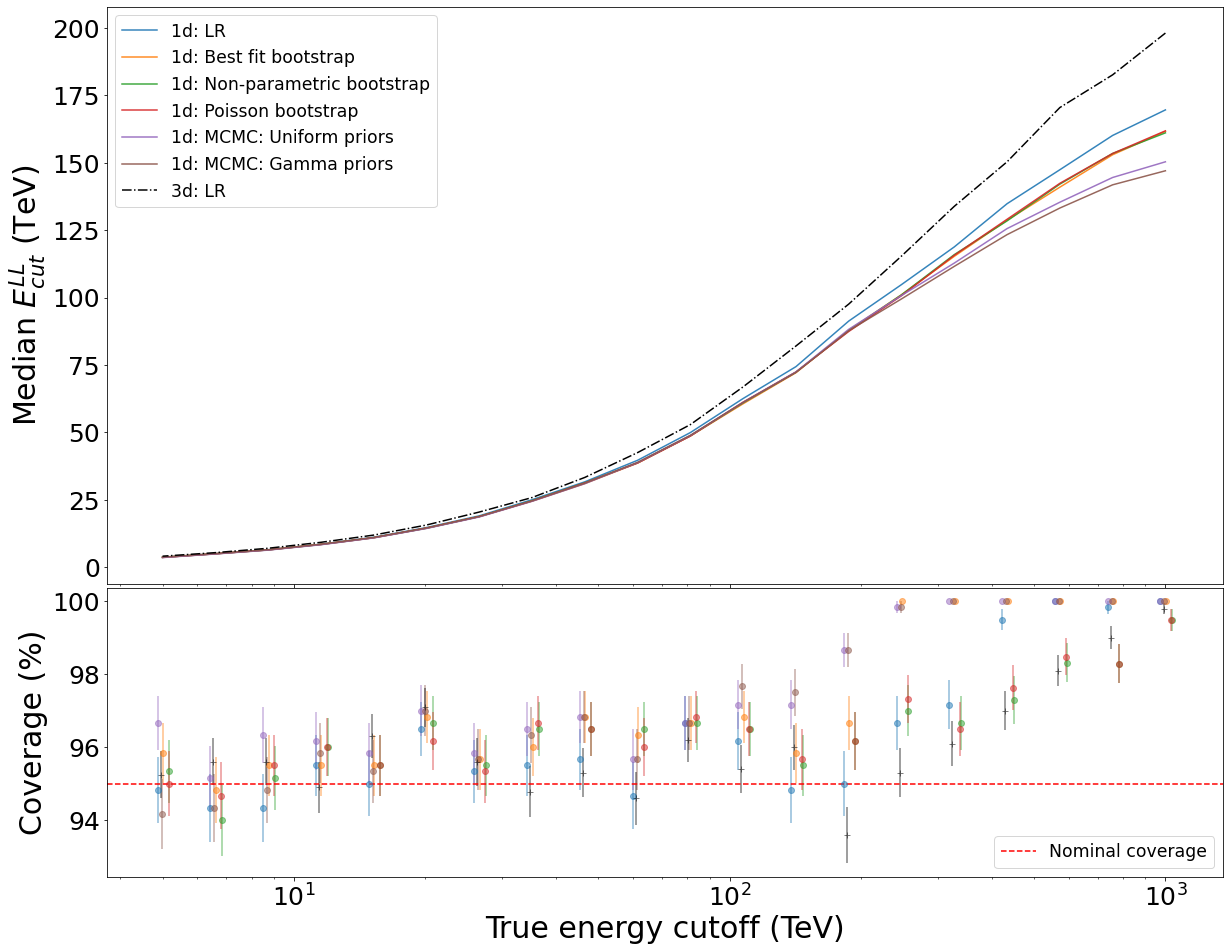}
\caption{Comparison of different methods to derive a lower limit on the energy cutoff for a $\gamma$-ray point-like source with power-law index $\Gamma=2.1$ and flux normalization $\phi_0=50$ mCrab. The instrument response functions of CTA South and $10$ h of observation time at a zenith angle of $20^\circ$ are assumed. Upper panel: median $95\%$ CL lower limit on the energy cutoff as a function of the true energy cutoff. Color codes for the different methods are indicated in the legend. Solid lines correspond to 1-dimensional analyses. Note that the green, orange and red solid lines overlap. The black dash-dotted line corresponds to profile likelihood limits derived with a 3-dimensional analysis. Lower panel: coverage of the interval $[E_\mathrm{c,\;\gamma}^\mathrm{LL},\infty)$ for different methods to derive $E_\mathrm{c,\;\gamma}^\mathrm{LL}=1/\lambda^\mathrm{UL}$ as a function of the true energy cutoff. The true energy cutoff is slightly shifted for each method to improve the visibility. Errorbars are approximate $68.3\%$ CL intervals on the coverage, derived under the assumption of the normal approximation to the binomial distribution. The figure and caption are taken from \citep{CTA_PeV}.}
\label{lower_limits_comp}
\end{figure}

Currently in the literature, there are two commonly used methods to provide evidence or make claims for the detection of PeVatron sources. These include using the detection significance of $\gamma$-ray emissions above 100~TeV, and the derived 95$\%$ CL lower limit of the hadronic energy cutoff. While both methods are useful for locating PeVatron sources, they encounter significant issues when used independently from each other. For instance, when using the 95$\%$ CL lower limit approach, the value of 1~PeV is often cited for "PeVatron candidates". However, it is worth noting that the 95$\%$ CL is considerably lower than the confidence level required for new discoveries, which corresponds to 5$\sigma$ statistical significance. Nevertheless, the 95$\%$ CL of 1~PeV proton cutoff energy is still a useful tool for PeVatron searches since the derivation of the lower limit takes into account the spectral shape. It is worth pointing out that there are different methods for deriving lower limits, and the profile likelihood, bootstrap, and Markov chain Monte Carlo (MCMC) are the most frequently used ones. These methods are discussed in detail in the appendix of \citep{CTA_PeV}, and a performance comparison is given in Figure~\ref{lower_limits_comp}. As shown in the figure, the profile likelihood method is the most sensitive for deriving spectral cutoff lower limits. For the derivation of spectral cutoff lower limits, the sensitivity of a method is measured with derived lower limit value being close to the true cutoff energy value. Consequently, the most sensitive method becomes the one minimizing the difference $E_{\mathrm{c}}^{\mathrm{True}}$ - $E_{\mathrm{c}}^\mathrm{LL}$. However, it is important to mention that the profile likelihood method may not always be applicable, particularly in complicated astrophysical environments. As the number of free parameters increases, for example in the case of fitting 2$-$3 sources simultaneously in the field of view which leads to more than 10 free parameters, it becomes extremely challenging for the profile likelihood method to converge, and it can get stuck at some local maxima. For these complicated cases, the MCMC method performs better and provides more reliable lower limits, albeit at the expense of time-consuming computational efforts.

The primary issue with the alternative method of searching for PeVatrons, which uses the detection significance of $\gamma$-ray emission above 100~TeV, is that the approach does not consider the spectral shape of the emission. This approach only provides cumulative information above 100~TeV, meaning that two different sources yielding identical E$>$100 significance values may have completely different 95$\%$ CL cutoff lower limits in their $\gamma$-ray spectrum. This is due to the fact that, even though they may have identical statistics above 100~TeV, the distribution of event statistics in energy can vary significantly. For instance, the source yielding lower $E_\mathrm{c,\;\gamma}^\mathrm{LL}$ may have events concentrated between 100~TeV and 200~TeV, while the other source with higher $E_\mathrm{c,\;\gamma}^\mathrm{LL}$ may have a wider distribution of events between 100~TeV and 500~TeV. Although a significant emission above 100~TeV is necessary for a reliable PeVatron detection, this criterion alone cannot guarantee that the detected source is indeed a PeVatron.

To make a reliable claim of PeVatron detection, it is essential to test both of the criteria mentioned above simultaneously. Recently, a new method called PeVatron Test Statistics (PTS) has been developed \citep{CTA_PeV} in order to combine these two approaches. As discussed by the authors, the PTS is essentially a likelihood ratio that measures the deviation of the proton energy cutoff, $E_\mathrm{cut,\,p}$, from a predetermined test proton cutoff energy threshold, $E_\mathrm{thr}$, and given by
\begin{equation}
\label{eq_PTS}
\mathrm{PTS}=-2\ln\frac{\hat
L(E_\mathrm{cut,\,p}= E_\mathrm{thr})}{\hat L(E_\mathrm{cut,\,p})}\,\mathrm{,}
\end{equation}\\
where $\hat L(E_\mathrm{cut,\,p}= E_\mathrm{thr})$ is the likelihood value  at the test energy threshold and $\hat L(E_\mathrm{cut,\,p})$ is the maximum likelihood over all $E_\mathrm{cut,\,p}$. The corresponding significance of the deviation from the threshold cutoff is 
\begin{equation}
\label{eq_S_PTS}
S_\mathrm{PTS}=\mathrm{sign}(E_\mathrm{cut,\,p}^*-E_\mathrm{thr})\sqrt{\mathrm{PTS}}\,
\end{equation}\\
where $E_\mathrm{cut,\,p}^*$ is the best-fit proton spectral cutoff energy value. By setting the energy threshold $E_\mathrm{thr}$ to 1~PeV, the PTS method can identify PeVatrons and provide a clear quantification through significance values. In other words, if $S_\mathrm{PTS}>5\sigma$, it indicates that the parent proton spectrum extends beyond the threshold energy of 1~PeV as a power-law, without showing any evidence of a spectral cutoff, with a CL of at least $5\sigma$. This implies that the source in question contributes to the CR spectrum at energies above 1~PeV, and a robust $5\sigma$ detection of an hadronic PeVatron spectral signature can be claimed. As given in \citep{CTA_PeV}, the interpretation of PTS significance, $S_\mathrm{PTS}$, for a given PeVatron energy threshold of 1~PeV is as follows:
\begin{itemize}
    \item $S_\mathrm{PTS} \geq 5\sigma$: A Pevatron detection can be claimed with a CL corresponding to at least $5\sigma$. This means a robust $5\sigma$ detection of an hadronic PeVatron spectral signature, indicating that the parent proton spectrum extends well beyond the threshold energy of 1~PeV.
    \item $S_\mathrm{PTS}<-5\sigma$: The association between a $\gamma$-ray source and a PeVatron can be excluded with a CL corresponding to at least $5\sigma$. This means a robust $5\sigma$ exclusion of an hadronic PeVatron spectral signature, indicating that the parent proton spectrum has a significant cutoff well below the threshold energy of 1~PeV.
    \item $|S_\mathrm{PTS}|<5\sigma$: There is insufficient data to differentiate between the PeVatron and the non-PeVatron hypotheses, and thus, further observational data are required.
\end{itemize}

\begin{figure}[ht!]
\centering
\includegraphics[width=17cm]{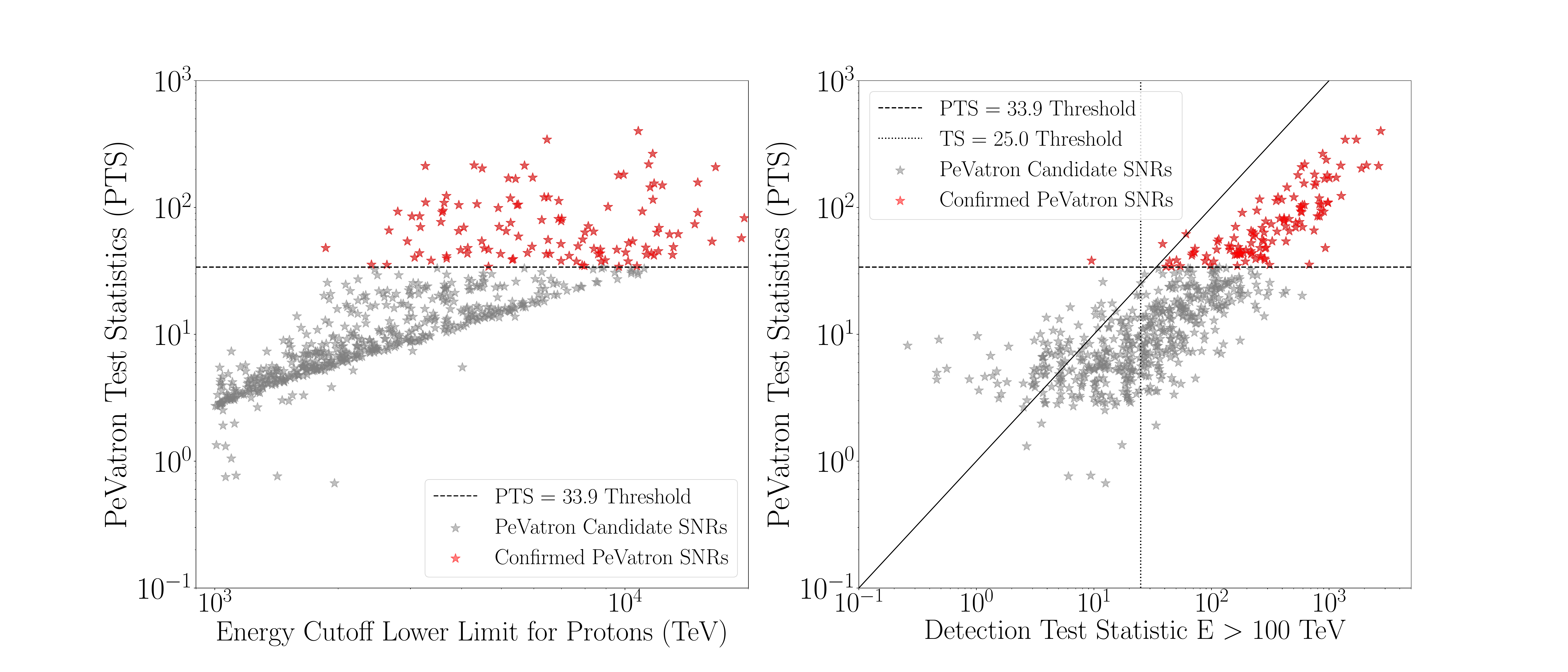}
\caption{Relation between the PTS and the lower 95\% CL limit on the proton spectral cutoff (left) and detection test statistic, TS$_\mathrm{Det}^\mathrm{100\; TeV}$, of the $\gamma$-ray flux above $100$ TeV (right). Confirmed PeVatrons are marked in red and PeVatron candidates are marked in grey. The PTS threshold of 33.9, corresponding to a global 5$\sigma$ significance for 100 independent tests, is indicated with horizontal dashed black lines. In the right panel, the vertical line indicates the $\gamma$-ray detection threshold TS$_\mathrm{Det}^{\mathrm{100\; TeV}}$=25 and the diagonal shows the equality PTS=TS$_\mathrm{Det}$. The figure and caption are taken from \citep{CTA_PeV}.}
\label{pts_ll}
\end{figure}

Figure~\ref{pts_ll} illustrates a comparison between the PTS and two other alternative PeVatron search criteria obtained from CTA simulations in \citep{CTA_PeV}, namely the 95$\%$ CL of proton spectral cutoff and the significance of $\gamma$-ray emission above 100~TeV. The figures show that the PTS exhibits a noticeable correlation with both criteria, and ensuring a $S_\mathrm{PTS}$~=~5$\sigma$ significance level satisfies both of the other criteria simultaneously.

It should also be noted that the threshold can be adjusted to investigate the contribution to the CR spectrum above a specific energy threshold. For instance, in order to examine the contribution of protons (and helium) assuming that the knee for protons is below 1~PeV, a threshold of 300$-$400~TeV can be chosen. In this case, the significance of PTS will reflect whether the source in question is a proton PeVatron. It is also possible to use leptonic emission models, i.e.~IC scattering, in PTS calculation to test leptonic PeVatrons. 

\subsection{Mapping spectral features: detection probability maps}

Examination across a broad range of spectral parameters can be beneficial in determining the detectability of spectral signatures from $\gamma$-ray sources showing various properties. In \citep{CTA_PeV}, 'detection probability maps' were introduced to represent 2D maps that cover a specific spectral phase space of a particular $\gamma$-ray model, such as the exponential-cutoff power-law (ECPL) model. These maps are constructed for a fixed spectral cutoff energy value of interest. Two types of probability maps are presented and explained in \citep{CTA_PeV} in detail. The first one is referred to as the 'spectral cutoff detection probability map', which quantifies the probability of detecting a high energy spectral $\gamma$-ray cutoff in an experiment. The second one, which is called 'PeVatron detection probability map', is a map that measures the probability of an experiment detecting a PeVatron source with 5$\sigma$ confidence, given a cutoff energy in the proton (hadron) spectrum. For example, the proton spectral cutoff value under consideration might be 3~PeV for the PeVatron detection maps, which corresponds to the knee feature observed in the CR spectrum. The study focused on the CTA experiment, with details provided in \citep{CTA_PeV}, and it was also discussed by the authors that such maps can be created for any experiment, taking into account the instrument response functions (IRFs), providing valuable insights into high energy capabilities, especially for future projects.

\begin{figure}[ht!]
\centering
\includegraphics[width=17cm]{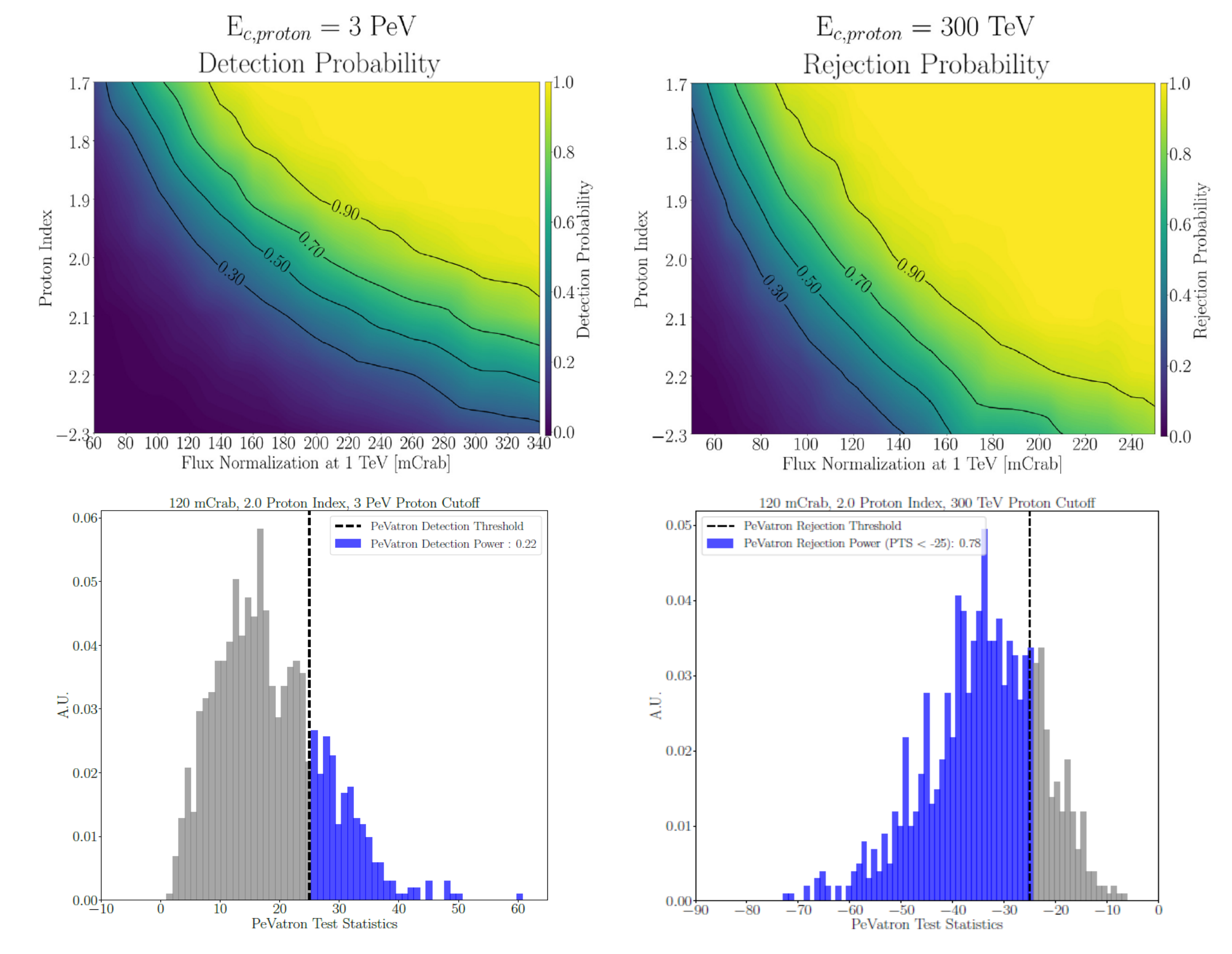}
\caption{Top panel: PeVatron detection (left) and rejection (right) probability maps for 10 h of CTA observations of point-like sources. The abscissa shows the true observed $\gamma$-ray flux normalization at 1 TeV originating from pp interactions observed from Earth, while the ordinate shows true proton spectral index. The color code shows the PeVatron detection (rejection) probabilities, while the black lines indicate the PeVatron detection (rejection) probability contours at 0.3, 0.5, 0.7, and 0.9 levels. PeVatron detection (PTS$>$25) probability map is given for a true proton energy cutoff at 3 PeV, while PeVatron rejection (PTS$<$-25) probability map for a true proton energy cutoff at 300 TeV. For comparison, 100 mCrab, 200 mCrab and 300 mCrab values correspond to differential flux level at 1 TeV of 3.84~$\times$ 10$^{-12}$ cm$^{-2}$ s$^{-1}$ TeV$^{-1}$, 7.68~$\times$ 10$^{-12}$ cm$^{-2}$ s$^{-1}$ TeV$^{-1}$ and 1.15~$\times$ 10$^{-11}$ cm$^{-2}$ s$^{-1}$ TeV$^{-1}$, respectively. The top figures and caption are taken from \citep{CTA_PeV}. Bottom panel: example of PTS distribution in a particular $\Phi_{0}$(1~TeV) = 120 mCrab and $\Gamma_{p}$ = 2.0 bin both for true proton energy cutoff at 3 PeV (left) and 300~TeV (right). The blue shaded regions show the corresponding detection and rejection probabilities, respectively.}
\label{cta_ptsmaps}
\end{figure}

The top panel of Figure~\ref{cta_ptsmaps} displays the PeVatron detection and rejection maps obtained from 10 h of CTA observations, whereas the bottom panel illustrates an example of the PTS distribution in a specific map bin with $\Phi_{0}$(1~TeV)~=~120~mCrab and $\Gamma_{p}$~=~2.0. The PTS distributions given in the lower panels allow the identification of robust 5$\sigma$ PeVatron detection and rejection thresholds, which correspond to PTS values of 25 and --25, respectively. The detection probability for a given bin is then obtained by computing the fraction of the distribution above the detection threshold of PTS~=~25 (as seen in the bottom left figure). Similarly, PeVatron rejection is defined by the fraction of the distribution below the detection threshold of PTS = --25. As explained in \citep{CTA_PeV}, repeating this process iteratively for multiple bins that cover the desired phase space results in the creation of PeVatron detection maps. The yellow areas of the maps shown in the top panel of Figure~\ref{cta_ptsmaps} indicate the phase space region where gamma-ray sources can be robustly detected (on the left) or rejected (on the right) as PeVatrons at a 5$\sigma$ level. It is worth noting that detection probabilities are strongly correlated to observation time, which means that as exposure increases, the black contour lines shift to lower flux normalization values as expected. However, as the performance degrades with increasing source extension, the black contour lines move to higher flux values, requiring the source to be more powerful for being detected as a PeVatron.

\begin{figure}[ht!]
\centering
\includegraphics[width=14cm]{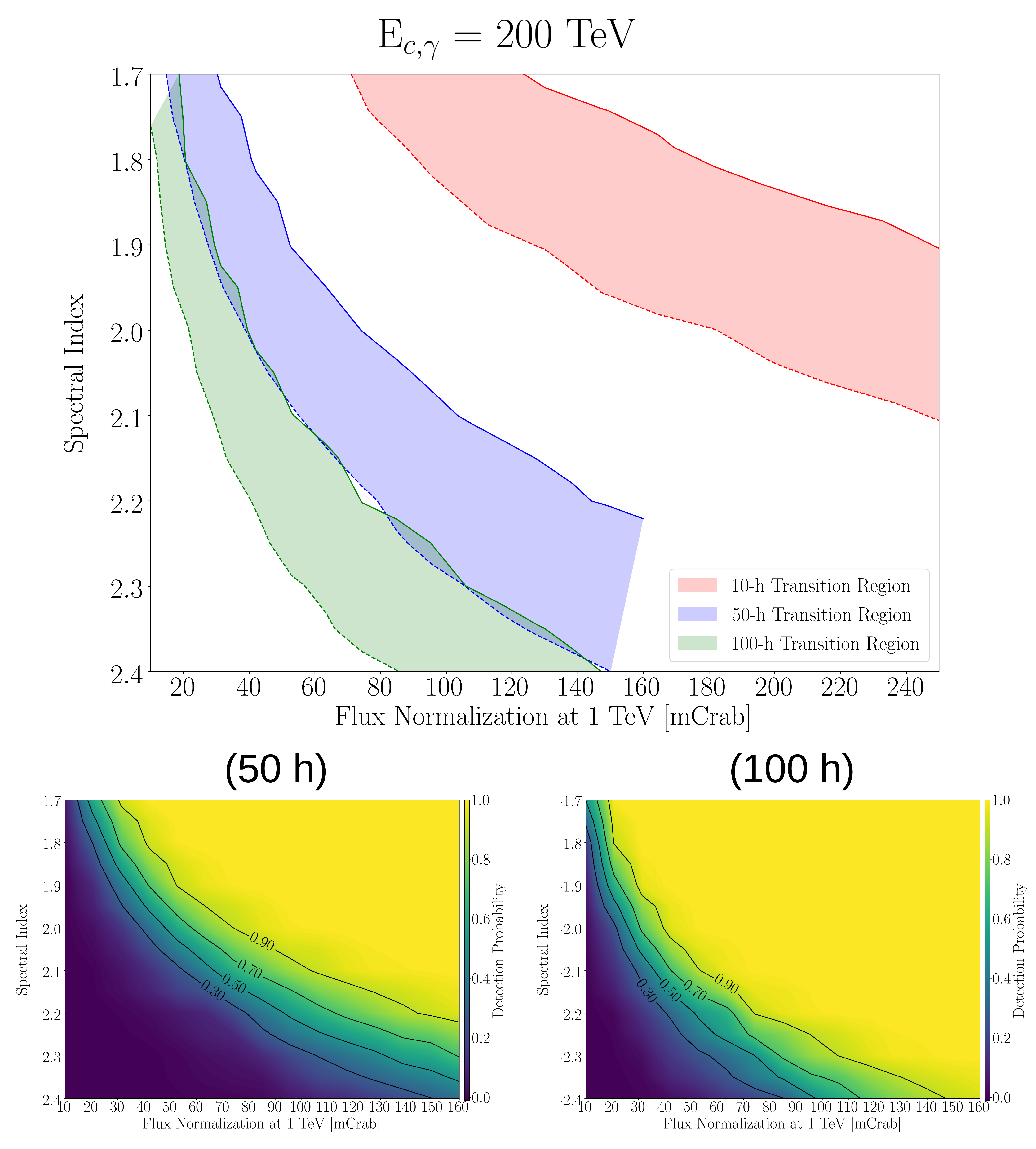}
\caption{Deep observation sensitivity of CTA to spectral cutoff features of point-like sources. The top panel shows the comparison of the "transition regions", i.e. the region between the detection probability contours of $30\%$ (dashed lines) and $90\%$ (solid lines), for the detection of a spectral cutoff with true value of $200$ TeV. The performance corresponding to 10 h, 50 h, and 100 h of CTA data is shown in red, blue and green. The corresponding spectral cutoff detection maps for 50 h and 100 h observations are given at the bottom left and right panels, respectively. The figures and caption are taken from \citep{CTA_PeV}.}
\label{cta_scdm_maps}
\end{figure}

To investigate the detection of high energy spectral $\gamma$-ray cutoffs, one can use a similar approach by replacing PTS with Test Statistics (TS) between the ECPL model and power-law model. By creating a distribution of TS values from many simulations, similar to those shown in Figure~\ref{cta_ptsmaps} (bottom), the probability of detecting a spectral cutoff at a particular $\gamma$-ray energy can be determined by calculating the fraction of the TS distribution above a given threshold, such as TS~=~25 (equivalent to a 5$\sigma$ level). The resulting spectral cutoff detection maps for a $\gamma$-ray cutoff energy of 200~TeV are shown in the bottom panel of Figure~\ref{cta_scdm_maps} for 50 and 100 h of deep CTA observations, taken from \citep{CTA_PeV}. Similarly to the PeVatron detection maps, the yellow areas on the maps represent the region of phase space where the 200~TeV $\gamma$-ray cutoff can be significantly detected by the experiment. To compare the different maps, one can use the transition region defined as the area between the detection probability contours of 30$\%$ and 90$\%$. The top panel of Figure~\ref{cta_scdm_maps} shows such a comparison for different exposure times of 10, 50, and 100 h. As the observation time increases, the transition region shifts towards lower fluxes, enabling the detection of 200 TeV $\gamma$-ray cutoffs even from fainter and softer sources. It is worth noting that such comparison can be made between i.e.~different proposed layout configurations of an experiment to compare performances.

The maps presented in this section are useful in the search for PeVatrons, but they also raise an important question regarding the relationship between the "detection or non-detection of a $\gamma$-ray cutoff" and "PeVatron detection," which is in general not straightforward. A strong indication of PeVatron detection, assuming that the $\gamma$-ray emission has a hadronic origin, is a robust detection of $\gamma$-ray spectral cutoffs, particularly well beyond 100~TeV after deep exposure, as shown in the 200~TeV map in Figure~\ref{cta_scdm_maps}.~On the other hand, not detecting a spectral cutoff in the instrument energy range means that the source spectrum appears as a power-law, which can have two possible explanations. The first possibility is that the underlying proton cutoff may be very high, causing the cutoff in the $\gamma$-ray spectrum to fall outside the instrument energy range, effectively making the source a PeVatron. Alternatively, a source can appear as a power-law because it is faint, and there are insufficient high energy statistics to detect the spectral cutoff, and in this case, it would be unclear whether underlying proton cutoff is high or low. For the latter case, further observations would be required to increase the statistics at high energies. The PTS concept is very useful in such cases, providing S$_{PTS} \geq$ 5$\sigma$ for the former PeVatron with a high proton cutoff, while giving $|S_\mathrm{PTS}|<5\sigma$ for the faint source exhibiting power-law spectrum. Consequently, the spectral cutoff detection maps provide limited information about PeVatron detection, whereas PeVatron detection maps, which are constructed based on the PTS approach, offer clearer insights. The sources with spectral parameter values within the yellow regions are robust PeVatrons at a S$_{PTS} \geq$ 5$\sigma$ level.

\subsection{Absorption of gamma rays in the Galaxy}
\label{gammagammaabs}

In principle, there is no theoretical upper limit on the amount of energy that $\gamma$ rays can have. However, the maximum energy of CRs that have been detected so far is $\sim$10$^{20}$~eV, or 100,000~PeV, as seen in Figure~\ref{cr_spect}. Based on the previous discussion in subsection~\ref{signatures_conn}, it is possible to produce $\gamma$ rays with energies of $\sim$10$^{19}$ eV through pp-interactions. Gamma-ray photons can interact with lower energy photons and even with each other, leading to the production of electron-positron pairs. Consequently, the propagation of $\gamma$ rays in the universe is limited by the attenuation of photons caused by the process of pair production, $\gamma\gamma$ $\rightarrow$ e$^{+}$ + e$^{-}$ process, also known as $\gamma\gamma$ absorption. Especially, the attenuation of $\gamma$ rays from the CMB photon field sets an effective horizon for $\gamma$-ray energies above 1 PeV \citep{abs1,abs2}.

\begin{figure}[ht!]
\centering
\includegraphics[width=14cm]{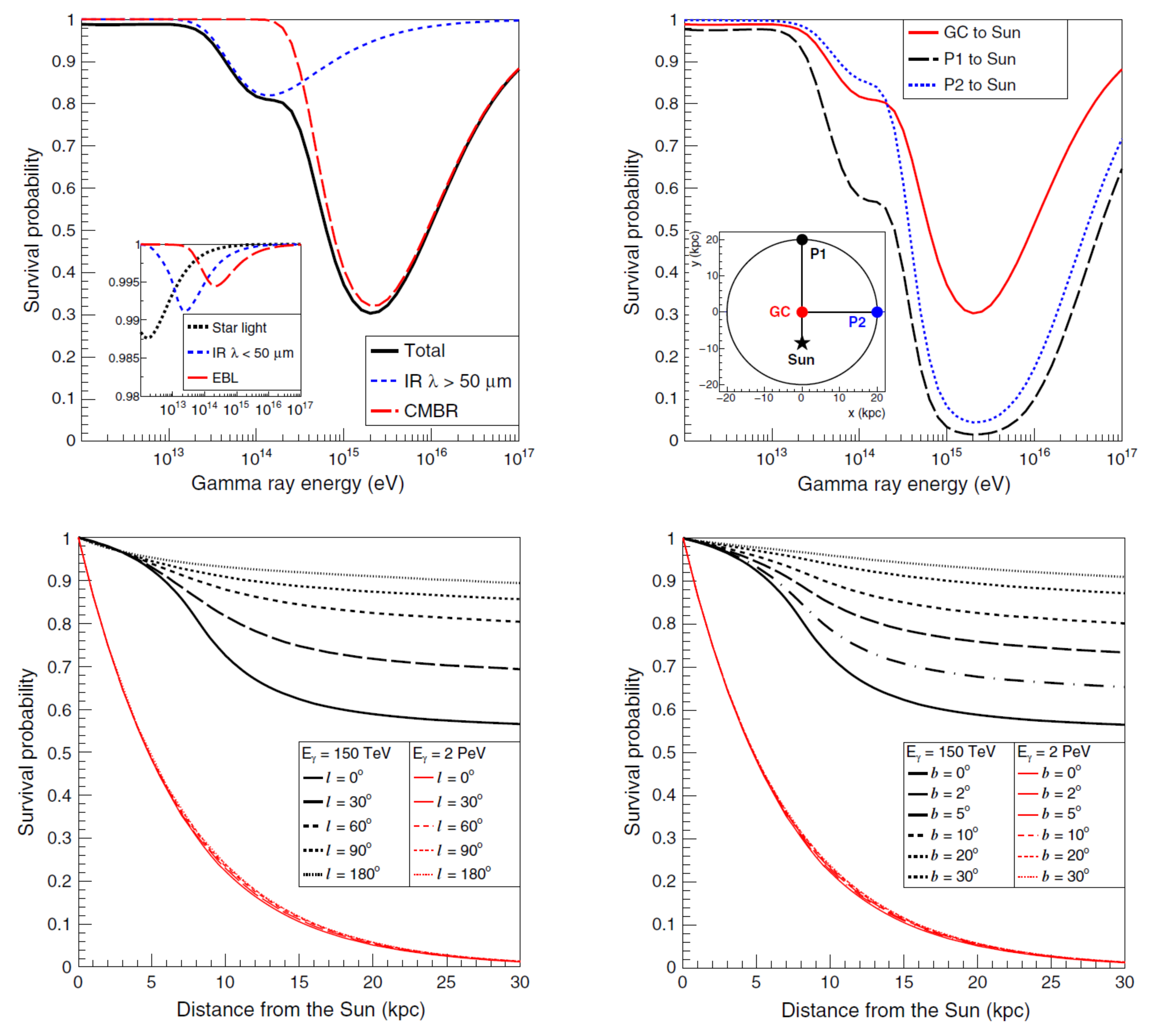}
\caption{Top panel: (left figure) survival probability of $\gamma$ rays for a trajectory from the GC to the Sun, plotted as a function of the $\gamma$-ray energy. The contributions of different radiation fields are shown. The inset shows the contributions of starlight, infrared radiation
with wavelength $\lambda$ $<$ 50 $\mu$m and EBL. (Right figure) Survival probability of $\gamma$ rays for three different trajectories in the Galactic plane, plotted as a function of the $\gamma$-ray energy. The inset shows the position of the sources. Bottom panel: survival probabilities of gamma rays of energy 150~TeV and 2~PeV as a function of the source distance, for lines of sight with different Galactic longitudes and fixed Galactic latitude of \textit{b} (left figure) and fixed Galactic longitude \textit{l} (right figure), respectively. The figures and captions are taken from  \citep{gamma_abs}}.
\label{absorption}
\end{figure}

It was extensively discussed in \citep{gamma_abs}, that there are four radiation field components within our Galaxy that are responsible for the absorption of $\gamma$ rays, including the extra-galactic background light (EBL) and the cosmic microwave background (CMB), which have universal, extra-galactic origins. The other two components, which have Galactic origins, are starlight and infrared (IR) radiation emitted by heated dust. To accurately estimate the absorption of $\gamma$ rays as they travel through the Galaxy, it is crucial to have a thorough understanding of the density, energy, and direction of photon fields at various locations within the Galaxy. Indeed, these different radiation fields have varying absorption effects at different energy ranges. The top left panel of Figure~\ref{absorption} illustrates how the survival probability of $\gamma$ rays, defined as P$_{\text{Surv}}$ = 1 - P$_{\text{Abs}}$, originating from the direction of the Galactic center changes as a function of $\gamma$-ray energy for a given absorbing photon field. As shown in the figure, $\gamma$-ray absorption due to radiation emitted by dust (dashed-blue line) reaches its maximum at energies around 100$-$150~TeV, while absorption due to the CMB field (dashed-red line) is particularly effective, especially above 200$-$300 TeV energies, and reaches its maximum around 2.2~PeV \citep{gamma_abs}. The sum of these two absorption effects (black solid line) creates a characteristic spectral absorption feature known as the 'shoulder' around 150~TeV energies. The top right panel of Figure~\ref{absorption} shows a comparison of the total absorption spectra in the case of $\gamma$ rays arriving from different locations of the Galaxy. As seen from the inset of this figure, PeV $\gamma$ rays arriving from the edges of our Galaxy (i.e.~20~kpc radius) are completely absorbed due to the CMB field, creating a PeV horizon at Galactic scales.

The bottom panels of Figure~\ref{absorption} nicely illustrate how the survival probabilities of $\gamma$ rays (P$_{\text{Surv}}$) are affected by the distance of their source from the Sun. The figures show the survival probabilities for two instances of $\gamma$-ray energies, 150~TeV and 2~PeV, calculated from fixed Galactic longitude (left panel) and latitude (right panel), respectively. Regardless of the line of sight, the figures reveal that nearly $80\%$ of 2~PeV $\gamma$ rays are absorbed at a distance of 10~kpc. Additionally, it is clear that 150~TeV $\gamma$ rays arriving from regions with dense IR field, which are expected to be most common at the center of the Galaxy (\textit{l}=0, \textit{b}=0), experience higher levels of absorption. This physical phenomenon should be taken into account in the search for the origin of CRs, particularly in the search for PeVatrons, which relies mostly on the detection of UHE photons that are most affected by absorption effects, as discussed in the framework of Figure~\ref{absorption}. The absorption effects due to dust and CMB radiation fields lead to decreased statistics, especially at the high energy end of the $\gamma$-ray spectra, causing the observed spectrum to appear "steepened" compared to the unabsorbed spectrum, as previously mentioned in subsection~\ref{CRKneeEx} and Figure~\ref{pev_illust}.

\begin{figure}[ht!]
\centering
\includegraphics[width=17cm]{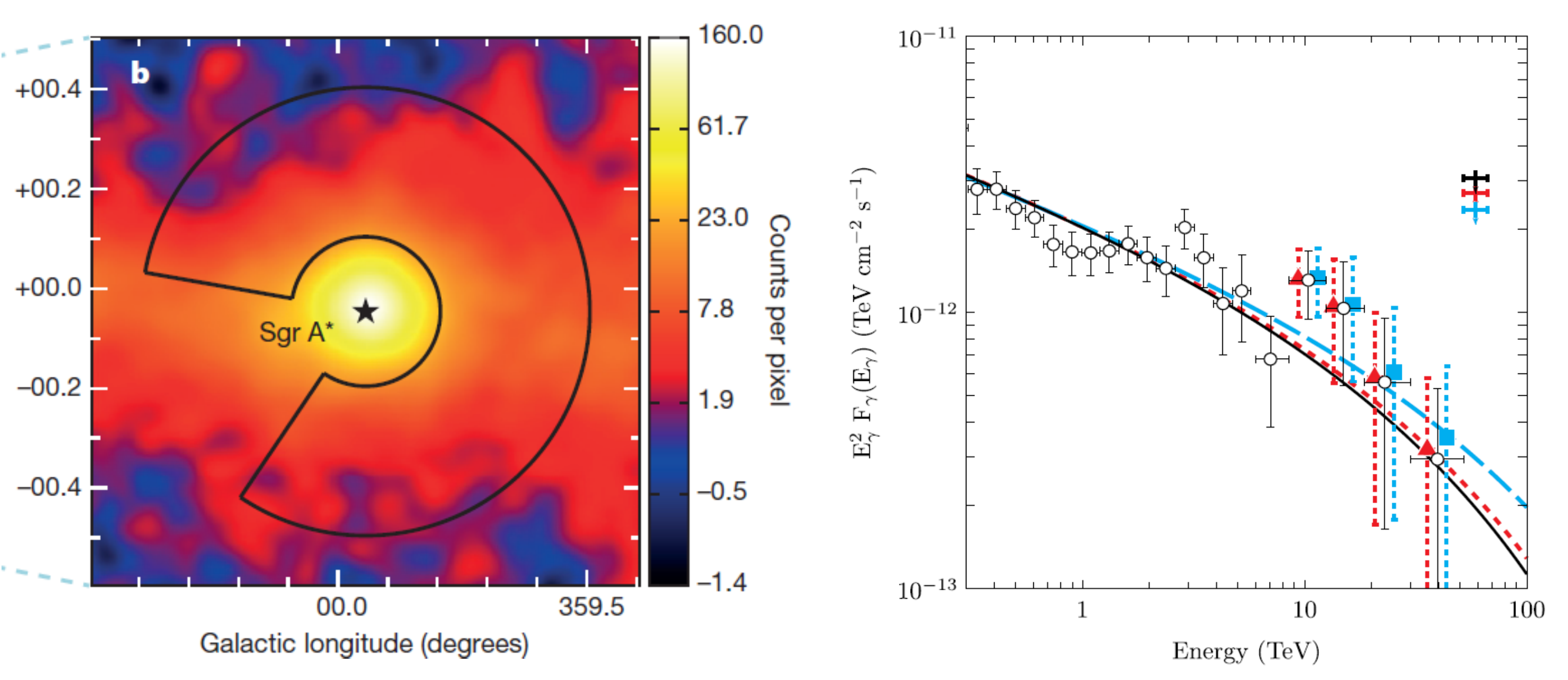}
\caption{Left figure: VHE $\gamma$-ray image of the the inner $\sim$70 pc of the Galactic center region. The color scale indicates counts per 0.02$^{\circ}$~×~0.02$^{\circ}$ pixel. The black star shows the location of Sgr A*, while the black contour lines show the region used to extract the spectrum of the diffuse emission. A section of 66$^{\circ}$, where a known VHE source exists, is excluded from the annuli. The left figure and caption are taken from \citep{hessGCNat}. Right figure: VHE $\gamma$-ray spectrum of the diffuse emission around the Galactic center measured by the HESS instrument (black hollow dots), extracted from the annulus region shown in the left figure,  together with absorption-corrected data flux points (red triangles and blue squares). The best-fit spectral models are shown for absorption uncorrected and corrected cases with black solid and red short-dashed and blue long-dashed curves, respectively. The right figure and caption are taken from \citep{pev_absorp}.}
\label{gc_hess_abs}
\end{figure}

In an interesting example of correcting spectral absorption effects, researchers in \citep{pev_absorp} applied a correction to the diffuse Galactic center data. The region of diffuse emission around the GC was the first region in our Galaxy where the existence of a PeVatron source was claimed \citep{hessGCNat}, based on a 95$\%$ C.L lower limit on the proton cutoff. Figure~\ref{gc_hess_abs} displays the VHE $\gamma$-ray image of the GC region (left figure) alongside the diffuse emission spectrum (right figure), extracted from an annulus-shaped region located in the vicinity of the GC. The best model describing the diffuse emission spectrum is a power-law with a photon index of 2.3, which does not show any clear signs of a spectral cutoff or break, at least up to tens of TeV energies. Assuming that the diffuse emission spectrum is generated by the interaction of CRs, injected from a central CR source, with the dense molecular clouds present in the GC region, the parent proton spectrum that gives rise to the observed $\gamma$-ray spectrum should also be a power-law, with a proton spectral index of $\sim$2.4 \citep{celli2020}. The authors have calculated the 68$\%$, 90$\%$, and 95$\%$ CL lower limits on the proton spectral cutoff energy to be 2.9~PeV, 0.6~PeV, and 0.4~PeV, respectively \citep{hessGCNat}.
\setcounter{footnote}{0}
The research presented in \citep{gamma_abs} primarily aimed to correct the absorption effects on the diffuse emission spectrum, specifically for energies above 10~TeV. The corrected flux points for energies above 10~TeV, accounting for both faint and strong interstellar radiation field models\footnote{The interstellar radiation field models are 'the F98 model', which gives an estimate for the strongest infrared emissions over this region, and 'the R12 model', which provides close to a lower bound infrared emissions. Consequently, the F98 model leads to higher level of $\gamma$ absorption with respect to the R12 model. See \citep{gamma_abs} for further details.}, are depicted in Figure~\ref{gc_hess_abs} as red triangles and blue squares, respectively, along with the best-fit $\gamma$-ray models represented by red and blue dashed curves. Using these $\gamma$-ray models, the 95$\%$ CL lower limits on the proton spectrum cutoff were determined as 670~TeV and 1180~TeV for faint and strong IR fields, respectively. Comparing the 95$\%$ CL lower limits on the proton cutoff between the absorption-uncorrected and corrected cases highlights the importance of considering such effects, particularly when the sources are situated in IR-dominated radiation fields and/or at greater distances.

In conclusion, to obtain a comprehensive understanding and accurate interpretation of PeVatron sources, it is crucial to consider and correct for absorption effects in the analysis of observational data, particularly for UHE sources with spectra spanning up to several hundred TeVs. For UHE sources, the CMB will be the primary absorbent radiation, and the degree of absorption will be closely linked to the source distance. Therefore, determining precise distances for these sources is essential for correcting the derived flux points, especially above energies of 200$-$300 TeV.

\section{The problem of source confusion in the search for PeVatrons} 
\label{source_conf}

The challenge of distinguishing between two (or more) nearby $\gamma$-ray sources is commonly known as 'problem of source confusion' in $\gamma$-ray astronomy. Typically, one of the VHE $\gamma$-ray sources is concealed beneath the tails of the bright VHE emission from another $\gamma$-ray source. While energy-dependent morphology studies may lead to new source discoveries, particularly in densely populated Galactic regions, this approach may not always provide a precise estimation of the degree of contamination. Although some attempts have been made using traditional background estimation models, e.g.~see \citep{j1826_130}, more advanced 3D analysis techniques \citep{Mohrmann_2019} would be necessary, in particular to reconstruct the confused spectra of overlapping sources. In a recent study reported in \citep{source_conf_cta}, authors used the likelihood ratio test to examine the ability of the CTA to identify and address source confusion. In the following section, a general overview of several cases of Galactic source confusion will be discussed in the context of PeVatron searches. It is worth mentioning that PeVatron candidates have been identified in all of these source confusion cases.

\subsection{The case of HESS~J1641$-$463: power of energy-dependent morphology analysis}

\begin{figure}[ht!]
\centering
\includegraphics[width=17cm]{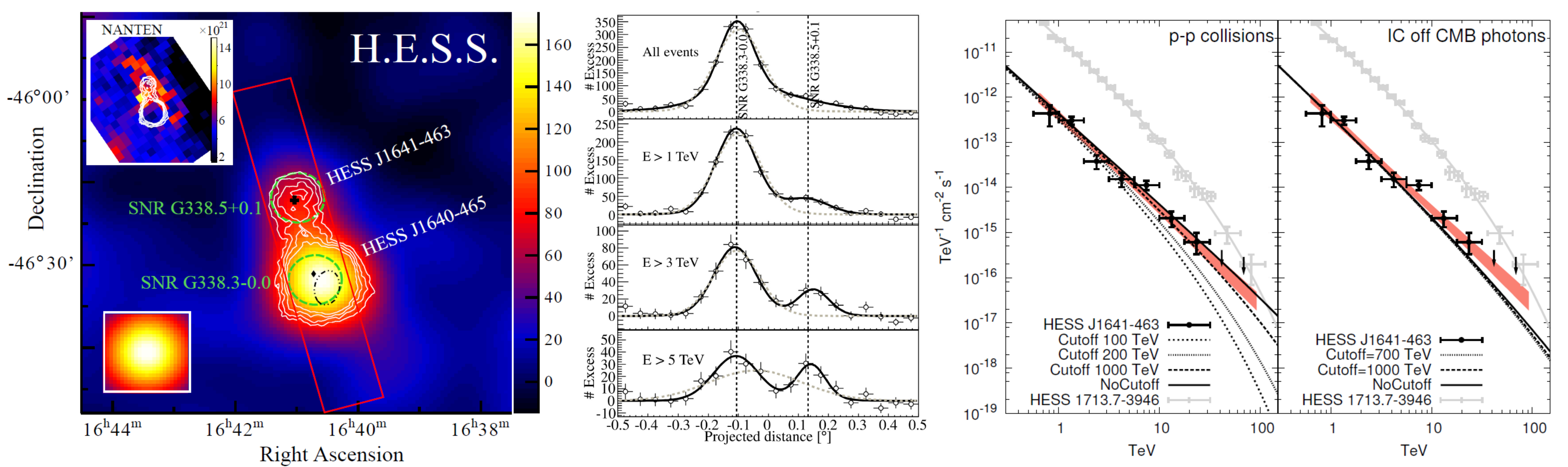}
\caption{Top panel: The left figure is the map of excess events with energies E$>$4~TeV for the region around HESS~J1641$-$463. The white contours indicate the significance of the emission at the 5, 6, 7, and 8$\sigma$ level. The black cross indicates the value and uncertainty of the best-fit position of the source, the green dashed circles show the positions and approximate extensions of the two nearby SNRs, the black diamond the position of PSR~J1640$-$4631. The red box indicates the area for the extraction of the profiles shown in the middle figure. Middle figure: distribution of VHE $\gamma$-ray excess profiles and Gaussian fits for the red rectangular slice shown in the left panel for different energy thresholds of 1~TeV, 3~TeV and 5~TeV. Vertical lines show the position of the SNR~338.3$-$0.0 and G338.5$+$0.1. Fits using a single and a double Gaussian function are shown in dashed and solid lines, respectively. Right figure: differential $\gamma$-ray spectrum of HESS~J1641$-$463 together with the expected emission from pp collisions (left) and IC off CMB photons (right). The pink area represents the 1$\sigma$ confidence region for the fit to a power-law model, the black data points the H.E.S.S. measured photon flux. The figures and captions are taken from \citep{j1641_463}.}
\label{jointj1641_SC}
\end{figure}

Figure~\ref{jointj1641_SC} illustrates the process of discovering new sources through energy-dependent morphology analysis. This analysis discussed here \citep{j1641_463} was carried out on a Galactic region which includes two VHE $\gamma$-ray sources, HESS~J1640$-$465 \citep{j1640} and HESS~J1641$-$463 \citep{j1641_463}. The former has an extended morphology with an extension of $\sim$0.11$^{\circ}$, while the latter exhibits a point-like morphology. The two sources are located at an angular distance of $\sim$0.28$^{\circ}$ away from each other. HESS~J1640$-$465 is a bright SNR with $\Phi_{0}$ (1~TeV) = (3.3 $\pm$ 0.1$_{stat}$ $\pm$ 0.6$_{sys}$) $\times$10$^{-12}$ TeV$^{-1}$~cm$^{-2}$~s$^{-1}$ and displays a significant spectral cutoff at E$_{cut}$ = 6.0$^{+2.0}_{-1.2}$, with a spectral index of $\Gamma$=2.11$\pm$ 0.09$_{stat}$ $\pm$ 0.10$_{sys}$ \citep{j1640}. On the other hand, HESS~J1641$-$463 is an unidentified source that shows a pure power-law spectrum up to at least 20~TeV without exhibiting any clear indication of a spectral cutoff. This source is nearly eight times fainter than HESS~J1640$-$465, with $\Phi_{0}$ (1~TeV) = (3.91 $\pm$ 0.69$_{stat}$ $\pm$ 0.78$_{sys}$)$\times$10$^{-13}$ TeV$^{-1}$~cm$^{-2}$~s$^{-1}$ and a power-law spectral index of $\Gamma$=2.07$\pm$ 0.11$_{stat}$ $\pm$ 0.20$_{sys}$ \citep{j1641_463}. At energies above 0.6~TeV, the total region of $\gamma$-ray emission appears as a single blob due to the difference in sources' relative brightness. In other words, HESS~J1641$-$463 was hidden under the exceptionally luminous VHE source HESS~J1640$-$465. The situation does not change dramatically at energies above 1~TeV, but becomes noticeable only at energies above 2$-$3 TeV. The top left panel of Figure~\ref{jointj1641_SC} shows the excess map of the region at energies above 4~TeV, in which two sources can be clearly distinguished from each other. It is important to note that HESS~J1641$-$463 emerges at high energy due to its power-law spectrum that goes up to at least 20~TeV without a spectral cutoff, while the spectrum of HESS~J1640$-$465 cuts off at 6.0~TeV. These two sources have nearly equal brightness above 5~TeV, as seen in the lower slice plot in the top middle panel of Figure~\ref{jointj1641_SC}. The improved angular resolution of H.E.S.S. telescopes at high energies is the main factor that makes it possible to distinguish between these two sources as the energy increases. The top right panel of Figure~\ref{jointj1641_SC} shows the spectrum of HESS~J1641$-$463 plotted together with emission models for hadronic pp-interactions (left) and leptonic IC off CMB photons (right) for different particle cutoff energies. It was discussed in \citep{j1641_463} that, assuming the VHE energy $\gamma$-rays are originated from hadronic interactions, HESS~J1641$-$463 becomes one of the most promising PeVatron candidates in the Southern sky and is expected to be observed at UHE due to its hard power-law spectrum. Here one can imagine that the UHE emission from this region, if it exists, will look like a single blob when observed with particle arrays, like i.e.~future SWGO. To fully understand the PeVatron nature of this intriguing source, UHE data above 100~TeV and high angular resolution VHE data taken with instruments such as HESS or future CTA would be necessary.

\subsection{Analysis of the complex HESS~J1825$-$137 and HESS~J1826$-$130 region: a dark component emerging at high energies}

\begin{figure}[ht!]
\centering
\includegraphics[width=17cm]{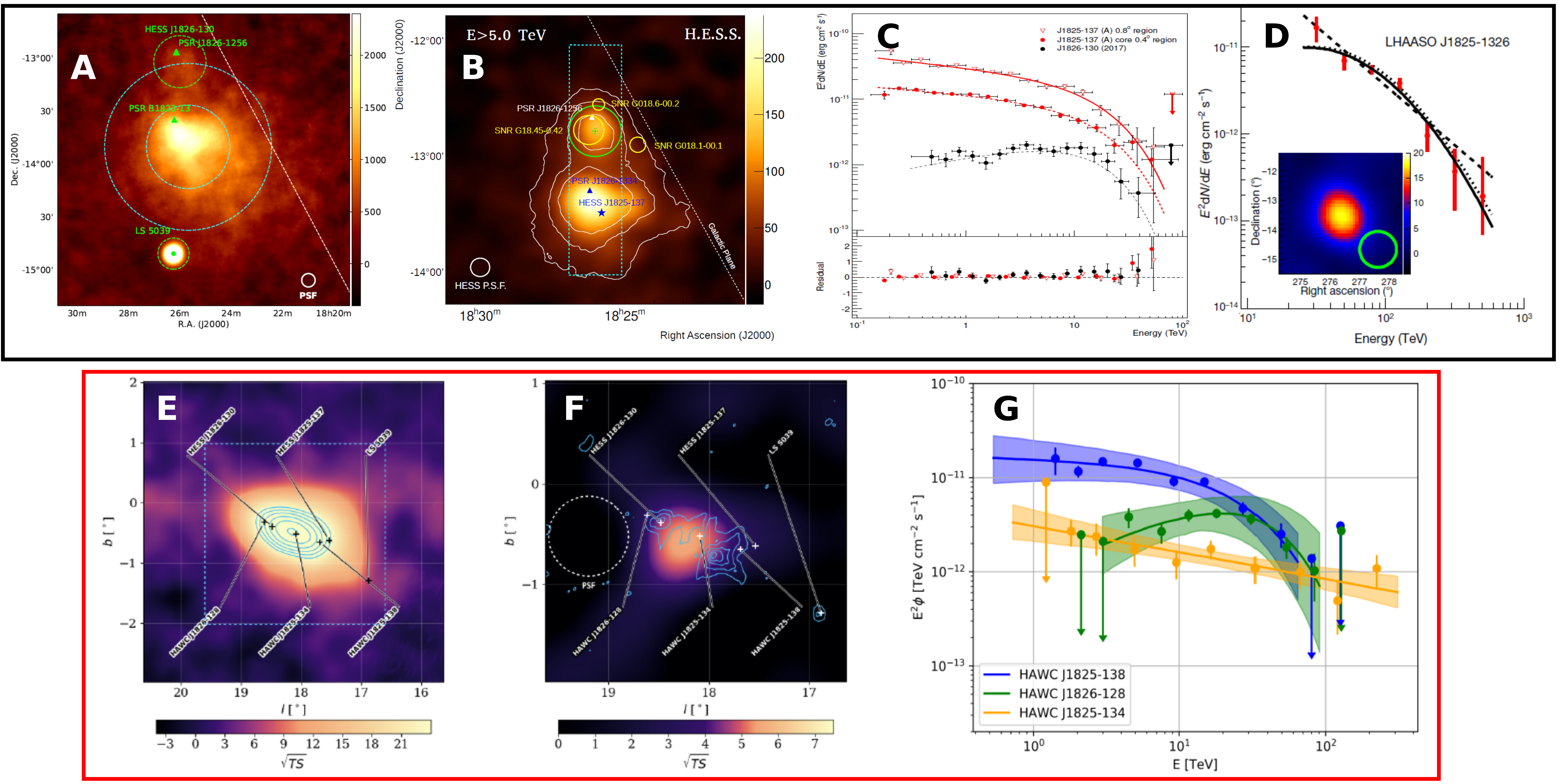}
\caption{Top panel: (A) the excess count map of the HESS~J1825$-$137 region above 0.2~TeV energies, with the locations of two energetic pulsars in the region indicated by green triangles. The two spectral extraction regions used in the right figure are overlaid. The larger region with a radius of 0.8$^{\circ}$, whereas the region with the smaller radius of 0.4$^{\circ}$ encompasses the core emission. (B) Map of excess counts with energies E$>$5~TeV for region around HESS~J1826$-$130. The white contours indicate the significance of the emission at the 5$\sigma$, 10$\sigma$, and 15$\sigma$ level. The green circle shows the integration region used for deriving the source spectrum (see inset C), while the green cross indicates the value and 1$\sigma$ uncertainty of the best-fit position of the source. The blue star shows the peak position of the bright nearby source HESS~J1825$-$137. The blue triangle indicates the position of PSR~J1826$-$1334. (C) Comparison of spectra extracted from the regions of 0.8$^{\circ}$ and 0.4$^{\circ}$ radii shown in (A) together with a spectrum of HESS J1826–130. All spectra are shown with a best-fit model of a power-law with exponential cut-off. (D) spectral energy distribution and significance map of LHAASO~J1825$-$1326. Spectral fits with a log-parabola function (solid lines) in the form of [$E$/(10~TeV)]$^{-a-blog[E/(10~\text{TeV})]}$ are compared with the power-law fits. Figures (A) and (C), and corresponding captions are taken from \citep{j1825_134}, while figure (B) and its caption are taken from \citep{j1826_130}. Figure (D) and its caption is taken from \citep{lhaaso2021}.\\
Bottom panel: (E) significance map around the eHWC~J1825-134 region for reconstructed energies greater than 1 TeV. The blue significance contours correspond to TS at 26, 28 30, 32 and 34. (F) zoom in of the blue dashed region (from E) including only reconstructed energies greater than 177 TeV. The blue contours corresponding to H.E.S.S. excess at 20, 35, 50, 60, and 70 counts with energies beyond 10 TeV. (G) spectral energy distribution  for the sources in the region of interest around eHWC~J1825$-$134 that were considered in the model. Figures (E), (F), (G) and corresponding captions are taken from \citep{hawc_j1825}.}
\label{jointj1825_SC}
\end{figure}

One of the most interesting source confusion example in the Galactic plane is between HESS J1825$-$137 \citep{j1825_134} and HESS~J1826$-$130 \citep{j1826_130}. HESS~J1825$-$137 is a well-known very bright PWN, which shows one of the most extended $\gamma$-ray emission regions in the Galaxy, exhibiting a clear energy-dependent morphology. Figures~\ref{jointj1825_SC}~(A) and \ref{jointj1825_SC}~(B) show the excess maps of this region above 0.2~TeV and 5~TeV, respectively. Both the total and core nebula spectra (see the green circles in Figure~\ref{jointj1825_SC}~(A) and corresponding spectra in Figure~\ref{jointj1825_SC}~(C)) show significant spectral cutoffs at around $\sim$19.0~TeV, together with spectral index values of $\Gamma_{\text{Total}}$=2.18~$\pm$~0.02$_{stat}$~$\pm$~0.02$_{sys}$ (for the total nebula) and $\Gamma_{\text{Core}}$=2.12~$\pm$~0.02$_{stat}$~$\pm$~0.03$_{sys}$ (for the core nebula), respectively \cite{j1825_134}. The total nebula is extremely bright and the flux above 1 TeV is (1.12~$\pm$~0.03$_{stat}$~$\pm$~0.25$_{sys}$)$\times$10$^{-11}$~cm$^{-2}$~s$^{-1}$, corresponding to $\sim$64$\%$ of the Crab Nebula flux. The unidentified source HESS~J1826$-$130 was completely hidden under the tails of bright VHE emission from HESS~J1825$-$137, and it could be seen clearly only above energies of $\sim$2$-$3~TeV \citep{j1826_130, hard_sources}. For comparison, integral flux above 1~TeV for HESS~J1826$-$130 is just $\sim$5$\%$ of the Crab Nebula flux,  more than 10 times fainter with respect to HESS~J1825$-$137. The emission from the direction of HESS~J1826$-$130 shows a clear spectral cutoff at E$_{cut}$ = $\sim$15.0~TeV, and the source has one of the hardest spectra ever observed in the VHE domain of $\Gamma$=1.78 $\pm$ 0.10$_{stat}$ $\pm$ 0.20$_{sys}$ \citep{j1826_130}. The comparison between the spectra of HESS~J1825$-$137 and HESS~J1826$-$130 is given in Figure~\ref{jointj1825_SC}~(C). Similar to the case of HESS~J1641$-$463, the energy-dependent morphology analysis of this particular Galactic plane region led to the discovery of HESS~J1826$-$130. It is worth mentioning that the discovery is mostly due to the very hard spectrum of HESS~J1826$-$130, which makes it possible for the source to emerge at high energies, although the nearby source is extremely bright. 

The location of this complex Galactic region is also visible for the LHAASO experiment. The total $\sim$1150 h of LHAASO observations led to detection of the UHE source, LHAASO~J1825$-$1326, at a significance level of 16.4$\sigma$ above 100~TeV energies \citep{lhaaso2021}. The significance map and the spectrum of LHAASO~J1825$-$1326 reconstructed from LHAASO observations is given in Figure~\ref{jointj1825_SC}~(D). As it can be seen from this figure, UHE emission coming from this region appears as a single blob, with 1$\sigma$ Gaussian extension of $\sim$0.62$^{\circ}$. The highest energy photon detected from the direction of LHAASO~J1825$-$1326 is 420~TeV \citep{lhaaso2021}. The discovery of LHAASO~J1825$-$1326 emitting UHE $\gamma$-rays was truly remarkable. As previously mentioned, the two known VHE $\gamma$-ray sources in this region, HESS~J1826$-$130 and HESS~J1825$-$137, exhibit significant cutoff features in their spectra. Therefore, it is challenging to account for the UHE emission observed above 100~TeV using only the spectral findings from the H.E.S.S. experiment.

The most interesting experimental results of this complex region came from HAWC observations. After the analysis of $\sim$1343 days of HAWC data, three distinct source components have been identified in this region \cite{hawc_j1825}, as it is shown in Figure~\ref{jointj1825_SC}~(E). Of these components, HAWC~J1826$-$128 and HAWC~J1825$-$138 are spatially coincident with the previously mentioned known VHE sources, HESS~J1826$-$130 and HESS~J1825$-$137, respectively, also exhibiting significant spectral cutoffs at comparable values of $\sim$24~TeV and $\sim$27~TeV. Therefore, these results suggest that they are essentially the same sources seen by different instruments. The new third component, HAWC~J1825$-$134, is located between HAWC~J1826$-$128 and HAWC~J1825$-$138, and surpisingly could not be detected at VHEs by the H.E.S.S. experiment. This new component is noteworthy because it emits UHE photons with energies up to 200~TeV, and its spectrum extends beyond 100~TeV without showing any indication of a spectral cutoff, having a power-law index of $\Gamma$~=~2.28~$\pm$~0.12$_{stat}$~$\pm$~0.10$_{sys}$ \citep{hawc_j1825}, as shown in Figures~\ref{jointj1825_SC}~(F) and \ref{jointj1825_SC}~(G). Furthermore, HAWC~J1825$-$134 and the LHAASO detected UHE source LHAASO~J1825$-$1326 are spatially coincident. This indicates that what LHAASO detects above 100~TeV is, in essence, this new third component, as the other two components have a sharp spectral cutoff at energies well below 100~TeV. One of the puzzling questions here is the non-detection of HAWC~J1825$-$134 by the H.E.S.S. experiment, which has in principle much better angular resolution with respect to HAWC. Furthermore, looking at the (yellow) source spectrum seen in Figure~\ref{jointj1825_SC}~(G), HAWC~J1825$-$134 has similar flux level as HESS~J1826$-$130, especially between 1$-$10~TeV, and the integral flux of HAWC~J1825$-$134 above 1~TeV is $\sim$2.4$\times$10$^{-12}$~cm$^{-2}$~s$^{-1}$, corresponding to $\sim$12$\%$ of the Crab Nebula flux, which is well within the range of H.E.S.S. sensitivity. It is also worth mentioning that HAWC~J1825$-$134 is also a point-like source for the HAWC experiment and the 95$\%$ CL upper limit on the extension is 0.18$^{\circ}$ \citep{hawc_j1825}, thus, degraded sensitivity due to the source extension cannot be the reason for the non-detection. Consequently, one expects that HESS should have detected this third component, especially after the extremely long exposure of more than 350 h \citep{j1825_134} taken on this complex region. One possible explanation is that HAWC~J1825$-$134 actually does not emit at E$<$10~TeV energies but only emerges above 10$-$20~TeV, and the E$<$10~TeV flux data seen in the (yellow) source spectrum in Figure~\ref{jointj1825_SC}~(G) can be the effect of the forward-folding method \citep{forward_folding} used in the analysis, which assumes a spectral shape, therefore could be misleading. The future observations of this region, especially with ground-based observatories effectively operating between 10~TeV  and 100~TeV with a superior angular resolution like CTA North and South or ASTRI Mini-Array \citep{astri_1,astri_2}, will shed light on the nature of this mysterious component, the very promising PeVatron candidate HAWC~J1825$-$134.

\subsection{HESS J1702$-$420: a promising dark PeVatron candidate at Southern sky}

\begin{figure}[ht!]
\centering
\includegraphics[width=15cm]{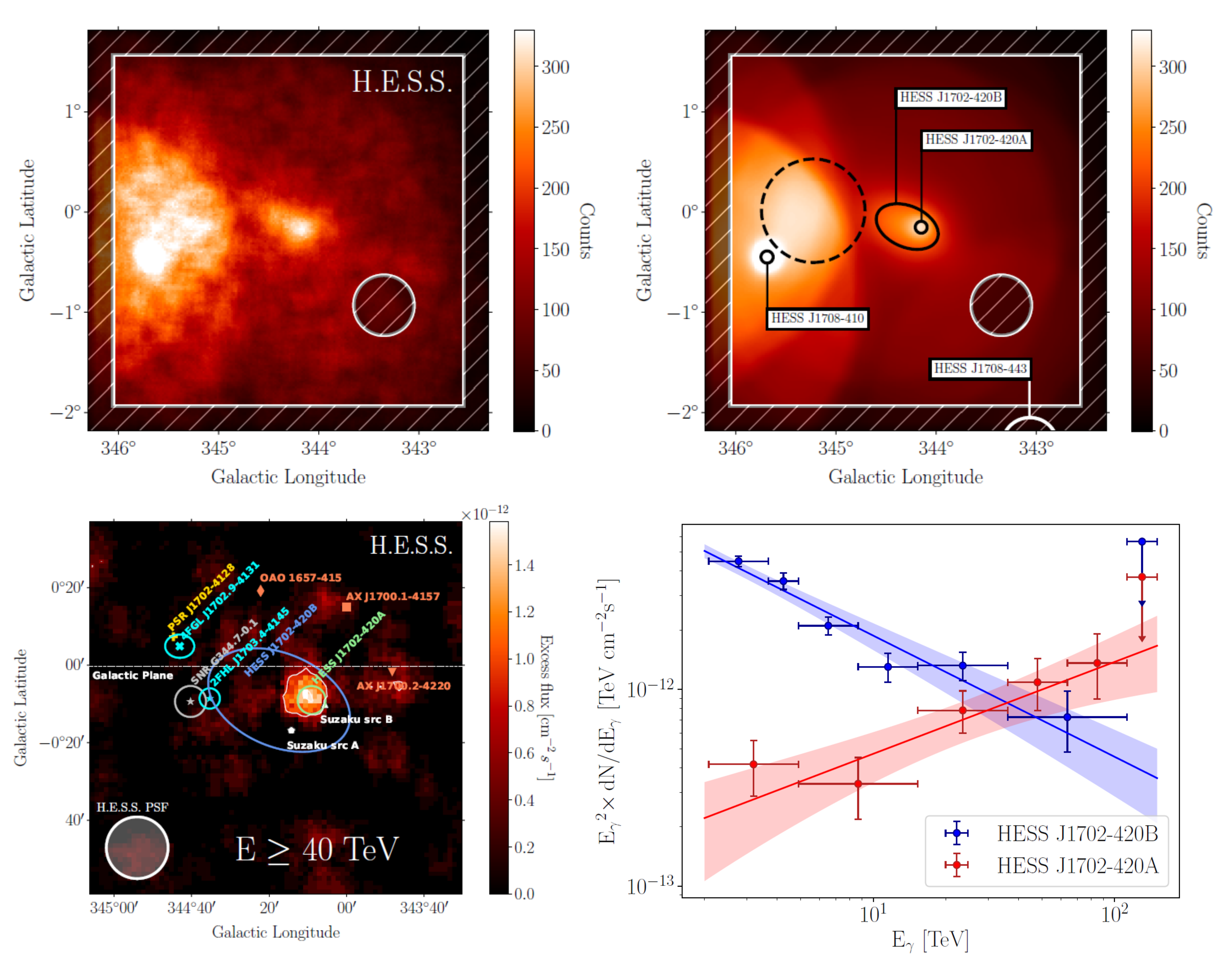}
\caption{Top panel: Left figure is the count map around HESS~J1702$-$420 region at energies above 2 TeV. The bright area around $\textit{l}\geq$345$^{\circ}$ results from deep observations of RX~J1713$-$3946 and HESS~J1708$-$410. Right figure: energy-integrated (E$>$2~TeV) map of model-predicted counts, with names and 1$\sigma$ shapes of all model components overlaid. The large-scale discarded component is indicated by the dashed circle. \\
Bottom panel: Left figure is the $\gamma$-ray flux map of the HESS~J1702$-$420 region, computed with the ring background method, above 40~TeV. The color code is in unit of $\gamma$-ray flux (cm$^{-2}$ s$^{-1}$) per smoothing area. The white contours indicate the 3$\sigma$ and 5$\sigma$ H.E.S.S. significance levels. Right figure: power-law spectra of HESS~J1702$-$420A (red solid line) and HESS~J1702$-$420B (blue solid line). The butterflies show the 1$\sigma$ statistical uncertainty on the spectral shapes. In the energy bins with less than 3$\sigma$ excess significance, the 3$\sigma$ confidence level upper limits are shown. The figures and captions are taken from \citep{j1702_420}.}
\label{j1702_SC}
\end{figure}

The unidentified VHE $\gamma$-ray source HESS~J1702$-$420 has been known for a long time since 2006 \citep{J1702_A,J1702_B}. The source was detected at 13$\sigma$ significance level due to its relatively bright flux level at 1~TeV of $\Phi_{0}$~=~9.1$\pm$1.1)$\times$10$^{-12}$~cm$^{-2}$~s$^{-1}$~TeV$^{-1}$ and hard spectrum with power-law index of $\Gamma$~=~2.07~$\pm$~0.08 \citep{J1702_B}, although it was observed for a relatively short total exposure of only $\sim$7 h. The morphology of the source was determined to be extended with a stretched Gaussian shape that had intrinsic semimajor and semiminor axes of 0.30$^{\circ}$ $\pm$ 0.02$^{\circ}$ and 0.15$^{\circ}$ $\pm$ 0.01$^{\circ}$, respectively. Although the source was promising from the perspective of PeVatron searches due to its hard VHE spectrum, no further H.E.S.S. observations were conducted on it until 2017. From 2017 to 2019, the source was observed mostly under large zenith angle observations between 37.5$^{\circ}$ and 59.1$^{\circ}$, which are known to increase the effective area \citep{magic_LZA_crab} and provide increased high energy statistics. The new data was combined with previous observations and analyzed using 3D likelihood analysis \citep{Mohrmann_2019}, which is known to be a more sensitive analysis method compared to traditional methods like the ring and reflected background estimation \citep{vhe_bkg}. The 3D analysis of this region led to remarkable results. The analysis was able to separate and distinguish two distinct components that were situated on top of each other as illustrated in the top right panel of Figure~\ref{j1702_SC}. The first component, HESS~J1702$-$420A, is a point-like source exhibiting one of the hardest spectrum observed at VHEs with a power-law index of $\Gamma$ = 1.53 $\pm$ 0.19$_{stat}$ $\pm$ 0.20$_{sys}$. The source spectrum extends up to 100~TeV without any clear indication of a spectral cutoff, making this source one of the most promising PeVatron candidates in the Southern sky \citep{j1702_420}. The excess flux map of HESS~J1702$-$420 region above 40~TeV is given in the bottom left panel of Figure~\ref{j1702_SC}. There are dense molecular clouds found along the line of sight, suggesting a plausible hadronic origin of the emission \citep{j1702_MC}. Assuming an hadronic origin, the 95$\%$ lower limit on the proton spectral cutoff is estimated to be $\sim$0.88~PeV by the authors \citep{j1702_420}. The other less interesting component HESS~J1702$-$420B displays a quite soft power-law spectrum with an index of $\Gamma$~=~2.62~$\pm$~0.10$_{stat}$~$\pm$~0.20$_{sys}$ exhibiting an extended morphology with an elongated Gaussian shape. The comparison between the spectra of detected A and B components is given in the bottom right panel of Figure~\ref{j1702_SC}. As it can be seen from the figure, the (red) spectrum of HESS~J1702$-$420A is extremely promising and exhibits spectral signatures of PeVatron sources. Nevertheless, UHE data above 100~TeV are required in order to understand the nature of this unidentified source. Especially, observations of HESS~J1702$-$420 region with particle arrays, like planned SWGO in the Southern hemisphere, will shed light on the PeVatron nature of this source.

\subsection{The sky-region including the Boomerang PWN and SNR G106.3+2.7: the power of high angular resolution observations in the hunt for Galactic PeVatrons}

\begin{figure}[ht!]
\centering
\includegraphics[width=15cm]{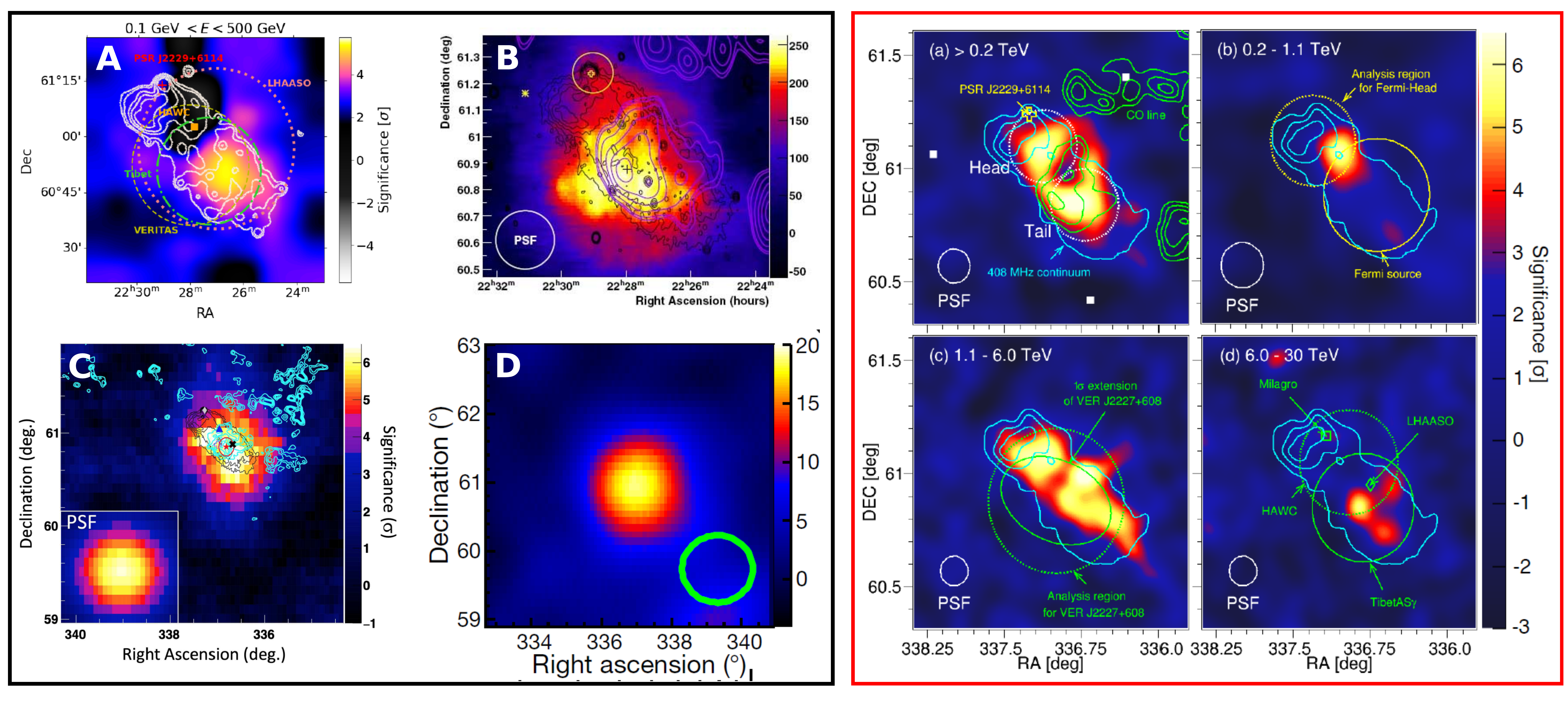}
\includegraphics[width=15cm]{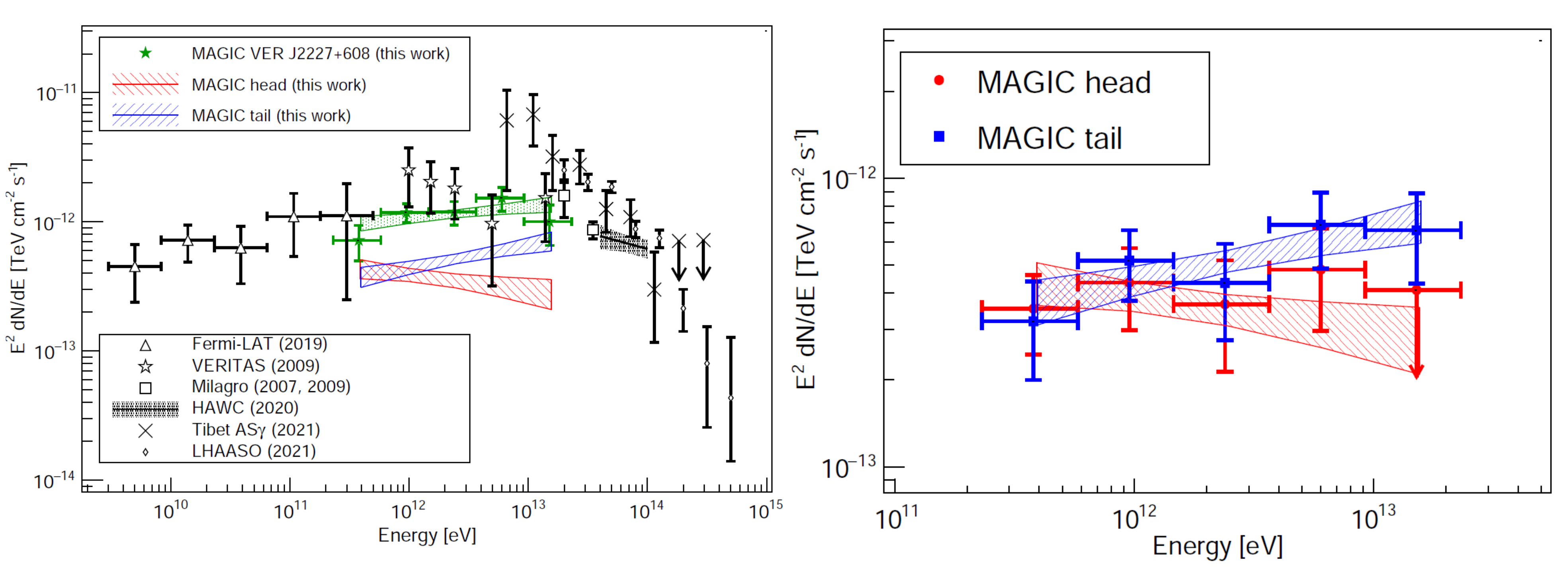}     \caption{Top panel: Left figure enclosed in black rectangle is the skymaps of the region including SNR~G106.3$+$2.7 and the Boomerang PWN. (A) Significance map obtained from Fermi-LAT observations \citep{fermi_j2227}, (B) excess map obtained from VERITAS observations \citep{veritas_j2227}, (C) significance map obtained from Tibet AS$\gamma$ observations \citep{tibet_boomerang}, (D) significance map obtained from LHAASO observations LHAASO\citep{lhaaso2021}. Right figure enclosed in red rectangle: energy-dependent significance maps of SNR~G106.3$+$2.7 observed with the MAGIC telescopes \citep{magic_boomerang}. (a) Map above 0.2~TeV. The PSF is 0.075$^{\circ}$. The white dotted circles show $\theta^{2}$ cut regions of the head and tail regions, respectively. (b) Map at 0.2$-$1.1 TeV. The PSF is 0.1$^{\circ}$. (c) Map at 1.1–6.0 TeV. The PSF is 0.065$^{\circ}$. (d) Map at 6.0$-$30 TeV. The PSF is 0.065$^{\circ}$.\\
Bottom panel: Left figure is the spectral energy distribution of the whole region of SNR~G106.3$+$2.7. Green data represent the spectrum of the VER~J2227$+$608 region as measured with the MAGIC telescopes. The zoom to shaded blue and red regions is shown in the left. The Fermi-LAT \citep{fermi_j2227}, VERITAS \citep{veritas_j2227}, Milagro \citep{mgro_j2227A,mgro_j2227B}, Tibet AS$\gamma$ \citep{tibet_boomerang}, and LHAASO measurements \citep{lhaaso2021} are shown. Right figure is the energy spectra of the head and tail regions. Red and blue data represent the spectra of the head and tail, respectively. The color 'bow-tie' areas show the result of fitting with a simple power-law function and 1$\sigma$ statistical uncertainties. The top right figure, bottom panel figures, and corresponding captions are taken from \citep{magic_boomerang}.}
\label{boomerang_joint}
\end{figure}

The UHE $\gamma$-ray source, LHAASO~J2226$+$6057, with the maximum detected photon energy of 570~TeV, is located in a very complex region of the Galactic plane \citep{lhaaso2021}. This region encompasses a SNR, SNR~G106.3$+$2.7, that has been detected at radio wavelengths. The SNR has a comet-shaped radio morphology, with a bright circular "head" region and a dimmer "tail" region. In addition, there is a PWN, which is powered by the energetic pulsar PSR~J2229$+$6114, situated at the north of the head region. This PWN is named as 'the Boomerang PWN', and the pulsar that powers it is a significant source of energy with its spin-down luminosity of $\dot{E}$=2.2$\times$10$^{37}$~erg~s$^{-1}$ and characteristic age of $\tau_{c}$=10 kyr. This intriguing complex region of the Galactic plane has been observed by Fermi$-$LAT \citep{fermi_j2227}, VERITAS \citep{veritas_j2227}, Milagro \citep{mgro_j2227A,mgro_j2227B}, Tibet AS$\gamma$ \citep{tibet_boomerang} and LHAASO \citep{lhaaso2021} experiments. The $\gamma$-ray skymaps of this region, reconstructed independently from Fermi$-$LAT, VERITAS, Tibet~AS$\gamma$ and LHAASO observations, are given in the top left panel of Figure~\ref{boomerang_joint}. As it can be seen from the skymaps, observation with all these instruments lead to the detection of an extended $\gamma$-ray emission, which is seen as a single component. Consequently, it was not clear whether the SNR G106.3$+$2.7, the Boomerang PWN or both at the same time were the source of the observed $\gamma$-ray emission. 

The MAGIC observations of this complex region with their superior angular resolution, reaching to 0.065$^{\circ}$ above 1.1~TeV energies \citep{magic_boomerang} could discriminate two different subcomponents, associated to 'head' and 'tail' regions, from each other. The confusion between 'head' and 'tail' components could be clearly resolved in an energy-dependent analysis. As it can be seen from the top right panel of Figure~\ref{boomerang_joint}, the $\gamma$-ray emission looks like a single blob-like component above 0.2~TeV energies. Investigating the emission in different energy bands shows that the 'head' component dominates the low energy part of the emission between 0.2~TeV and 1.1~TeV, while the 'tail' component becomes prominent above the energies of $\sim$6~TeV. In between 1.1~TeV and 6.0~TeV, both components produce detectable $\gamma$-ray emission that are confused and appear as a blob like single extended component (see inset C of Figure~\ref{boomerang_joint} right) \citep{magic_boomerang}. The resolved spectra of the 'head' and 'tail' components are shown at the bottom panel of Figure~\ref{boomerang_joint}. The left figure shows the comparison of spectra reconstructed from MAGIC observations with the previous spectral measurements obtained with different instruments, while the right figure zooms only on the 'head' and 'tail' spectra. Both components exhibit power-law spectra without showing any signs of a cutoff. The 'head' component has a relatively soft spectral index of $\Gamma_{\text{HEAD}}$=2.12$\pm$ 0.12$_{stat}$ $\pm$ 0.15$_{sys}$, while the spectral index is harder ($\Gamma_{\text{TAIL}}$=1.83$\pm$ 0.10$_{stat}$ $\pm$ 0.15$_{sys}$) for the tail component, and makes it a promising PeVatron candidate source \citep{magic_boomerang}. These results suggest the interpretation that the UHE $\gamma$-ray emission seen by the LHAASO experiment is actually connected to the 'tail' component of the SNR G106.3$+$2.7. The future observations of this complex region with instruments providing both high angular resolution and sensitivity, especially above 10~TeV, like CTA North or ASTRI Mini-Array will help understanding the PeVatron nature of the 'tail' region.

\section{Summary and Conclusion}
\label{conclusions}

This review article presents an overview of the search for Galactic PeVatrons through ground-based $\gamma$-ray observations. Recent advancements in $\gamma$-ray astronomy have provided evidence for the presence of PeVatrons in our Galaxy. Specifically, this review article offers clear definitions for hadronic and leptonic PeVatrons as astrophysical objects capable of accelerating 'individual' hadrons and leptons to at least 1 PeV energies and beyond, respectively. Meanwhile, "CR knee PeVatrons" are defined as astrophysical objects capable of accelerating 'bulk of hadrons' at least up to the knee energies observed in the CR spectrum. In this regard, the quest for the origin of Galactic CRs is directly linked to CR knee PeVatrons. A preliminary estimate of the minimum effective area required for future particle arrays to detect at least one PeV photon after one year of observation time is provided in this article. Based on the published LHAASO spectra of three UHE sources (LHAASO~J2226$+$6057, LHAASO~J1908$+$0621 and LHAASO~J1825$-$1326), the minimum effective area required is estimated to be $\sim$4$-$5~km$^{2}$. Additionally, it was shown that a preliminary estimate of the integral diffuse $\gamma$-ray flux above 100 TeV of F$_{\gamma, \text{diff}}$($>$100~TeV) = (2.48 $\pm$ 0.91)$\times$10$^{-13}$ cm$^{-2}$s$^{-1}$sr$^{-1}$, which is derived from Tibet AS$\gamma$ observations, can be used to estimate the background event counts obtained from 12 LHAASO UHE sources with an accuracy better than 10$\%$. The article also reviews the spectral signatures of hadronic PeVatrons resulting from pp interactions followed by subsequent $\pi^{0}$ decay, presenting a 2D plot that illustrates the relationship between the parent proton spectrum parameters ($\beta_{p}$ and $\Gamma_{p}$) and the resulting $\gamma$-ray spectral parameters ($E_\mathrm{cut,\,\gamma}$/$E_\mathrm{cut,\,p}$ and $\beta_{\gamma}$). The available analysis methods used to search for PeVatron signatures are discussed, as well as the effects of $\gamma\gamma$ absorption in the search for PeVatrons, particularly for UHE sources that are located at far distances within our Galaxy. A detailed discussion on the issue of source confusion in the search for PeVatrons is provided, focusing on four complex regions in the Galactic plane: HESS~J1640$-$465~/~HESS~J1641$-$463, HESS~J1825$-$137~/~HESS~J1826$-$130, HESS~J1702$-$420 and Boomerang~PWN~/~SNR~G106.3$+$2.7 regions. The importance of high angular resolution observations in the search for PeVatrons is highlighted, with examples of the effectiveness of energy-dependent morphology analysis. In conclusion, the quest for the origin of Galactic CRs is an ongoing pursuit and will remain a major focus in the field of Galactic $\gamma$-ray astronomy in the upcoming years. To achieve breakthroughs, synergies between VHE and UHE $\gamma$-ray astronomy, as well as joint data analyses from IACTs and particle arrays will play a crucial role for future investigations. By utilizing the superior angular and energy resolutions provided by IACTs in conjunction with the unprecedented sensitivities of particle arrays above 100 TeV, it can be potentially possible to resolve this long-standing enigma and shed light on the origin of Galactic cosmic rays in the next decade.

\section*{Acknowledgment}
E.O.~Angüner acknowledges financial support by TÜBİTAK Research Institute for Fundamental Sciences. I express my sincere gratitude to Elena Amato and Heide Costantini for their useful and constructive comments.

\newpage

\bibliographystyle{ieeetr}
\bibliography{PeVReviewTJP_Arxiv}

\end{document}